\documentclass[vecphys]{svmult}

\usepackage{makeidx}
\usepackage{graphicx}
\usepackage{multicol}
\usepackage{cite}
\usepackage[bottom]{footmisc}

\usepackage{amsmath}
\usepackage{amssymb}
\usepackage{url}

\makeindex

\begin{document}

\title{The Composite Operator Method (COM)}
\author{Adolfo Avella\inst{1,2,3,4} \and Ferdinando Mancini\inst{1,3,5}}
\authorrunning{Adolfo Avella and Ferdinando Mancini}
\institute{Dipartimento di Fisica ``E.R. Caianiello'', Universit\`a degli Studi di Salerno, 84084 Fisciano (SA), Italy \and 
CNR-SPIN, UoS di Salerno, 84084 Fisciano (SA), Italy \and Unit\`a CNISM di Salerno, Universit\`a degli Studi di Salerno, 84084 Fisciano (SA), Italy \and \texttt{avella@physics.unisa.it} \and \texttt{mancini@physics.unisa.it}}
\maketitle

\begin{abstract}
The Composite Operator Method (COM) is formulated, its internals illustrated in detail and some of its most successful applications reported. COM endorses the emergence, in strongly correlated systems (SCS), of composite operators, optimally deals with their unusual features and implements algebra constraints, and other relevant symmetries, in order to properly compute the unconventional properties of SCS.
\end{abstract}

\section{Strong Correlations and Composite Operators\label{sec:Correlations}}
In the last decades, a large part of the research activity in solid state and condensed matter physics has been devoted to the study of electronic systems with very unconventional properties (i.e. with properties not consistent with the Fermi liquid theory). It is commonly believed that the origin of such anomalous behaviors should be traced back to the presence of strong electronic correlations in such systems.

\textbf{Weakly correlated\index{system!weakly correlated}} systems are described by Hamiltonians where the multi-body terms, and the interactions they describe, feature so little coupling constants with respect to the average energies of the one-body terms (e.g. wide-band $s$ and $p$ metals) that we can successfully resort to perturbation theories. In particular, in such cases, mean-field-like approximations allow to diagonalize the Hamiltonian in terms of new independent particles, described by \textbf{canonical\index{operator!canonical} operators}, that is, operators bounded to satisfy canonical commutation relations. On the contrary, in \textbf{strongly correlated\index{system!strongly correlated}} systems (e.g. narrow-band $d$ and $f$ metals) \cite{Hubbard_63,Anderson_72}, the interactions are sufficiently intense to make completely useless any perturbation scheme that will be just bound to fail. In some relevant cases, even more than the strength of the interactions, it is their true nature, that is, their actual operatorial form, to bring all troubles (e.g. Kondo model \cite{Hewson_97} ).

In order to tackle this problem, we need to change perspective and relax some of the constraints implied by the quest for new independent particles. In particular, we should not expect to be able to correctly describe strongly correlated systems only by means of canonical operators. Strong interactions modify dramatically the properties of the original particles. What are observed are new particles with new peculiar properties entirely determined by the dynamics and by the boundary conditions (i.e. all elements characterizing the physical situation under study). These new objects appear as the final result of the modifications imposed by the interactions on the original particles and embed, since the very beginning, the effects of correlations. Thus, the choice of new fundamental operators, whose properties will be self-consistently determined by dynamics, symmetries and boundary conditions, becomes relevant.

Let us consider a very simple, but very pedagogical, system in order to concretely illustrate this occurrence. A lattice of ions sited sufficiently far apart to avoid that the electrons can hop from one to another. For the sake of simplicity, we can also imagine that only one $s$-like orbital is active per ion and that the Coulomb repulsion is sufficiently intense only between same-site, that is, same-orbital, electrons:
\begin{equation}
H = - \mu \sum_{\mathbf{i}\sigma} c^\dagger_\sigma(i) c_\sigma(i) + U \sum_\mathbf{i} n_\uparrow(i) n_\downarrow(i)\;,
\label{eq:atomicHam}
\end{equation}
where $\mu$ is the chemical potential, $\mathbf{i}$ is a vector of the lattice, $c_\sigma(i)$ is the destruction operator in the Heisenberg picture ($i=(\mathbf{i},t)$) of an electron of spin $\sigma$ in a Wannier state centered at the site $\mathbf{i}$, $U$ is the strength of the on-site Coulomb repulsion and $n_\sigma(i) = c^\dagger_\sigma(i) c_\sigma(i)$ is the electronic number operator.

Although the Hamiltonian (\ref{eq:atomicHam}) looks very simple, there is neither known canonical transformation capable to diagonalize this Hamiltonian nor perturbation scheme, at any order, getting any closer to the exact solution or even just grasping its main features. This is simply due to the impossibility to describe this system in terms of independent particles represented by canonical operators.

If we analyze the equation of motion of the electronic operator $c_\sigma(i)$
\begin{equation}
\I \frac{\partial}{\partial t} c_\sigma(i) = \left[ c_\sigma(i) , H \right] = - \mu   c_\sigma(i) + U   n_{\bar{\sigma}}(i) c_\sigma(i)\;,
\label{eq:EMc}
\end{equation}
we encounter a \textbf{fermionic\index{operator!fermionic} operator}, that is, an operator constituted by an odd number of electronic operators, describing an electron of spin $\sigma$ dressed by the charge fluctuations of the electrons of opposite spin ($\bar{\sigma}$). Let us name this new operator $\eta_\sigma(i) = n_{\bar{\sigma}}(i) c_\sigma(i)$. The equation of motion of $\eta$ reads as
\begin{equation}
\I \frac{\partial}{\partial t} \eta_\sigma(i) = \left[ \eta_\sigma(i) , H \right] = (U - \mu) \eta_\sigma(i) \;.
\label{eq:EMeta}
\end{equation}
It is immediate to see that there exists a companion of $\eta$, which we named $\xi_\sigma(i) = c_\sigma(i) - \eta_\sigma(i)$, whose equation of motion reads as
\begin{equation}
\I \frac{\partial}{\partial t} \xi_\sigma(i) = \left[ \xi_\sigma(i) , H \right] = - \mu   \xi_\sigma(i) \;.
\label{eq:EMxi}
\end{equation}
Equations (\ref{eq:EMeta}) and (\ref{eq:EMxi}) resemble that of independent particles as the \textbf{current\index{operator!current of}} (i.e. the r.h.s of the equation of motion) is directly proportional to the operator itself. We define as \textbf{\mbox{eigenoperator}\index{eigenoperator}\index{operator!eigenoperator|see{eigenoperator}}} of a given Hamiltonian, an operator whose current is just proportional to the operator itself. We will call \textbf{\mbox{eigenenergy}\index{eigenoperator!eigenenergy}\index{eigenenergy}} the proportionality constant between the eigenoperator and its current (e.g. $(U-\mu)$ for $\eta$).

$\xi$, together with $\eta$, constitutes a \textbf{fermionic closed basis\index{operator!basis!closed}} for this system. In our definition, a closed basis is the smallest set of operators necessary to build up the original operators as a linear combination (e.g. $c = \xi + \eta$) and which closes the equations of motion. That is, a set of operators $(O_1,O_2,\ldots,O_n)$ constitutes a closed basis if $\I \frac{\partial}{\partial t} O_p = \sum_q c_{pq} O_q$. 

\begin{table}[tbp]
\centering
\caption{Anti-commutation relations for $\xi$ and $\eta$}
\renewcommand{\arraystretch}{1.2}
\setlength\tabcolsep{5pt}
\begin{tabular}{@{}ll@{}}
$\{\xi^\dagger_\sigma , \xi_{\sigma'}\} = \delta_{\sigma\sigma'}(1-c^\dagger_{\bar{\sigma}}c_{\bar{\sigma}})+\delta_{\sigma\bar{\sigma}'}c^\dagger_\sigma c_{\bar{\sigma}} $ &
$[\xi^\dagger_\sigma \xi_{\sigma} , \xi_{\sigma'}] = - \delta_{\sigma\sigma'} \xi_{\sigma}$ \\
$\{\eta^\dagger_\sigma, \eta_{\sigma'}\}=\delta_{\sigma\sigma'}c^\dagger_{\bar{\sigma}}c_{\bar{\sigma}}-\delta_{\sigma\bar{\sigma}'}c^\dagger_\sigma c_{\bar{\sigma}}$ &
$[\xi^\dagger_\sigma \xi_{\sigma}, \eta_{\sigma'}] =\delta_{\sigma\bar{\sigma}'} \eta_{\bar{\sigma}}$ \\
$\{\xi_\sigma,\eta_{\sigma'}\}=\delta_{\sigma\bar{\sigma}'}c_\sigma c_{\bar{\sigma}}$ &
$[\eta^\dagger_\sigma \eta_{\sigma}, \eta_{\sigma'}] = -\eta_{\sigma'}$ \\
$\{\xi_\sigma , \xi_{\sigma'}\} = \{\eta_\sigma , \eta_{\sigma'}\} = \{\xi^\dagger_\sigma , \eta_{\sigma'}\} = 0$ &
$[\eta^\dagger_\sigma \eta_{\sigma}, \xi_{\sigma'}] = 0$
\end{tabular}
\label{tab:comm_eta_xi}
\end{table}

Often, by means of a fermionic closed basis, it is possible to \textbf{diagonalize} the related Hamiltonian, that is, to rewrite the Hamiltonian only in terms of number-like operators of elements of the operatorial basis. For instance, the Hamiltonian (\ref{eq:atomicHam}) can be rewritten as
\begin{equation}
H = - \mu \sum_{\mathbf{i}\sigma} \xi^\dagger_\sigma(i) \xi_\sigma(i) + \left(\frac{1}{2} U - \mu\right) \sum_{i\sigma} \eta^\dagger_\sigma(i) \eta_\sigma(i)\;.
\label{eq:atomicHamII}
\end{equation}
This diagonalization, although exact, is formal as $\xi$ and $\eta$ are not canonical operators and they close commutation relations with their \textbf{number-like\index{operator!number-like} operators} $\xi^\dagger \xi$ and $\eta^\dagger \eta$ very different from that closed by the electronic operator $c$ with $n=c^\dagger c$ (see Table~\ref{tab:comm_eta_xi}; site indexes have been neglected: they would simply lead to Kronecker deltas).

Equation (\ref{eq:atomicHamII}) shows that, owing to the presence of the interactions, the original electrons $c$ are no more observables and new stable elementary excitations, described by the operators $\xi$ and $\eta$, appear. The electronic bare energy level $E = -\mu$ splits into two levels ($E_1 = - \mu$ and $E_2 = U - \mu$) and the original electrons turn out to be exactly the worst place to start: no perturbative scheme leads to the level splitting, which is the main and only feature of this very simple system. On the basis of this evidence, one is induced to move the attention from the original electronic operator to the new operators generated by the interaction. Such new operators can be written in terms of the original ones and are known as \textbf{composite\index{operator!composite} operators}. A formulation which would treat composite operators as fundamental objects, instead of the original electronic operators, would be very promising. As a matter of fact, such a formulation would allow to set up approximation schemes where some amount of the interaction is already exactly taken into account through the chosen operatorial basis, and would permit to overcome the hoary problem of finding an appropriate expansion parameter.

However, one price must be paid. In general, composite operators are neither Fermi nor Bose operators, since they do not satisfy canonical (anti)commutation relations (see, for instance, Table~\ref{tab:comm_eta_xi}). This very simple fact makes a tremendous difference with respect to the case in which one deals with the original electronic operators, which satisfy a canonical algebra. In developing perturbation schemes where the building blocks are propagators of composite operators, one cannot use any more the consolidated scheme: diagrammatic expansions, Wick's theorem and many other standard tools are no more valid or applicable. The formulation of the whole Green's function method must be revisited \cite{Mancini_00,Mancini_04} and new frameworks of calculations have to be formulated. In the following section, we will formulate the Composite Operator Method (COM) \cite{Mancini_04} and illustrate its internals; COM systematically endorses the emergence, in strongly correlated systems, of composite operators and optimally deals with their unusual features.

\section{The Composite Operator Method (COM)\label{sec:COM}}

\subsection{Basis $\psi$\label{sec:Basis}}

Given a Hamiltonian $H$ describing a solid state system, first we have to choose a \textbf{basis\index{operator!basis}} $\psi$ constituted of a finite number of composite operators, either fermionic or bosonic. A fermionic basis will be used if the original interacting particles are electrons and we want to analyse thermodynamic and single-particle properties of the system. A bosonic basis can be either the \emph{natural} basis for systems constituted by bosons (e.g. spins, phonons, \ldots) or a set of bosonic operators, constituted by an even number of electronic operators, representing the electronic number operator (charge), the electronic spin operator, \ldots and necessary to study the charge, spin, \ldots response of the system. In this latter case, the dynamics described by the bosonic basis is strongly coupled to that described by the related fermionic basis and viceversa. As a matter of fact, this is the case even if we do not explicitly open the bosonic \emph{sector} as the self-consistent parameters appearing in the fermionic sector are just thermal averages of the related bosonic operators. A fully simultaneous self-consistent solution of both sectors is usually required. This consequence of the non-canonical algebra satisfied by composite operators is a manifestation of the relevance (actually, the dominance) of spin, charge, \ldots correlations in all properties exhibited by strongly correlated systems.

Obviously, a closed basis is the best basis one can adopt. It is possible to obtain a closed basis just crossing the entire hierarchy of the equations of motion of the relevant canonical operator (e.g. the electronic operator, its number or spin operator, \ldots). Namely, one has to: (i) compute high-order time derivatives of the relevant canonical operator by repeatedly commuting it with the Hamiltonian; (ii) acquire in the basis all distinct composite operators appearing in the high-order currents; (iii) terminate as soon as no new composite operator appears. This well-defined procedure, being equivalent to compute all spectral moments (i.e. the moments of the relevant canonical operator Green's function) \cite{Kalashnikov_69,Nolting_72,Mancini_98b}, catches, in principle, all scales of energy featured by the system under analysis.

Following this procedure, one can see that there is a large class of systems (finite systems \cite{Avella_01,Mancini_00}, bulk systems with interacting localized electrons \cite{Mancini_05,Mancini_08,Mancini_10}, Ising-like systems \cite{Mancini_05b,Mancini_06a,Avella_05,Mancini_08a}, \ldots) where it is possible to obtain a closed basis with a finite number of elements and one can aim at the exact solution. In all other bulk systems, this procedure will lead to a closed basis with an infinite number of components. In such cases, one can: rewrite $H$ as $H=H_0+H_I$, where $H_0$ is the most relevant part whose finite closed basis is known; adopt the closed basis of $H_0$ as a truncated basis for $H$; treat $H_I$ as a \emph{perturbation} to $H_0$. If the system under study features only analytical scales of energy (i.e. scales of energy whose mathematical expressions are expandable in power series of Hamiltonian model parameters), or if the range of temperatures/frequencies of interest does not involve not-analytical scales of energy, the adoption of a truncated basis can lead to an accurate description of the system under analysis. Such a description can be systematically improved adding to the truncated basis more and more elements according to the above described procedure. For instance, in the Hubbard model, we need at least a $2$-element basis to correctly describe the Hubbard on-site Coulomb repulsion $U$ and at least a $4$-element basis to correctly describe the virtual exchange energy $J=4t^2/U$ \cite{Avella_01,Odashima_05}.

Not analytical scales of energy can be present in some models. For example, in the single-impurity Anderson model \cite{Hewson_97}, the emergence of a not-analytical Kondo energy scale is directly connected to the virtual exchange processes triggered by the hybridization between conduction and impurity electrons and dominated by on-site Coulomb repulsion at the impurity site. This occurrence forces to explore different routes to construct a proper operatorial basis. In fact, in such cases, the adoption of any truncated basis would simply lead to a description of the system under analysis completely lacking any reference to not-analytical scales of energy. Such a description would turn out really poor at low enough temperatures/frequencies where neglected energy scales usually manifest themselves triggering quite unconventional behaviours (impurity spin screening, superconductivity, \ldots). Actually, in such cases, one has to properly tweak the equations of motion of specific composite operators and smartly exploit self-consistency in order to get sensible results by means of a basis with a finite number of elements \cite{Villani_00,Avella_02}.

Summarizing, many recipes can assure a correct and controlled description of relevant aspects of the dynamics. One can put in the basis:
\begin{itemize}
\item the higher-order composite operators emerging from the hierarchy of the equations of motion: the conservation of a definite number of spectral moments is guaranteed; polynomial and analytical scales of energy can be resolved with desired accuracy \cite{Mancini_98b,Mancini_04}
\item the eigenoperators of Hamiltonian terms describing specific interactions or virtual processes: the latter are correctly taken into account \cite{Avella_01,Avella_02e}
\item the eigenoperators of the problem reduced to a small cluster: all polynomial and analytical energy scales and their competitions/interplays are just exactly taken into account at short distances/large momenta \cite{Avella_02a,Bak_02,Avella_01,Avella_03b}
\item the composite operator describing the Kondo-like singlet \cite{Villani_00,Avella_02d,Avella_02f,Avella_02}
\end{itemize}

\subsection{Equations of motion of $\psi$\label{sec:EM}}

Once the basis $\psi(i)$, constituted of $n$ composite operators, has been chosen, we adopt a matricial notation and write
\begin{equation}
\psi(i)=\left(
\begin{array}{c}
\psi_1(i) \\
\vdots \\
\psi_n(i)
\end{array}
\right)\;,
\label{eq:psi}
\end{equation}
where we have not specified the nature, fermionic or bosonic, of the basis. In case of fermionic operators, it is understood that we use the spinorial representation
\begin{equation}
\psi_m(i)=\left(
\begin{array}{c}
\psi_{m\uparrow}(i) \\
\psi_{m\downarrow}(i)
\end{array}
\right)\;.
\label{eq:psi_spinor}
\end{equation}

The current $J$ of $\psi$ can be exactly rewritten as
\begin{equation}
J(i) = \I \frac{\partial}{\partial t} \psi(i) = \left[ \psi(i) , H \right] = \sum_\mathbf{j} \varepsilon(\mathbf{i},\mathbf{j}) \psi(\mathbf{j},t) + \delta J(i)\;,
\label{eq:EMpsi}
\end{equation}
where the matrix $\varepsilon$ is known as the \textbf{energy matrix\index{energy matrix}}. $\delta J$ is named \textbf{residual current\index{operator!residual current of}} and describes the component of the current $J$ \emph{orthogonal} to the basis $\psi$
\begin{equation}
\left\langle \left[ \delta J(\mathbf{i},t), \psi^\dagger(\mathbf{j},t) \right]_\eta \right\rangle = 0 \Rightarrow \varepsilon(\mathbf{i},\mathbf{j}) = \sum_\mathbf{l} m(\mathbf{i},\mathbf{l}) I^{-1}(\mathbf{l},\mathbf{j})\;.
\label{eq:dJ}
\end{equation}
It is absolutely worth noticing that iff $\psi$ is a closed basis, $\delta J$ is identically zero.

$m(\mathbf{i},\mathbf{j}) = \left\langle \left[ J(\mathbf{i},t), \psi^\dagger(\mathbf{j},t) \right]_\eta \right\rangle$ is simply named $\mathbf{m}$\textbf{-matrix\index{$m$-matrix}} and $I(\mathbf{i},\mathbf{j}) = \left\langle \left[ \psi(\mathbf{i},t), \psi^\dagger(\mathbf{j},t) \right]_\eta \right\rangle$ is known as the \textbf{normalization matrix\index{normalization matrix}}. The matrix $I$ is hermitian by definition, while the matrix $m$ is hermitian because of the time independence of the matrix $I$: $0 = \I \frac{\partial}{\partial t} I = m - m^\dagger$. $\left\langle \cdots \right\rangle $ stands for the quantum mechanical average in the (grand-)canonical ensemble. $\eta=1,-1$ marks anti-commutators $\lbrace , \rbrace $ and commutators $[ , ]$, respectively. For fermionic (bosonic) composite operators, the choice $\eta=1$ ($\eta=-1$) is much more convenient (i.e. it makes the calculations much simpler), although the opposite choice is also possible.

Once a basis has been chosen, either closed or truncated, we need to calculate the normalization $I$ and the $m$ matrices in order to construct the energy matrix $\varepsilon$, which will soon turn out to be a fundamental quantity. The calculation of the relevant (anti)commutators is straightforward, but it can be lengthy and cumbersome if the basis contain many elements. The thermal averaging procedure is instead much less mechanical as it involves choosing the various phases to be studied. All correlation functions present in the normalization $I$ and in the energy $\varepsilon$ matrices need to be computed self-consistently and form a first block of unknowns in the theory.

\subsubsection{Weight and orthogonality of composite operators\label{sec:I}}

The entries of the normalization matrix $I$ give a measure of both the \emph{weight\index{operator!weight}} of a composite operator (diagonal entries) and the degree of \emph{orthogonality\index{operator!orthogonality}} between two of them (off-diagonal entries). These two concepts (weight and orthogonality) have obvious meanings for canonical electronic operators whose anti-commutation relations are just $\mathbb{C}$-numbers ($\lbrace c_l, c_{l'}^\dagger \rbrace = \delta_{ll'}$): an electronic operator always \emph{weights} just $1$ and is \emph{orthogonal} by definition to any other electronic operator, clearly describing a single-particle state with different quantum numbers. As for composite operators, owing to the non-canonical commutation relations they obey, these concepts become more relevant and less obvious: both properties strongly depend on the actual operatorial form of the composite operators as well as, through the thermal averaging, on the Hamiltonian describing the system under analysis and the boundary conditions. Such a dependence manifests itself through the emergence, in the $I$ matrix entries, of thermal averages of bosonic composite operators, whose determination is not always straightforward. Actually, it is just the presence of these parameters in the $I$ matrix, to leave sometimes a problem unsolved although a closed basis has been individuated. If the basis is not closed ($\delta J \neq 0$), the same kind of averages appears in the $\varepsilon$ matrix too.

These features of composite operators (non-trivial weights and imperfect orthogonality) are distinctive landmarks of strong correlations. They are fundamental to the occurrence of well-known effects such as the transfer of spectral weight (directly connected to composite operator weights) between (sub-)bands on changing model or external (temperature, filling, pressure, external fields, \ldots) parameters, and the interplay among scales of energies (connected to the \emph{overlapping} of composite operators).

For instance, if we choose $\psi=(\xi,\eta)$ as (closed) fermionic basis for the toy model (\ref{eq:atomicHam}), we would find the following $I$ and $\varepsilon$ matrices:
\begin{equation}
\begin{split}
& I_{\sigma\sigma'}(\mathbf{i},\mathbf{j}) = \delta_\mathbf{ij} \left[ \delta_{\sigma\sigma'}\left( 
\begin{array}{cc}
1-\langle n_{\bar{\sigma}}(\mathbf{i}) \rangle & 0 \\
0 & \langle n_{\bar{\sigma}}(\mathbf{i}) \rangle
\end{array}
\right) + \delta_{\sigma\bar{\sigma}'} \langle c^\dagger_\sigma(\mathbf{i}) c_{\bar{\sigma}}(\mathbf{i}) \rangle \left(
\begin{array}{cc}
1 & 0 \\
0 & -1
\end{array}
\right)
\right] \\
& \varepsilon_{\sigma\sigma'}(\mathbf{i},\mathbf{j}) = \delta_\mathbf{ij} \delta_{\sigma\sigma'} \left(
\begin{array}{cc}
-\mu & 0 \\ 
0 & U-\mu
\end{array}
\right)\;.
\label{eq:Ie}
\end{split}
\end{equation}
As the chosen basis is closed, the energy matrix is free of unknown parameters, while the normalization matrix contains anyway some of them to be computed self-consistently according to the chosen phases to be investigated (e.g. paramagnetic, ferromagnetic, Ne\'el, CDW, SDW, ...).

\subsection{Dyson equation for $G$\label{sec:Dyson}}

We can now define a generalized matricial Green's function $G$ for the basis $\psi$
\begin{equation}
G^\mathcal{Q}(i,j) = \left\langle \mathcal{Q}[ \psi(i) \psi^\dagger(j) ]_\eta \right\rangle = \left\lbrace
\begin{array}{ll}
\theta(t_i-t_j) \left\langle [\psi(i),\psi^\dagger(j)]_\eta \right\rangle & \mathrm{for} \quad \mathcal{Q} = \mathcal{R} \\
-\theta(t_j-t_i) \left\langle [\psi(i),\psi^\dagger(j)]_\eta \right\rangle & \mathrm{for} \quad \mathcal{Q} = \mathcal{A} \\
\begin{array}{l}
\theta(t_i-t_j) \left\langle \psi(i) \psi^\dagger(j) \right\rangle \\ 
- \eta \theta(t_j-t_i) \left\langle \psi^\dagger(j) \psi(i) \right\rangle
\end{array}
& \mathrm{for} \quad \mathcal{Q} = \mathcal{C} 
\end{array}
 \right.\;,
\label{eq:GQ}
\end{equation}
where $\mathcal{R}$, $\mathcal{A}$ and $\mathcal{C}$ stand for the usual retarded, advanced and causal operators, respectively.

For the sake of simplicity, we pick up a phase with spatial homogeneity and move to momentum and frequency spaces through Fourier transform. Then, according to (\ref{eq:EMpsi}), $G$ satisfies the following equation of motion
\begin{equation}
G^\mathcal{Q}(\mathbf{k},\omega) = G_0^\mathcal{Q}(\mathbf{k},\omega) + G_0^\mathcal{Q}(\mathbf{k},\omega) T^\mathcal{Q}(\mathbf{k},\omega) G_0^\mathcal{Q}(\mathbf{k},\omega)\;,
\label{eq:EMGkw}
\end{equation}
where $G_0^\mathcal{Q}(\mathbf{k},\omega)$, determined by the equation
\begin{equation}
\left[\omega - \varepsilon(\mathbf{k})\right]G_0^\mathcal{Q}(\mathbf{k},\omega) = I(\mathbf{k})\;,
\label{eq:G0}
\end{equation}
will play the role of fundamental building block and the scattering matrix $T^\mathcal{Q}(\mathbf{k},\omega)$ is defined as
\begin{equation}
T^\mathcal{Q}(\mathbf{k},\omega)= I^{-1}(\mathbf{k}) B^\mathcal{Q}(\mathbf{k},\omega) I^{-1}(\mathbf{k})\;,
\label{eq:T}
\end{equation}
with $B^\mathcal{Q}(i,j) = \left\langle \mathcal{Q}[\delta J(i) \delta J^\dagger(j)]_\eta \right\rangle$.

Now, we can introduce the irreducible self-energy $\Sigma^\mathcal{Q}(\mathbf{k},\omega)$, defined through the relation $T^\mathcal{Q}(\mathbf{k},\omega) G_0^\mathcal{Q}(\mathbf{k},\omega) = I^{-1}(\mathbf{k}) \Sigma^\mathcal{Q}(\mathbf{k},\omega) G^\mathcal{Q}(\mathbf{k},\omega)$, and get
\begin{equation}
G^\mathcal{Q}(\mathbf{k},\omega) = \frac{1}{\omega - \varepsilon(\mathbf{k}) - \Sigma^\mathcal{Q}(\mathbf{k},\omega)} I(\mathbf{k}) = \frac{1}{[G_0^\mathcal{Q}(\mathbf{k},\omega)]^{-1} - I^{-1}(\mathbf{k})\Sigma^\mathcal{Q}(\mathbf{k},\omega)}\;.
\label{eq:Dyson}
\end{equation}

Equation (\ref{eq:Dyson}) generalizes the well-known Dyson equation to the case of Green's functions of composite operators \cite{Mancini_00,Mancini_04}. Together with the apparent similarities, it is absolutely worth noting the striking differences:
\begin{itemize}
\item all ingredients are matrices either $n \times n$ (bosonic) or $2n \times 2n$ (fermionic);
\item the equivalent of the usual non-interacting Green's function, $G_0$, is, instead, fully interacting and, in principle, exact (if the basis $\psi$ is closed then we have $\delta J = 0$, consequently $\Sigma = 0$, and finally $G=G_0$);
\item the normalization matrix $I$ evidently assumes the role of overall weight of the matricial Green's function $G$ ($I$ is its $0^{th}$ moment: $\lim_{\omega \to \infty} G \rightsquigarrow I/\omega$);
\item the eigenvalues of the energy matrix $\varepsilon$ clearly play the fundamental role of poles, which will be more or less severely modified by the self-energy $\Sigma$.
\item the self-energy $\Sigma$ describes both the dynamics \emph{orthogonal} to that described by the chosen basis $\psi$, and the \emph{residual} interactions among the elements of the latter. Smarter is the choice of $\psi$, up to a closed basis, less relevant is the knowledge of $\Sigma$ to fully understand the system under study. Hence, we decided to rename the generalized irreducible self-energy $\Sigma$ as residual self-energy: it is just the irreducible propagator of the residual current $\delta J$.
\end{itemize}

\subsection{Propagator $G_0$\label{sec:G0}}

Let us take a step back, the propagator $G_0$ satisfies the following equation
\begin{equation}
\left[ \omega - \varepsilon(\mathbf{k})\right ] G_0^\mathcal{Q}(\mathbf{k},\omega) = I(\mathbf{k})\;,
\label{eq:EMG0}
\end{equation}
whose most general solution is \cite{Mancini_00,Mancini_04}
\begin{equation}
G_0^\mathcal{Q}(\mathbf{k},\omega) = \sum_{l=1}^n \left\lbrace \mathcal{P}\left[ \frac{\sigma^{(l)}(\mathbf{k})}{\omega-\omega_l(\mathbf{k})} \right] - \I \pi \delta\left[ \omega-\omega_l(\mathbf{k}) \right] g^{(l)\mathcal{Q}}(\mathbf{k}) \right\rbrace\;.
\label{eq:G0gen}
\end{equation}
$\sigma^{(l)}(\mathbf{k})$ are the spectral density functions in matrix form, fully determined by the energy matrix $\varepsilon$ and the normalization matrix $I$ as
\begin{equation}
\sigma^{(l)}_{ab}(\mathbf{k}) = \Omega_{al}(\mathbf{k}) \sum_c \Omega^{-1}_{lc}(\mathbf{k}) I_{cb}(\mathbf{k})\;,
\label{eq:sigma}
\end{equation}
where $\Omega(\mathbf{k})$ is the matrix whose columns are the eigenvectors of the energy matrix $\varepsilon$. $\omega_l(\mathbf{k})$ are the eigenvalues of this latter. $g^{(l)\mathcal{Q}}(\mathbf{k})$ are unknown momentum functions, in matrix form, not fixed by the equations of motion (i.e. they correspond to constants in time), to be determined taking into account the boundary conditions specific of the type of propagator under study (retarded, advanced or causal).

For both fermionic ($\eta=1$) and bosonic ($\eta=-1$) basis, the boundary conditions turn out to be sufficient to fully determine $g^{(l)\mathcal{R,A}}(\mathbf{k})$ and we obtain a Lindhard-like representation for $G_0^\mathcal{R,A}(\mathbf{k},\omega)$
\begin{equation}
G_0^\mathcal{R,A}(\mathbf{k},\omega) = \sum_{l=1}^n \frac{\sigma^{(l)}(\mathbf{k})}{\omega-\omega_l(\mathbf{k}) \pm \I \delta} \;.
\label{eq:G0AR}
\end{equation}

Contrarily, $g^{(l)\mathcal{C}}(\mathbf{k})$ and, consequently, the causal Green's function $G_0^\mathcal{C}(\mathbf{k},\omega)$ can be fully determined by the boundary conditions only for fermionic basis
\begin{equation}
G_0^\mathcal{C}(\mathbf{k},\omega) = \sum_{l=1}^n \sigma^{(l)}(\mathbf{k}) \left[ \frac{1 - f_\mathrm{F}(\omega)}{\omega - \omega_l(\mathbf{k}) + \I \delta} + \frac{f_\mathrm{F}(\omega)}{\omega - \omega_l(\mathbf{k}) - \I \delta} \right] \;,
\label{eq:G0Cfer}
\end{equation}
where $f_\mathrm{F}(\omega)$ is the Fermi distribution function.

We conclude the discussion about fermionic basis reporting the expression for the correlation function $C(i,j) = \left\langle \psi(i) \psi^\dagger(j) \right\rangle$
\begin{equation}
C(\mathbf{k},\omega) = 2 \pi \sum_{l=1}^n \left[1 - f_\mathrm{F}(\omega_l(\mathbf{k})) \right] \sigma^{(l)}(\mathbf{k}) \delta\left[\omega - \omega_l(\mathbf{k})\right] \;.
\label{eq:Cf0}
\end{equation}
It is necessary noticing that the correlation function $C$ can be computed by means of (\ref{eq:Cf0}) only if the basis $\psi$ is closed or if we neglect the residual self-energy. Otherwise, we should use the more general expression
\begin{equation}
C(\mathbf{k},\omega) = \mp 2 \left[1 - f_\mathrm{F}(\omega) \right] \Im \left[G^\mathcal{R,A}(\mathbf{k},\omega) \right] \;.
\label{eq:Cf}
\end{equation}

For bosonic basis, instead, $g^{(l)\mathcal{C}}(\mathbf{k})$ cannot be fully determined by the boundary conditions, but we can still obtain a modified Lindhard-like representation for $G_0^\mathcal{C}(\mathbf{k},\omega)$ \cite{Mancini_00,Mancini_04}
\begin{equation}
G_0^\mathcal{C}(\mathbf{k},\omega) = -2 \pi \I \Gamma(\mathbf{k}) \delta(\omega) + \sum_{l=1}^n \sigma^{(l)}(\mathbf{k}) \left[ \frac{1 + f_\mathrm{B}(\omega)}{\omega-\omega_l(\mathbf{k}) + \I \delta} - \frac{f_\mathrm{B}(\omega)}{\omega-\omega_l(\mathbf{k}) - \I \delta} \right] \;,
\label{eq:G0Cbos}
\end{equation}
where $f_\mathrm{B}(\omega)$ is the Bose distribution function, provided that we explicitly keep in the final expression an unknown zero frequency momentum function in matrix form $\Gamma(\mathbf{k})$.

The actual value of $\Gamma(\mathbf{k})$ is directly related to the degree of ergodicity of the dynamics of the basis $\psi$ driven by the Hamiltonian $H$
\begin{equation}
\Gamma(\mathbf{i-j}) = \lim_{\mathsf{T} \to \infty} \frac{1}{\mathsf{T}} \int_0^\mathsf{T} dt \left\langle \psi(\mathbf{i},0) \psi^\dagger(\mathbf{j},t) \right\rangle \;.
\label{eq:GammaErgI}
\end{equation}
If we would know, by any source of information, that the dynamics of $\psi$ driven by $H$ is fully ergodic, the last equation would simply give
\begin{equation}
\Gamma(\mathbf{i-j}) = \left\langle \psi(\mathbf{i}) \right\rangle \left\langle \psi^\dagger(\mathbf{j}) \right\rangle \;,
\label{eq:GammaErgII}
\end{equation}
leading to an effective removal of the unknown $\Gamma$ function from the theory. Unfortunately, although many people just overlook this deliberately, there is absolutely no way to asses ergodicity in advance and $\Gamma(\mathbf{k})$ forms another block of unknowns in the theory. In particular, ergodicity cannot be supposed a priori for finite systems treated in statistical ensembles different from the micro-canonical one and for bosonic operators commuting with the Hamiltonian (i.e. for constants of motion). 

$\Gamma(\mathbf{k})$ also appears in the expression for the correlation function $C(i,j) = \left\langle \psi(i) \psi^\dagger(j) \right\rangle$
\begin{equation}
C(\mathbf{k},\omega) = 2 \pi \Gamma(\mathbf{k}) \delta(\omega) + 2 \pi \sum_{l=1}^n \left[1 + f_\mathrm{B}(\omega_l(\mathbf{k})) \right] \sigma^{(l)}(\mathbf{k}) \delta\left[\omega - \omega_l(\mathbf{k})\right] \;,
\label{eq:Cb0}
\end{equation}
but this is absolutely not the only issue with the above expression (\ref{eq:Cb0}).

If $\sigma^{(l)}(\mathbf{k})$ is not identically zero for all values of $\mathbf{k}$ where $\omega_l(\mathbf{k})$ vanishes, the correlation function $C(\mathbf{k},\omega)$ diverges as $1 / \beta \omega_l(\mathbf{k})$ for the same values of $\mathbf{k}$ in the limit $\omega \to 0$ (i.e. at all times). Such a behaviour of $C(\mathbf{k},\omega)$ usually manifests the establishing of long-range spatial correlations in the system, but it is admissible iff the corresponding correlation function in real space $C(\mathbf{r},t)$ stays always finite. Accordingly, the divergence should be integrable; this immediately excludes the possibility to have long-range order in finite systems and, at finite temperatures, in infinite systems (i.e. in the thermodynamic limit) with too low spatial dimension. Actually, the lowest spatial dimension allowed to host long-range order will simply depend on the actual functional form of the vanishing $\omega_l(\mathbf{k})$ (e.g. if $\omega_l(\mathbf{k}) \propto |\mathbf{k}|^\alpha$ then the spatial dimension has to be strictly larger than $\alpha$). No restriction applies to infinite systems of any dimension at zero temperature. This is just the content of Mermin-Wagner theorem \cite{Mermin_66}.

For bosonic basis too, it is necessary noticing that the correlation function $C$ can be computed by means of (\ref{eq:Cb0}) only if the basis $\psi$ is closed or if we neglect the residual self-energy. Otherwise, we should use the more general expression
\begin{equation}
C(\mathbf{k},\omega) = - 2 \frac{1 + f_\mathrm{B}(\omega)}{1 + 2 f_\mathrm{B}(\omega)} \Im \left[G^\mathcal{C}(\mathbf{k},\omega) \right] \;.
\label{eq:Cb}
\end{equation}
The use of the casual Green's function, for a bosonic basis, is strictly necessary in order to properly take into account the ergodicity issue that does not manifest in neither the advanced nor the retarded propagators, but only in the causal propagator and in the correlation function.

\subsection{Residual self-energy $\Sigma$\label{sec:Sigma}}

Obviously, we do not need to compute the residual self-energy at all (it is just identically zero) if we choose a closed basis; this is one of the main reasons why a closed basis is the best one you can choose. Accordingly, as much as a truncated basis is \emph{large} enough less relevant will be the contribution of the residual self-energy to the description of the system under analysis. It is worth noticing that even if we would completely neglect the residual self-energy, the Green's function of the original interacting particles constituting the system, expressed in terms of the relevant entries of the propagator $G_0$, will anyway feature a fully momentum and frequency dependent irreducible self-energy with a $(n-1)$-polar structure, but sufficient to describe all scales of energy caught by the chosen basis

Nevertheless, one cannot neglect the residual self-energy without being aware of the main drawback: one is actually promoting to the rank of \emph{true} particles (i.e. with an infinite life-time) objects that, being still subject to all virtual processes not properly taken into account by a truncated basis, have finite life-times (i.e. they are quasi-particle) roughly inversely proportional to the largest neglected energy scale involving them (i.e. the transition described by a composite operator can or cannot be one of those necessary to construct the relevant virtual process). Clearly, this can be systematically controlled by enlarging the basis, although only on a quantitative level. Again, not-analytical energy scales can change so dramatically the properties of particles and quasi-particles to require a qualitative change of perspective also regarding the residual self-energy determination and role, as described in the Sect.~\ref{sec:Basis}.

As a matter of fact, it is sometimes convenient keeping the operatorial basis somewhat simpler than what is actually handleable in order to get simpler expressions for the residual currents too and effectively compute, starting from the latter, the residual self-energy. Such a procedure can lead to a description of the system under analysis featuring both some of the relevant energy scales and the decay effects inherent to the quasi-particle nature of composite operators belonging to a truncated basis.

In these years, we have been developing and testing different ways to compute residual self-energy; among others: the Two-site resolvent approach \cite{Matsumoto_96,Matsumoto_97} and the Non Crossing Approximation \cite{Avella_03c,Krivenko_04,Avella_07,Avella_07a,Avella_07b,Avella_07c,Avella_08,Avella_09}. Any of them has its pros and cons and specific problems can be better tackled by one or the other (see Sect.~\ref{sec:Hubbard}).

\subsection{Self-consistency\label{sec:Self-cons}}

The intrinsic complexity of the operatorial algebra closed by composite operators (for instance, see Table~\ref{tab:comm_eta_xi}) hides a noteworthy possible exploitation of the same algebra. Composite operators, whose lattice sites overlap (any composite operator may span a certain number of lattice sites), obey algebra constraints directly coming from the Pauli exclusion principle. For instance, $\xi$ and $\eta$ satisfy the following exact relation: $\xi_\sigma(\mathbf{i}) \eta_{\sigma'}^\dagger(\mathbf{i}) = 0$ that can be easily traced back to $c_\sigma(\mathbf{i}) c_\sigma(\mathbf{i}) = 0$.

The relevance of such algebra constraints resides, on one side, in their capability to enforce the Pauli principle and its derivatives, and, on the other side, in the not-so-trivial request that they should be obeyed at the level of thermal averages too. In fact, the related thermal averages (e.g. $\langle \xi_\sigma(\mathbf{i}) \eta_{\sigma'}^\dagger(\mathbf{i}) \rangle = 0$), through the well-known relation existing between correlation functions and Green's functions (fluctuaction-dissipation theorem; see Sect.~\ref{sec:G0}), depend on all unknown parameters appearing in the relevant Green's function (in $I$, $\varepsilon$, and $\Sigma$ plus $\Gamma(\mathbf{k})$ for bosonic basis) and, in turn, can be used to fix them:
\begin{equation}
\langle \psi(\mathbf{i}) \psi^\dagger(\mathbf{j}) \rangle = C(\mathbf{i-j},t=0)\;.
\label{eq:Pauli}
\end{equation}
The l.h.s of (\ref{eq:Pauli}), for $\mathbf{i}$ and $\mathbf{j}$ such that elements of the basis $\psi$ span on, at least, one common site, would be fixed by the algebra, which imposes \emph{contractions} (e.g. $\xi_\sigma(\mathbf{i})\eta_{\sigma'}^\dagger(\mathbf{i})=0$, $n(\mathbf{i})n(\mathbf{i})=n(\mathbf{i})+2\eta^\dagger_\uparrow(\mathbf{i})\eta_\uparrow(\mathbf{i})$). The r.h.s. of (\ref{eq:Pauli}) is given by the actual expression of the Green's function and contains all unknowns of the theory (for bosonic basis, it also contains $\Gamma(\mathbf{i-j})$).

This deceptively simple conclusion has enormous implications on both the capability to solve, either exact or approximately, strongly correlated systems, and the \emph{quality} of the solution. Not only algebra constraints allow to find a solution, but they make this solution as closer as possible to the exact one because they embody the primary cause of electronic correlations: the Pauli principle. 

It is also worth noticing that the symmetries enjoined by the Hamiltonian imply the existence of constants of motion and the possibility to formulate relations among matrix elements of the relevant Green's functions known as Ward-Takashi identities \cite{Ward_50,Takahashi_57}. These identities can/should also be used to fix the unknowns and to constrain the theory.

\subsection{Summary\label{sec:Sum}}

Summarizing, COM framework envisages four main steps:
\begin{enumerate}
\item Choose a composite operator basis according to the system under analysis and all information we can gather from relevant numerical and exact solutions (see Sect.~\ref{sec:Basis})
\item Compute $I$ and $m$ matrices and obtain $\varepsilon$ matrix and propagator $G_0$ in terms of unknown correlators and unknown $\Gamma$ function (see Sect.~\ref{sec:Basis} and \ref{sec:G0}).
\item Choose a recipe to compute $\Sigma$ or just neglect it (see Sect.~\ref{sec:Sigma}).
\item Self-consistently compute the unknowns through Algebra Constraints and Ward-Takashi identities (see Sect.~\ref{sec:Self-cons}).
\end{enumerate}

In the last fifteen years, COM has been applied to several models and materials: Hubbard \cite{Matsumoto_97,Avella_98e,Mancini_00,Avella_00b,Avella_00,Avella_03,Avella_01,Mancini_04,Odashima_05,Avella_07a,Avella_09}, $p$-$d$ \cite{Ishihara_94,Fiorentino_01,Avella_11}, $t$-$J$ \cite{Matsumoto_96,Avella_01}, $t$-$t'$-$U$ \cite{Avella_98d,Avella_99,Avella_03g}, extended Hubbard ($t$-$U$-$V$) \cite{Avella_04,Avella_06b}, Double-exchange \cite{Bak_02}, Kondo \cite{Villani_00}, Anderson \cite{Avella_02}, Kondo-Heisenberg \cite{Matsumoto_94}, two-orbital Hubbard \cite{Plekhanoff_11}, Kondo lattice \cite{Mancini_00e}, Ising \cite{Mancini_05,Mancini_05b,Mancini_06a,Mancini_08,Mancini_08a,Mancini_10}, BEG \cite{Mancini_10a}, Heisenberg \cite{Allega_94}, $J_1-J_2$ \cite{Plekhanoff_06a,Avella_08a,Plekhanoff_10}, Hubbard-Kondo \cite{Avella_06a}, Singlet-hole \cite{Mancini_96a}, Cuprates \cite{Avella_98c,Mancini_98c,Avella_03a,Avella_07} \ldots A comparison with the results of numerical simulations has been systematically carried on. The interested reader may refer to the works cited in Ref.~\cite{Mancini_04} and, for the last years, at the web page: http://scs.physics.unisa.it.

In the second part of this manuscript, as relevant application of the formalism, we will consider the Hubbard model and we will go through the different approximation schemes illustrated in the previous Sections in a systematic way. A comprehensive comparison with the results of numerical simulations and with the experimental data for high T$_c$ cuprates will be also reported.

\section{Case study: The Hubbard model\label{sec:Hubbard}}

\subsection{The Hamiltonian\label{sec:Hub-Ham}}

The Hubbard model reads as
\begin{equation}
H = \sum_\mathbf{ij} \left(-\mu \delta_\mathbf{ij} - 2dt \alpha_\mathbf{ij} \right) c^\dagger(i) c(j) + U \sum_\mathbf{i} n_\uparrow(i) n_\downarrow(i)\;.
\label{eq:Hub-Ham}
\end{equation}
The notation is the same used in Sect.~\ref{sec:Correlations} with the following additions: $t$ is the hopping and the energy unit; $d$ is the dimensionality of the system; $\alpha_\mathbf{ij}$ is the projector on the nearest-neighbour sites, whose Fourier transform, for a $d$-dimensional cubic lattice with lattice constant $a$, is $\alpha(\mathbf{k}) = \frac{1}{d} \sum_{n = 1}^d \cos(k_n a)$. The electronic operators $c(i)$ and $c^\dagger(i)$, as well as all other fermionic operators, are expressed in the spinorial notation [e.g. $c^\dagger(i) = \left(c_\uparrow^\dagger(i) \quad c_\downarrow^\dagger(i) \right)$. A detailed and comprehensive summary of the properties of Hamiltonian (\ref{eq:Hub-Ham}) are given in Sect.~1 of Ref.~\cite{Mancini_04}. We here report a study of this model by means of the Composite Operator Method as formulated in Sect.~\ref{sec:COM}. In order to proceed in a pedagogical way, we will go through different stages. In Sect.~\ref{sec:Hub-2pole-Fer}, we consider a truncated fermionic basis (2-pole approximation) given by the two Hubbard operators $\xi(i)$ and $\eta(i)$. In Sect.~\ref{sec:Hub-2pole-Bos}, we complement the analysis considering the related bosonic sector (charge and spin). In Sect.~\ref{sec:Hub-Sigma}, we implement the study by considering the contribution of the residual self-energy. In Sect.~\ref{sec:Hub-4pole}, we enlarge the truncated basis by including higher-order composite fields (4-pole approximation). At all these stages, we will present COM results and compare them with experimental and/or simulation data. Due to the pedagogical nature of this manuscript, we will restrict the analysis to the paramagnetic state. Ordered phases (ferro and antiferromagnetic phases) have been also analyzed and the related results can be found in Ref.~\cite{Mancini_04}. Finally, in Sect.~\ref{sec:Hub-dwave}, we will discuss the superconducting solution of the model in the $d$-wave channel and compare COM results with high-T$_c$ cuprates experimental data.

\subsubsection{Two-pole solution - Fermionic Sector \label{sec:Hub-2pole-Fer}}

On the basis of the discussions reported in Sects.~\ref{sec:Correlations} and \ref{sec:Basis}, we will adopt as fermonic basis
\begin{equation}
\psi (i) = \left(
\begin{array}{c}
\xi(i) \\
\eta(i)
\end{array}
\right)\;,
\label{eq:Hub-Bas}
\end{equation}
where $\xi(i)$ and $\eta(i)$ are the Hubbard operators defined in Sect.~\ref{sec:Correlations}. This field satisfies the equation of motion
\begin{equation}
\I \frac{\partial}{\partial t} \psi(i) = J(i) = \left(
\begin{array}{l}
-\mu \xi(i) - 2dt c^\alpha(i) - 2dt \pi(i) \\
(U - \mu) \eta(i) + 2dt \pi(i)
\end{array}
\right)\;,
\end{equation}
with $n_\mu(i) = c^\dagger(i) \sigma_\mu c(i)$ and $\pi(i) = \frac{1}{2} \sigma^\mu n_\mu(i) c^\alpha(i) + c(i) c^{\dagger\alpha}(i) c(i)$.
$\sigma^\mu=(-\mathbf{1},\vec{\sigma})$ and $\sigma_\mu=(\mathbf{1},\vec{\sigma})$, where $\vec{\sigma}$ are the Pauli matrices. Hereafter, given a generic operator $\Phi(i)$, we will use the notation $\Phi^\alpha(i) = \sum_\mathbf{j} \alpha_\mathbf{ij} \Phi(\mathbf{j},t)$. The current $J(i)$ is projected on the basis (\ref{eq:Hub-Bas}) and the residual current $\delta J(i)$ is neglected (truncated basis). The normalization and energy matrices have the expressions
\begin{equation}
I(\mathbf{k}) = \left(
\begin{array}{cc}
I_{11} & 0 \\
0 & I_{22}
\end{array}
\right) = \left(
\begin{array}{cc}
1 - n/2 & 0 \\
0 & n/2
\end{array}
\right)
\quad \quad \varepsilon(\mathbf{k}) = \left(
\begin{array}{cc}
m_{11}(\mathbf{k}) I_{11}^{-1} & m_{12}(\mathbf{k}) I_{22}^{-1} \\
m_{12}(\mathbf{k}) I_{11}^{-1} & m_{22}(\mathbf{k}) I_{22}^{-1}
\end{array}
\right)\;,
\label{eq:Hub-I&Eps}
\end{equation}
where the entries of the $m$-matrix are given by
\begin{align}
& m_{11}(\mathbf{k}) = - \mu I_{11} - 2dt \left[ \Delta + \alpha(\mathbf{k}) (1 - n + p) \right] \\
& m_{12}(\mathbf{k}) = 2dt \left[ \Delta + \alpha(\mathbf{k}) (p - I_{22}) \right] \\
& m_{22}(\mathbf{k}) = (U - \mu) I_{22} - 2dt \left[ \Delta + \alpha(\mathbf{k}) p \right] \;,
\label{eq:Hub-M}
\end{align}
$n = 1/N \sum_\mathbf{i} \langle n(i) \rangle$ is the particle number per site; the parameters $\Delta$ and $p$ cause a constant shift of the bands and a bandwidth renormalization, respectively, and are defined as	
\begin{align}
& \Delta = \langle \xi^\alpha(i) \xi^\dagger(i) \rangle - \langle \eta^\alpha(i) \eta^\dagger(i) \rangle \\
& p = \frac{1}{4}\langle n_\mu^\alpha(i) n_\mu(i) \rangle - \langle \left[ c_\uparrow(i) c_\downarrow(i) \right]^\alpha c_\downarrow^\dagger(i) c_\uparrow^\dagger(i) \rangle \;.
\label{eq:Delta-p}
\end{align}
$\Delta$ is the difference between upper and lower intra-subband contributions to kinetic energy, while $p$ is a combination of the nearest-neighbor charge-charge, spin-spin and pair-pair correlation functions.

The retarded $G^\mathcal{R}(i,j) = \left\langle \mathcal{R} \left\{ \psi(i) \psi^\dagger(j) \right\} \right\rangle$ and the correlation $C(i,j) = \left\langle \psi(i) \psi^\dagger(j) \right\rangle$ functions are given by
\begin{align}
& G^\mathcal{R}(\mathbf{k},\omega) = \sum_{n = 1}^2 \frac{\sigma^{(n)}(\mathbf{k})}{\omega - E_n(\mathbf{k}) + \I\delta} \\
& C(\mathbf{k},\omega) = \sum_{n = 1}^2 \left[1 - f_\mathrm{F}(\omega)\right] \sigma^{(n)}(\mathbf{k}) \delta[\omega - E_n(\mathbf{k})] \;.
\end{align}

The energy spectra $E_n(\mathbf{k})$ read as
\begin{equation}
E_1 (\mathbf{k}) = R(\mathbf{k}) + Q(\mathbf{k}) \quad \quad E_2 (\mathbf{k}) = R(\mathbf{k}) - Q(\mathbf{k}) \;,
\end{equation}
where
\begin{equation}
R(\mathbf{k}) = \frac{1}{2} \left[U - 2\mu - 4dt\alpha(\mathbf{k})\right] - \frac{m_{12}(\mathbf{k})}{2I_{11} I_{22}} \quad \quad Q(\mathbf{k}) = \frac{1}{2} \sqrt{g^2(\mathbf{k}) + \frac{4m_{12}^2(\mathbf{k})}{I_{11} I_{22}}} \;,
\end{equation}
with $g(\mathbf{k}) = - U + \frac{1 - n}{I_{11} I_{22}} m_{12}(\mathbf{k})$. The spectral functions $\sigma^{(n)} (\mathbf{k})$ have the following expressions
\begin{equation}
\begin{array}{c}
\sigma_{11}^{(1)}(\mathbf{k}) = \frac{I_{11}}{2}[1 + \frac{g(\mathbf{k})}{2Q(\mathbf{k})}] \hfill \\
\sigma_{12}^{(1)}(\mathbf{k}) = \frac{m_{12}(\mathbf{k})}{2Q(\mathbf{k})} \hfill \\
\sigma_{22}^{(1)}(\mathbf{k}) = \frac{I_{22}}{2}[1 - \frac{g(\mathbf{k})}{2Q(\mathbf{k})}] \hfill
\end{array}
\quad \quad \quad \quad
\begin{array}{c}
\sigma_{11}^{(2)}(\mathbf{k}) = \frac{I_{11}}{2}[1 - \frac{g(\mathbf{k})}{2Q(\mathbf{k})}] \hfill \\
\sigma_{12}^{(2)}(\mathbf{k}) = -\frac{m_{12}(\mathbf{k})}{2Q(\mathbf{k})} \hfill \\
\sigma_{22}^{(2)}(\mathbf{k}) = \frac{I_{22}}{2}[1 + \frac{g(\mathbf{k})}{2Q(\mathbf{k})}] \hfill 
\end{array} \;.
\end{equation}

The determination of $G^\mathcal{R} (\mathbf{k},\omega)$ and $C(\mathbf{k},\omega)$ requires the knowledge of the chemical potential and of the two bosonic correlators $\Delta$ and $p$. These quantities are self-consistently determined by means of the following set of coupled equations
\begin{align}
& n = 2 (1 - C_{11} - C_{22}) \\
& \Delta = C_{11}^\alpha - C_{22}^\alpha \\
& C_{12} = \langle \xi (i)\eta ^\dagger (i) \rangle = 0 \;,
\end{align} 
where $C_{ab} = \langle \psi_a(\mathbf{i}) \psi_b^\dagger(\mathbf{i}) \rangle$ and $C_{ab}^\alpha = \langle \psi_a^\alpha(\mathbf{i}) \psi_b^\dagger(\mathbf{i}) \rangle$. The first equation fixes the chemical potential in terms of all other parameters; the second comes from the definition of $\Delta$ [cfr.~(\ref{eq:Delta-p})]; the third comes from the constraint (\ref{eq:Pauli}). Once this set of coupled self-consistent equations has been solved, we can calculate all relevant single-particle and thermodynamical properties of the model. The double occupancy per site: $D = 1/N \: \sum_\mathbf{i} \langle n_\uparrow(i) n_\downarrow(i) \rangle = I_{22} - C_{22}$; the energy bands $E_n (\mathbf{k})$; the Fermi surface by means of the equation $E_n (\mathbf{k}) = 0$; the momentum distribution function $n(\mathbf{k})$ and the density of states $N(\omega)$:
\begin{align}
& n(\mathbf{k}) = 2 \int_{-\infty}^{+\infty} d\omega f_\mathrm{F}(\omega)\left[-\frac{1}{\pi}\Im [G_{cc}^\mathcal{R}(\mathbf{k},\omega)]\right] \\
& N(\omega) = \frac{a^d}{(2\pi )^d} \int_{\Omega_\mathrm{B}} d^dk \left[-\frac{1}{\pi}\Im [G_{cc}^\mathcal{R}(\mathbf{k},\omega)]\right] \;,
\end{align} 
where $\Omega_\mathrm{B}$ is the volume of the first Brillouin zone, $G_{cc}^\mathcal{R} (\mathbf{k},\omega) = \sum_{a,b = 1}^2 G_{ab}^\mathcal{R} (\mathbf{k},\omega)$; the internal energy $E = \langle H \rangle = 4dt \sum_{a,b = 1}^2 C_{ab}^\alpha + UD$; the specific heat $C = dE/dT$; the free energy $F(n,T) = \int_0^n \mu (n',T) dn'$; the entropy $S = (E - F)/T$.

Once the fermionic propagator is known, there are several ways to compute response functions (i.e. the retarded propagators of the two-particle excitations: charge, spin, pair, \ldots). These techniques are usually based on diagrammatic expansions of the two-particle propagators in terms of the single-particle one. However, when operators with non-canonical commutations are involved, the complicated algebra invalidates the Wick theorem and, consequently, does not allow any simple extension of decoupling schemes and more involved diagrammatic approximations \cite{Matsumoto_85,Izyumov_90} are needed. Another technique \cite{Mancini_95b,Mancini_04}, the one-loop approximation for composite operators, has been developed by means of the equations of motion approach. By using this technique it is possible to calculate the causal function $\langle \mathcal{T}[n_\mu(i) n_\mu(j)] \rangle$ and then the charge and spin susceptibilities. We do not report the details of the calculations, which can be found in Ref.~\cite{Mancini_04}, and summarize the results. For the causal propagators we have
\begin{align}
& \langle \mathcal{T}[n(i)n(j)] \rangle = n^2 - \frac{n(2 - n)}{n - 2D} \sum_{a,b,c = 1}^2 I_{aa}^{-1} Q_{abac}^\mathcal{C}(i,j) \\
& \langle \mathcal{T}[n_3(i)n_3(j)] \rangle = -\frac{n(2 - n)}{n + 2D - n^2} \sum_{a,b,c = 1}^2 I_{aa}^{-1} Q_{abac}^\mathcal{C}(i,j) \;,
\end{align}
where $Q_{abcd}^	\mathcal{C} (i,j) = G_{ab}^	\mathcal{C} (i,j) G_{cd}^	\mathcal{C} (j,i)$ is the fermionic loop constructed from the propagator $G_{ab}^	\mathcal{C} (i,j) = \langle \mathcal{T}[\psi _a (i)\psi _b^\dagger (j)] \rangle$. For the charge and spin susceptibilities, we have
\begin{align}
& \chi _c (\mathbf{k},\omega ) = \frac{n(2 - n)}{n - 2D} \sum_{a,b,c = 1}^2 I_{aa}^{-1} Q_{abac}^\mathcal{R}(\mathbf{k},\omega ) \\
& \chi _s (\mathbf{k},\omega ) = \frac{n(2 - n)}{n + 2D - n^2} \sum_{a,b,c = 1}^2 I_{aa}^{-1} Q_{abac}^\mathcal{R}(\mathbf{k},\omega ) \;,
\end{align}
where $Q_{abcd}^\mathcal{R} (\mathbf{k},\omega )$ is the retarded part of $Q_{abcd}^	\mathcal{C} (\mathbf{k},\omega )$ and has the expression
\begin{multline}
Q_{abcd}^	\mathcal{R} (\mathbf{k},\omega ) = \\
= \frac{a^d }{(2\pi )^d} \sum_{n,m = 1}^2 \int d^dp \frac{\{f_\mathrm{F}[E_m(\mathbf{p})]-f_\mathrm{F}[E_n(\mathbf{k} + \mathbf{p})]\} \sigma _{ab}^{(n)} (\mathbf{k} + \mathbf{p}) \sigma _{cd}^{(m)} (\mathbf{p})}
{\omega + E_n (\mathbf{k} + \mathbf{p}) - E_m(\mathbf{p}) + \I\delta }
\end{multline}

\paragraph{Results and comparisons \label{sec:Hub-2pole-Fer-Res}}

\subparagraph{Algebra constraints \label{sec:Hub-2pole-Fer-Res-AC}}

\begin{figure}[tbp]
\centering
\includegraphics[width=.49\textwidth]{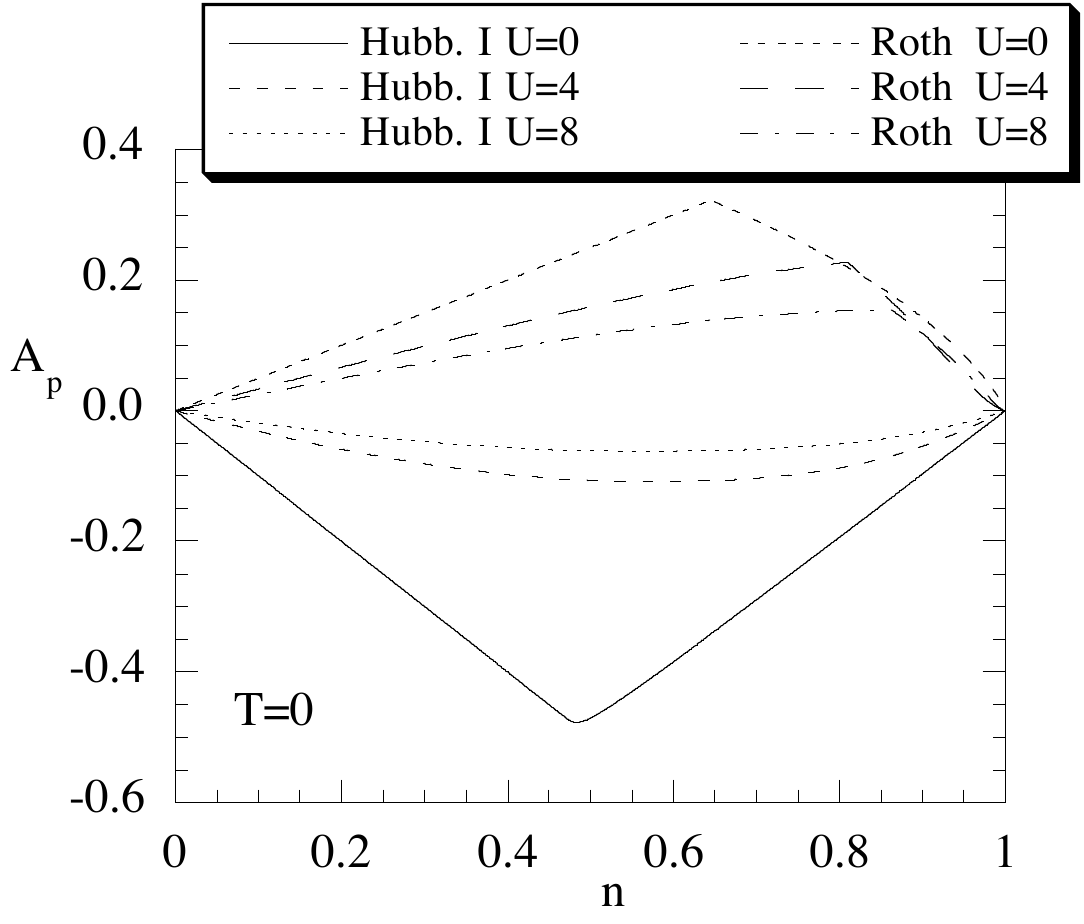}
\includegraphics[width=.49\textwidth]{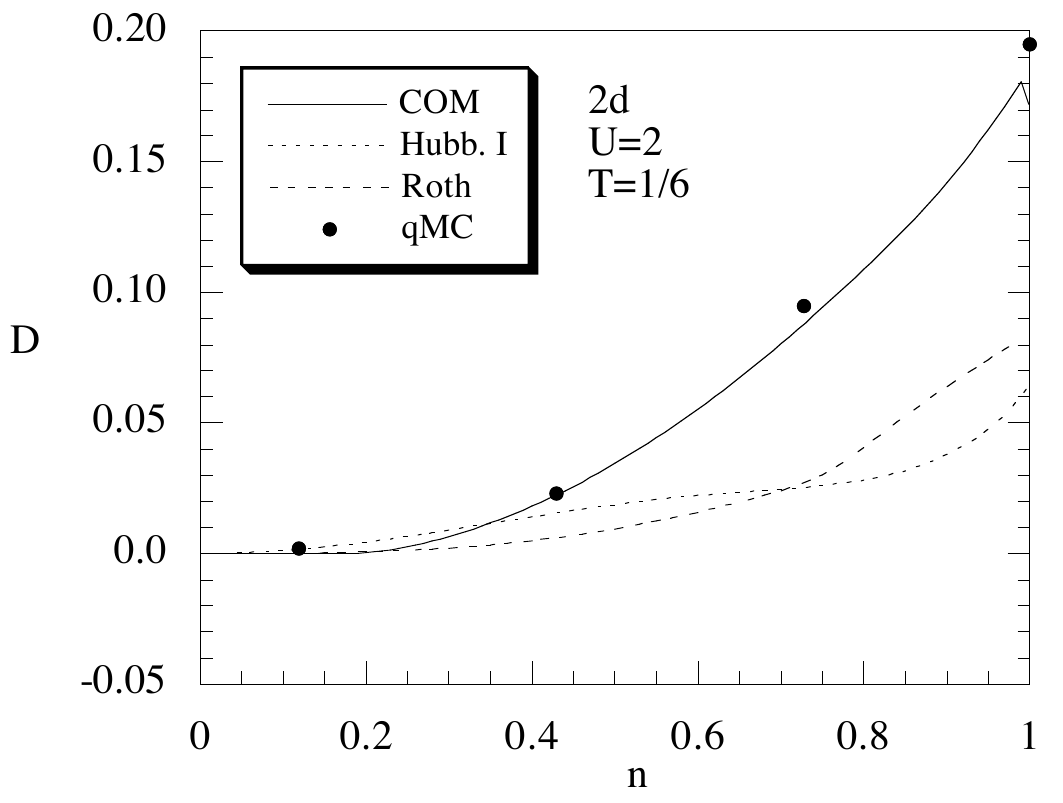}
\caption[]{(left) Pauli amplitude $A_p=C_{12}/C_{22}$ versus filling $n$ at $T=0$ for several values of $U$. (right) Double occupancy $D$ versus filling $n$ for $T=1/6$ and $U=2$. The solid, dotted and dashed lines stand for COM, Hubbard I and Roth, respectively. qMC data refer to a $12\times12$ cluster \cite{Moreo_90}.}
\label{fig:Pauli}
\end{figure}

The violation of the Algebra Constraints in the Roth and Hubbard I approximations is analyzed in Fig.~\ref{fig:Pauli} (left panel), where the normalized Pauli amplitude $A_p=C_{12}/C_{22}$, which is identically zero in COM (as it should be due to the Pauli principle), is reported versus $n$ at $T=0$ for various values of $U$. We see that in the Roth \cite{Roth_69,Roth_69a} and Hubbard I \cite{Hubbard_63} schemes the Pauli principle is satisfied only in the cases $n=0$ and $n=1$. The deviation increases by decreasing $U$, reaching its maximum value in the noninteracting limit (i.e. $U=0$). On the contrary, the Pauli principle is recovered in the limit $U\to \infty $ for any value of $n$. At any rate, the Hubbard I solution violates several other sum rules \cite{Gebhard_97}. As a consequence of the fact that the Pauli principle is not satisfied, the particle-hole symmetry enjoined by the Hubbard Hamiltonian is also violated \cite{Avella_98,Mancini_04}. In order to give a measure of how relevant is fulfilling Algebra Constraints, and in particular those directly coming from Pauli principle, in the right panel of the same Fig.~\ref{fig:Pauli}, we show COM results for the double occupancy in comparison with the Roth and Hubbard I ones, and the data obtained on finite size clusters by quantum Monte Carlo method \cite{Moreo_90}.

\subparagraph{Chemical potential \label{sec:Hub-2pole-Fer-Res-mu}}

\begin{figure}[tbp]
\centering
\includegraphics[width=.49\textwidth]{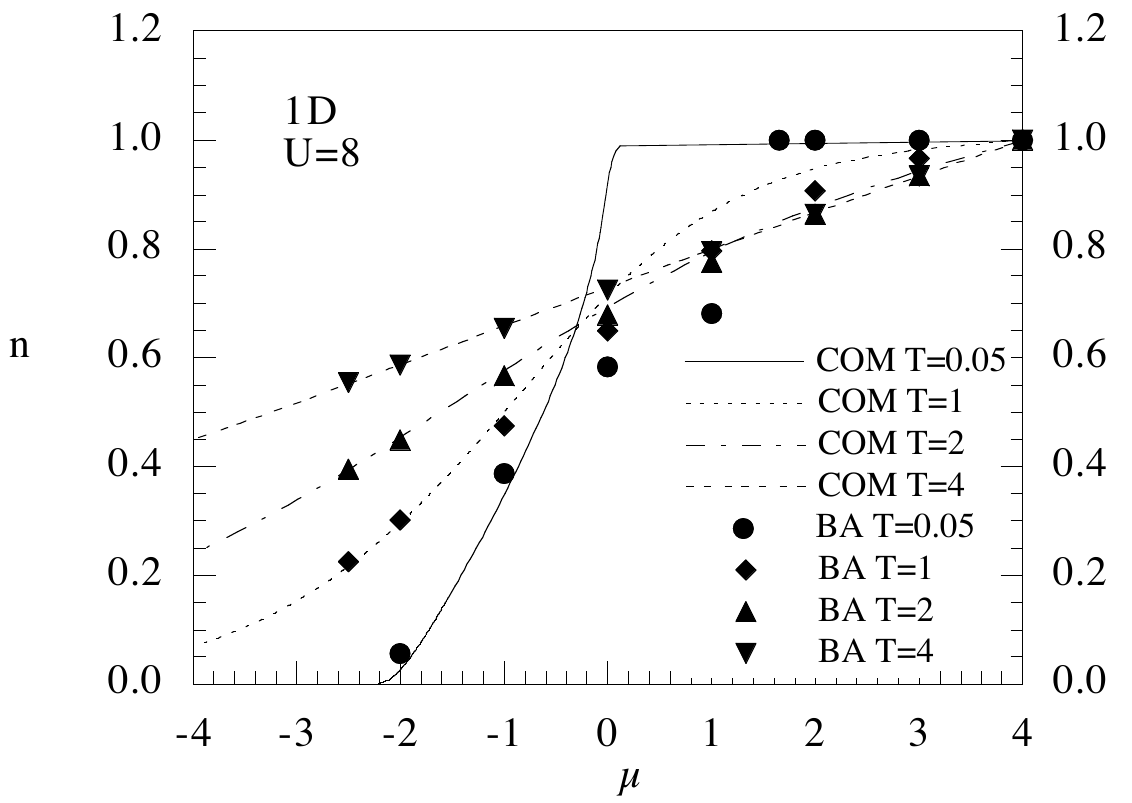}
\includegraphics[width=.49\textwidth]{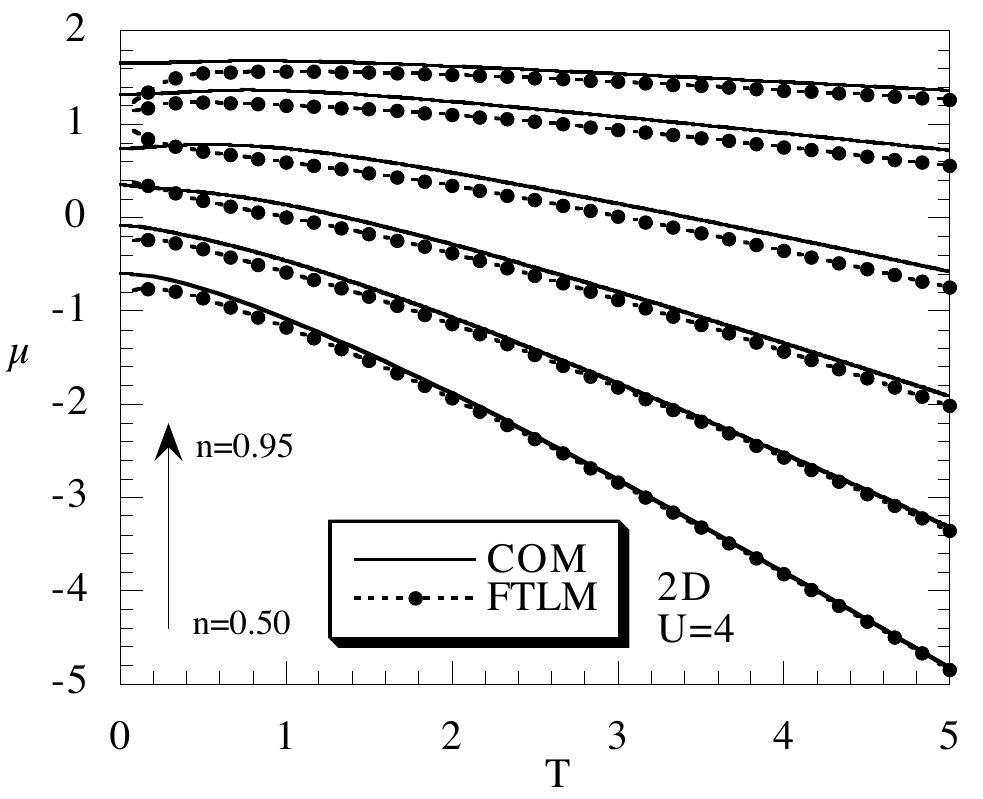}
\caption[]{(left) Filling $n$ versus chemical
potential $\mu $ for $T=0.05$, $1$, $2$ and $4$ and $U=8$. BA result is taken from Ref.~\cite{Kawakami_89}. (right) Chemical potential $\mu$ versus temperature $T$ for $U=4$ and $n=0.5 \to 0.95$. FTLM result is taken from Ref.~\cite{Bonca_02}.}
\label{fig:mu}
\end{figure}

In Fig.~\ref{fig:mu} (left panel), we show the particle density versus chemical potential $\mu$ for various temperatures. COM results are compared with the Bethe Ansatz (BA) ones \cite{Kawakami_89}. The agreement is very good at low temperatures for densities smaller than $0.55$. In the half-filled chain, $T\approx t$ is a relevant temperature as it signs the border between $T$-regions dominated by either spin or charge correlations \cite{Shiba_72a,Schulte_96}. The agreement between COM and BA at $T\ge t$ is very good for the whole range of filling. Of course, at high temperatures, COM result reaches an excellent agreement with BA since the effect of correlations is completely suppressed. In Fig.~\ref{fig:mu} (right panel), we show COM results for the chemical potential as a function of the temperature $T$ and of the filling for $U=4$ in comparison with numerical results obtained by means of the Lanczos technique on a $4 \times 4$ cluster \cite{Bonca_02}. The agreement is rather good in particular for low filling and high temperature where the numerical results are more reliable and the finite size of the cluster is less effective.

\subparagraph{Double occupancy \label{sec:Hub-2pole-Fer-Res-D}}

\begin{figure}[tbp]
\centering
\includegraphics[width=.49\textwidth]{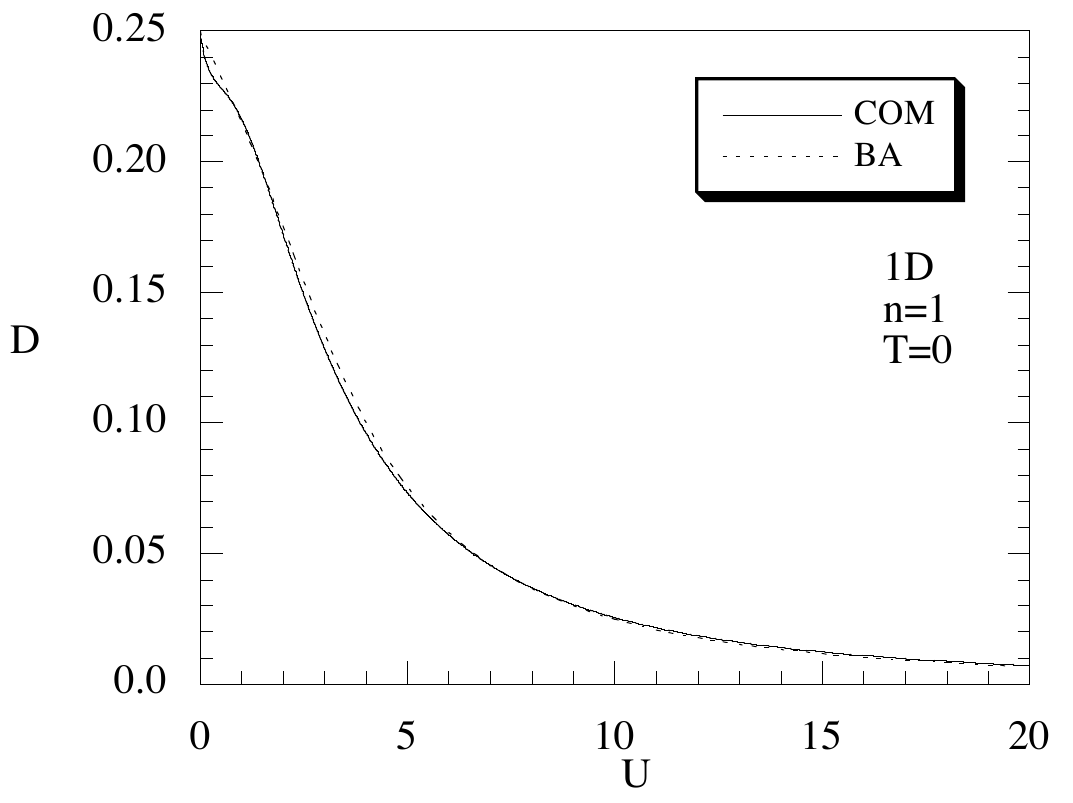}
\includegraphics[width=.49\textwidth]{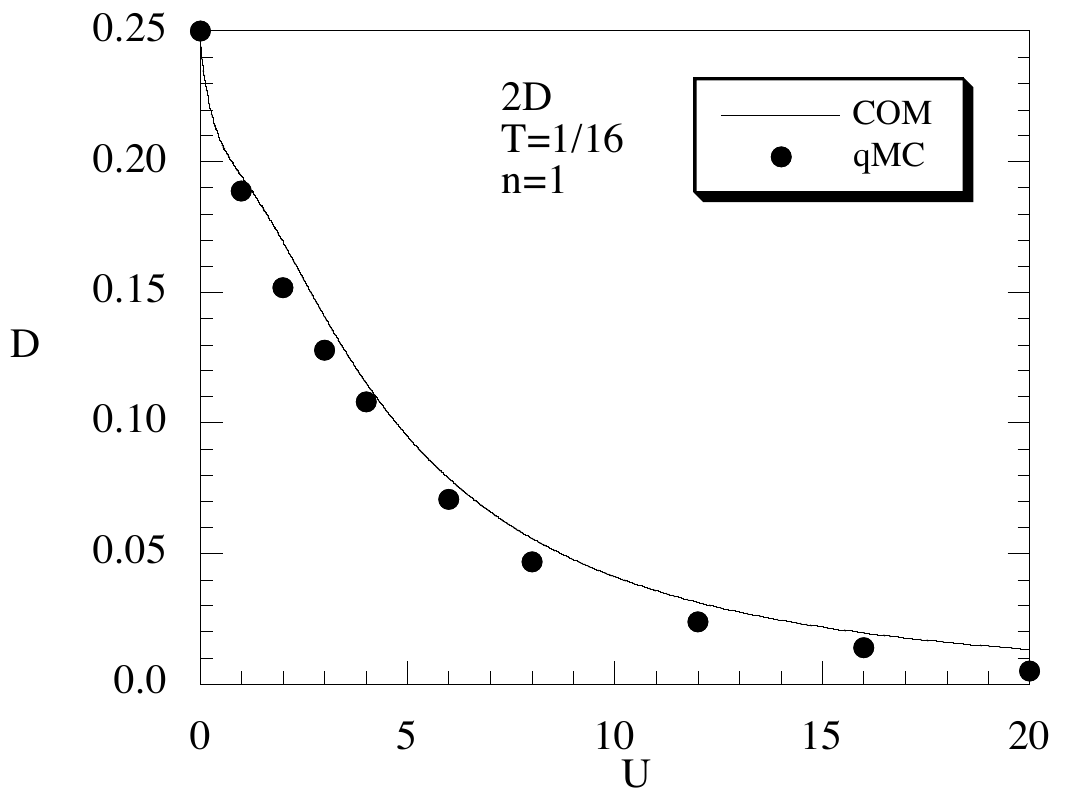}
\caption[]{(left) Double occupancy $D$ versus Coulomb interaction $U$ at $T=0$ and $n=1$. Dotted lines stand for BA results. (right) Double occupancy $D$ versus Coulomb interaction $U$ for $T=1/6$ and $n=1$, solid symbols stand for qMC data ($4\times4$) \cite{White_89}.}
\label{fig:D}
\end{figure}

The double occupancy for the one-dimensional case is shown in Fig.~\ref{fig:D} (left panel), where COM is plotted versus Coulomb interaction $U$ at $T=0$ and for $n=1$. The results are compared with the exact ones by Bethe ansatz. The agreement with BA is excellent. Such a good agreement is not reached by any other analytical approach, neither by the Gutzwiller or the ladder and the self-consistent ladder approximations \cite{Metzner_87,Buzatu_95,Hirsch_80}. In particular, these approaches fail to reproduce the correct asymptotic behavior.  As shown in Fig.~\ref{fig:D} (left panel), the double occupancy goes to zero as $U\rightarrow\infty$: the electrons localize only at infinite $U$, where the half-filled Hubbard chain is equivalent to the spin-$1/2$ AF Heisenberg chain. The double occupancy for the two-dimensional case is shown in Fig.~\ref{fig:D} (right panel), where COM solution is plotted versus Coulomb interaction $U$ at $T=0$ and for $n=1$. The results are compared with the data of numerical simulation by qMC \cite{White_89}. The agreement is excellent.

\subparagraph{Internal Energy \label{sec:Hub-2pole-Fer-Res-U}}

\begin{figure}[tbp]
\centering
\includegraphics[width=.49\textwidth]{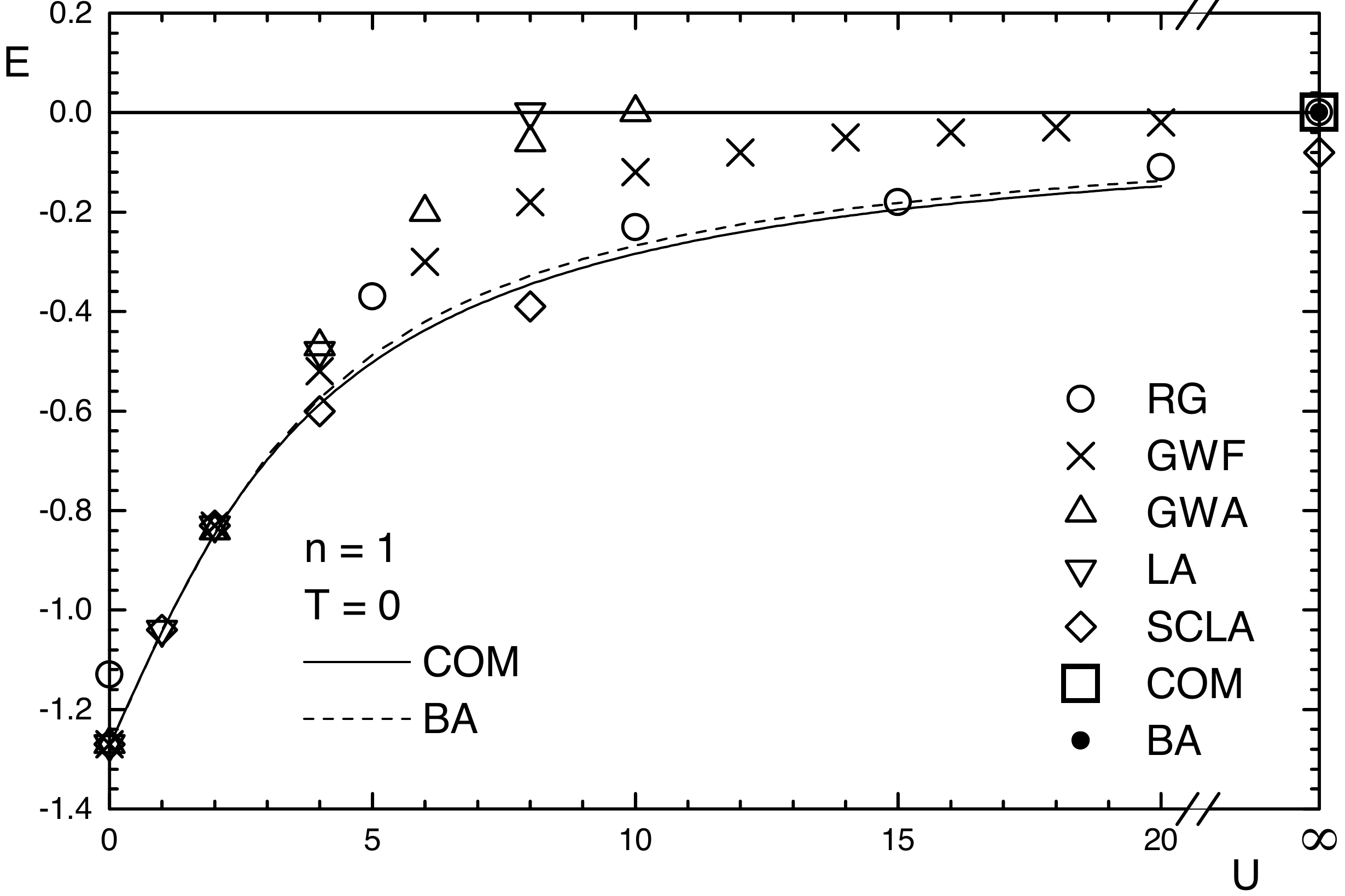}
\includegraphics[width=.49\textwidth]{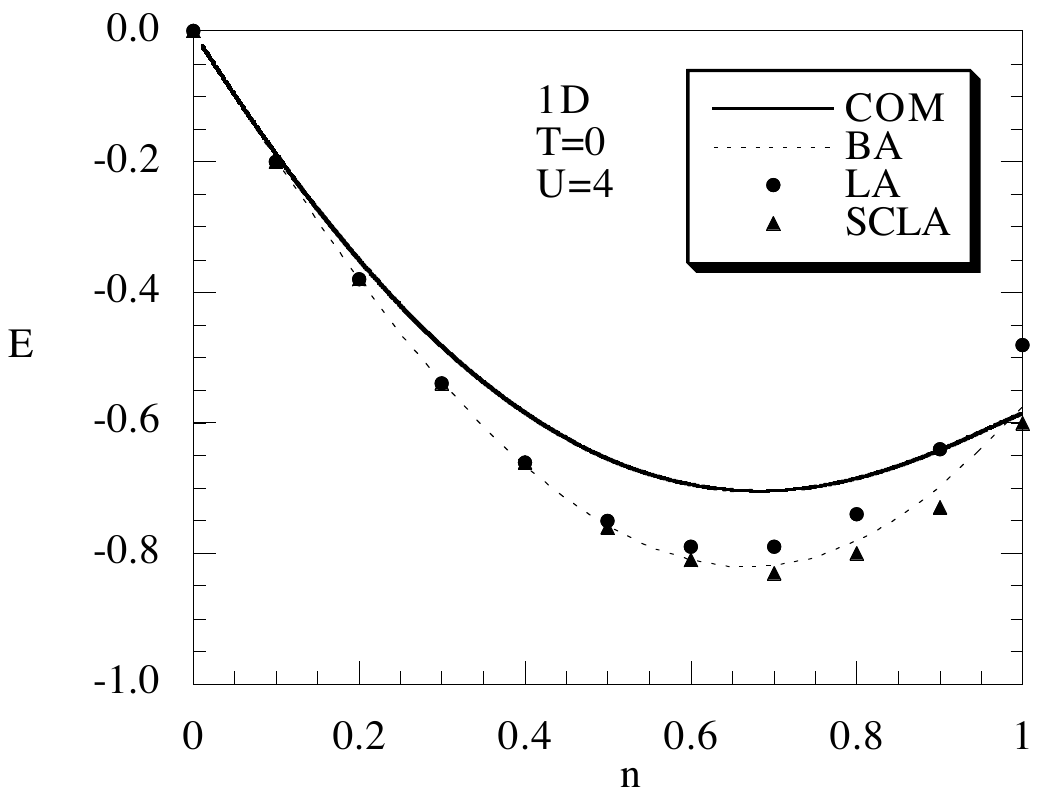}
\caption[]{Internal energy $E$ versus: (left) Coulomb interaction $U$ for $T=0$ and $n=1$; (right) filling $n$ for $T=0$ and $U=4$.}
\label{fig:U1}
\end{figure}

\begin{figure}[tbp]
\centering
\includegraphics[width=.49\textwidth]{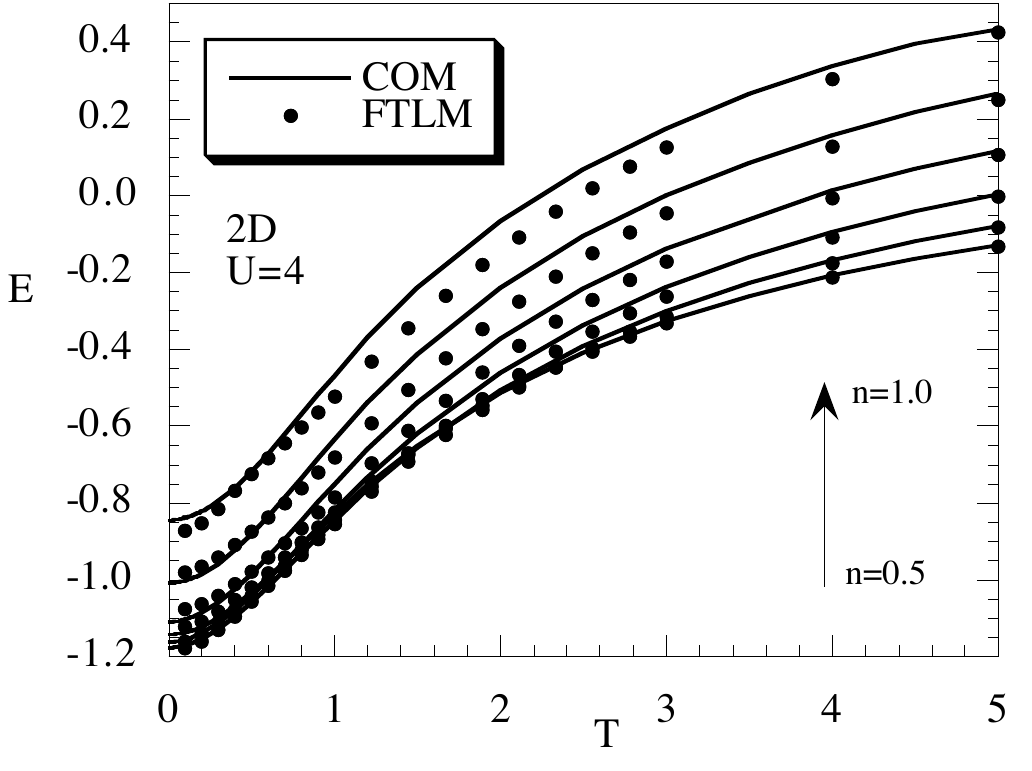}
\includegraphics[width=.49\textwidth]{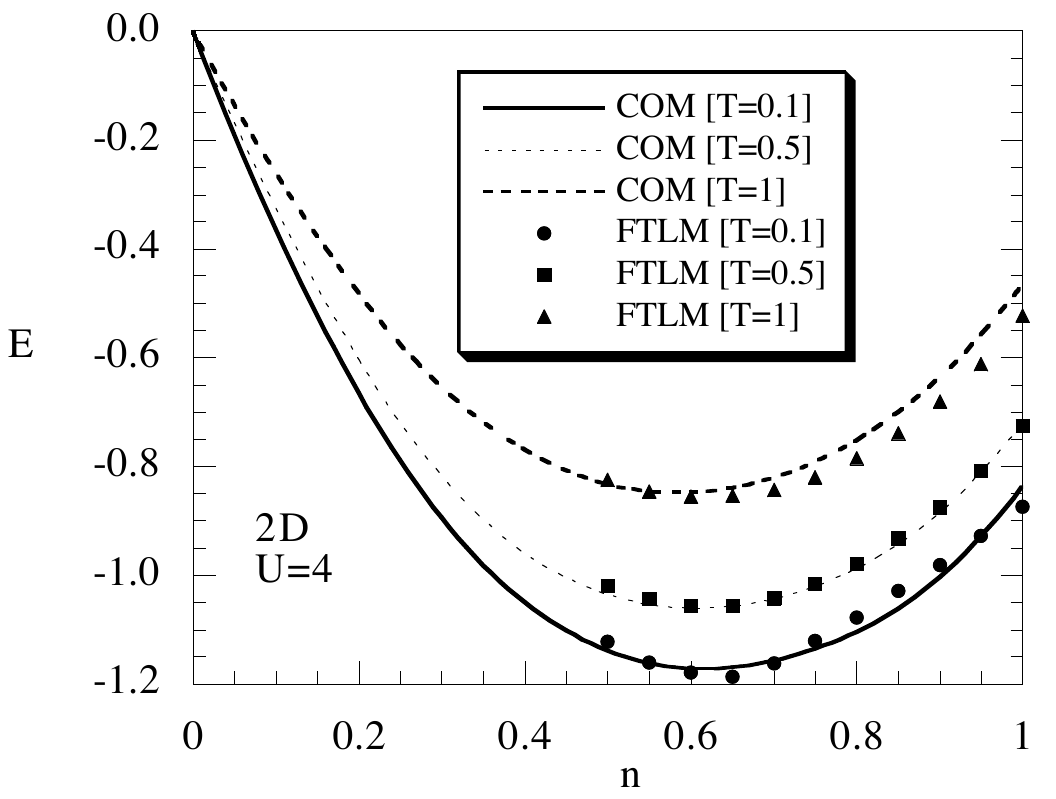}
\caption[]{Internal energy $E$ versus: (left) temperature $T$ for $U=4$ and $n=0.5\to 1$; (right) filling $n$ for $U=4$ and various values of temperature $T$. Lanczos data ($4\times 4$) are taken from Ref.~\cite{Bonca_02}.}
\label{fig:U2}
\end{figure}

The internal energy at $n=1$ and $T=0$ is shown as a function of $U$ in Fig.~\ref{fig:U1} (left panel). The results obtained by means of the BA \cite{Shiba_72} and other analytical approaches \cite{Hirsch_80,Metzner_87,Buzatu_95} are also reported. The agreement between COM and BA is excellent. The self-consistent Ladder approximation (SCLA) \cite{Buzatu_95} shows also a very good agreement for all values of the coupling, but it does not have the correct behavior for an infinite value of the Coulomb interaction. Moreover, both the Ladder (LA) \cite{Buzatu_95} and the Gutzwiller (GWA and GWF) \cite{Metzner_87} approximations go to zero at finite $U$, whereas the Renormalization Group (RG) \cite{Hirsch_80} has the right asymptotic behavior for $U\to \infty $, but it does not reproduce the non-interacting limit. The doping dependence of the internal energy is shown in Fig.~\ref{fig:U1} (right panel) for $U=4$. For comparison, we also report the BA results \cite{Shiba_72} and LA and SCLA approaches \cite{Hirsch_80,Buzatu_95}. COM agrees reasonably with the BA, reaching the best agreement at half filling. The ladder approximation \cite{Buzatu_95} deviates more and more from the BA as approaching half filling; the self-consistent ladder approximation \cite{Buzatu_95} probes excellently at any doping. In Fig.~\ref{fig:U2} (left panel) we present the internal energy versus temperature in the range $0\le T\le 5$ and $U=4$ for several values of the filling. The results are compared with with finite-temperature Lanczos method (FTLM) data \cite{Bonca_02}. The agreement is quite remarkable in the entire range of temperature and for all studied dopant concentrations. In Fig.~\ref{fig:U2} (right panel), we report the energy per site as a function of the particle density for different values of $T$ and $U$ and compare COM results with Lanczos data on a $4\times 4$ lattice \cite{Bonca_02}. The agreement is very good on the whole range of filling and for all values of temperature. It is worth noticing that at $T=0.1$ the numerical data report a strange kind of oscillations, not present in our results, that will be probably absent in the bulk system. It is worth noticing that many more comparisons, all showing a very good agreement, between COM results and numerical simulation data have been presented in Refs.~\cite{Mancini_95a,Mancini_99a}. The overall picture emerging from such comparisons is that COM generally gives a very accurate description of the behavior of the internal energy as a function of the external parameters ($T$, $U$ and $n$) over a very wide range of their values.

\subparagraph{Specific heat \label{sec:Hub-2pole-Fer-Res-Cv}}

\begin{figure}[tbp]
\centering
\includegraphics[width=.49\textwidth]{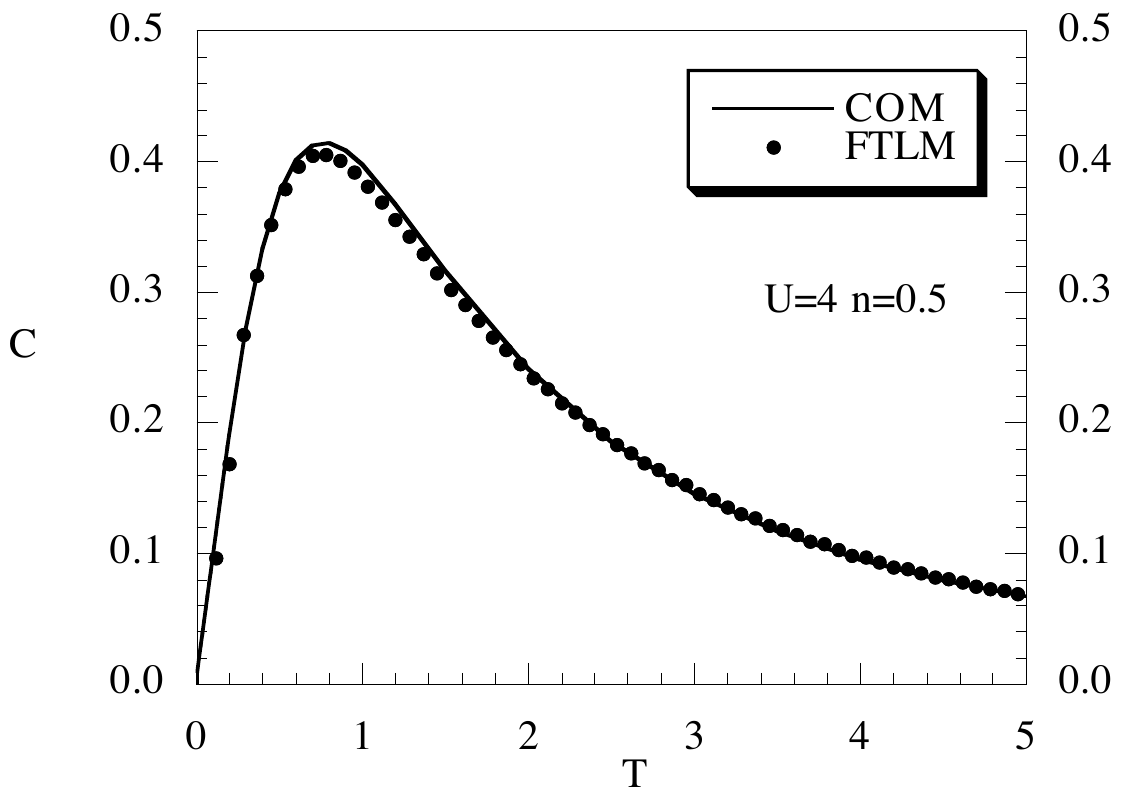}
\includegraphics[width=.49\textwidth]{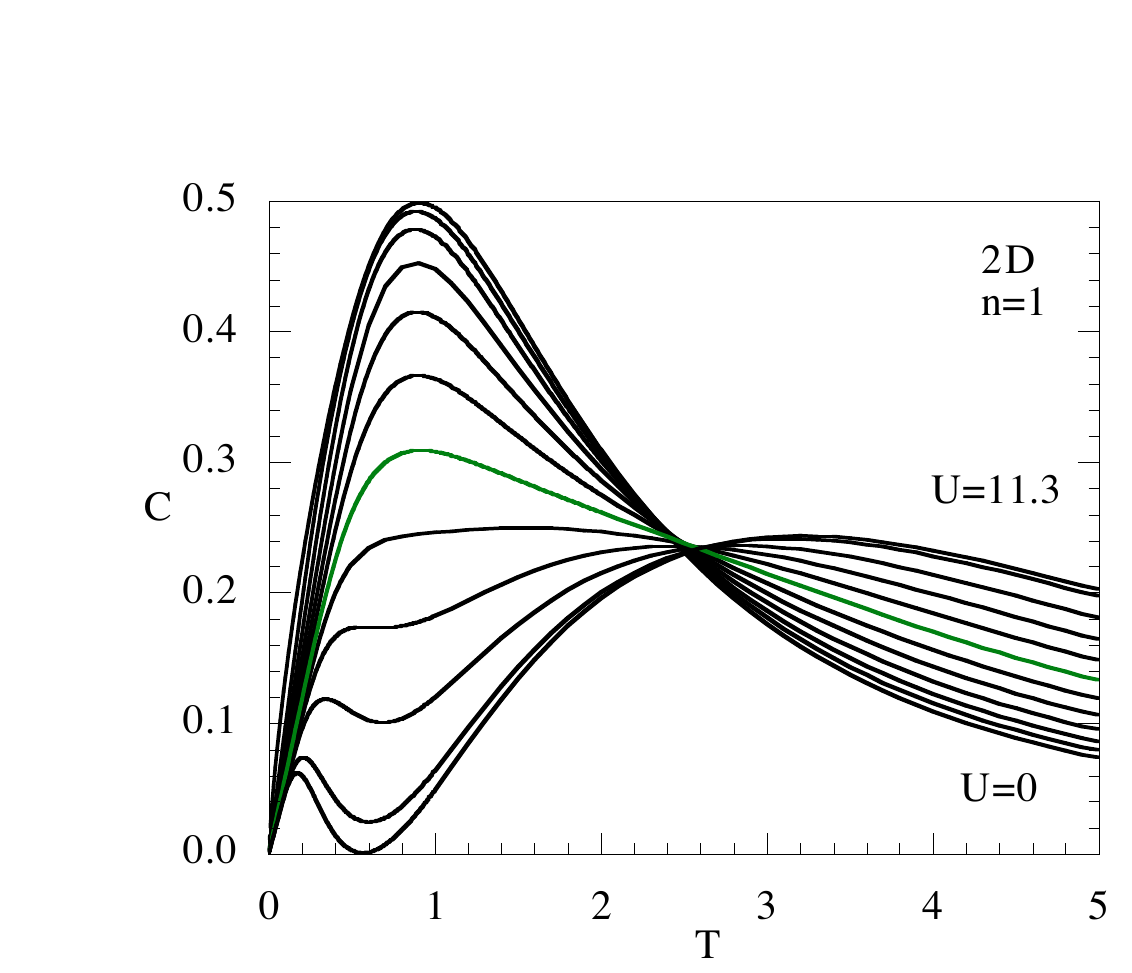}
\caption[]{Specific heat $C$ versus temperature $T$: (left) at $n=0.5$ and $U=4$; (right) at $n=1$ and various values of Coulomb interaction $U$.}
\label{fig:Cv}
\end{figure}

In Fig.~\ref{fig:Cv} (left panel), we report the specific heat as a function of the temperature for $n=0.5$ and $U=4$. The agreement with FTLM data \cite{Prelovsek} is excellent in the whole range of temperatures. The presence of two peaks in the specific heat has been attributed to the spin and charge excitations. Our results show a double peak structure too. When $U$ is weak the two peaks overlap and there is no resolution. By increasing $U$ the position of the charge peak moves to higher temperatures and we distinguish the two contributions as shown in Fig.~\ref{fig:Cv} (right panel), where $C$ is plotted as a function of $T$ at half-filling for several values of $U$. A peak appears at low temperatures when $U$ is rather large ($U\ge 8$). This behavior qualitatively reproduces the qMC results \cite{Duffy_97}. It is also relevant to observe the presence of a crossing point at a some definite temperature. Such crossing points have been experimentally observed in several materials and obtained by means of analitical and numerical studies (see Section 3.4.2 of Ref.~\cite{Mancini_04} for references and a detailed discussion).

\subparagraph{Entropy \label{sec:Hub-2pole-Fer-Res-S}}

\begin{figure}[tbp]
\centering
\includegraphics[width=.49\textwidth]{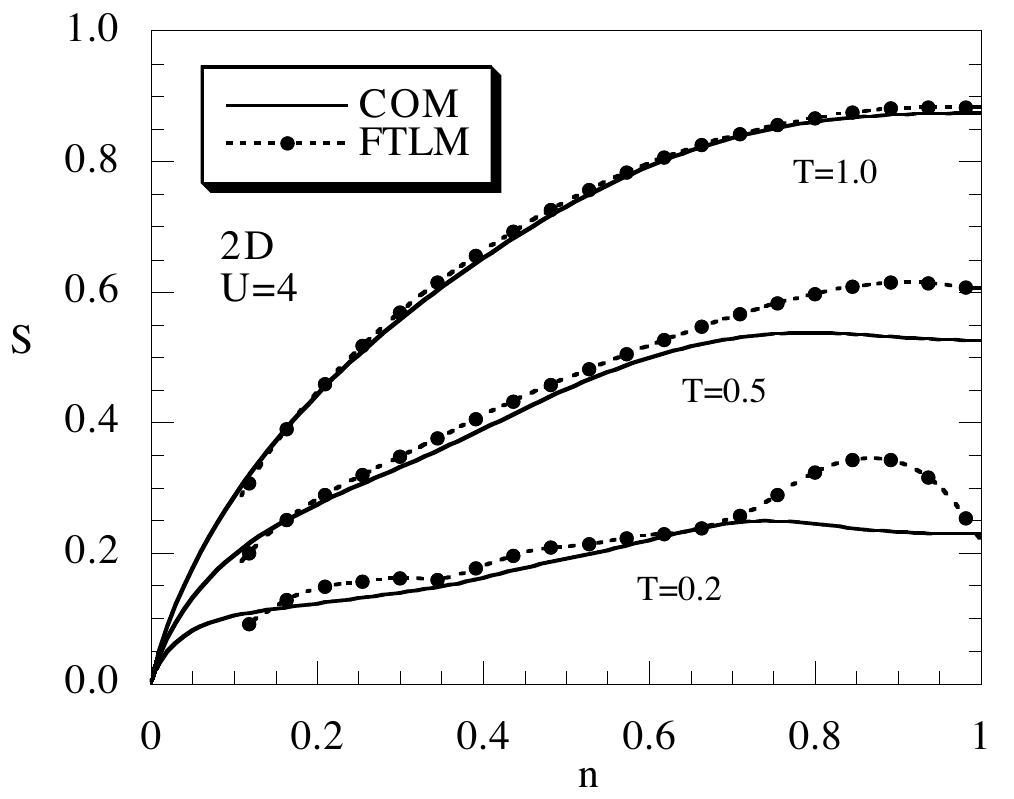}
\includegraphics[width=.49\textwidth]{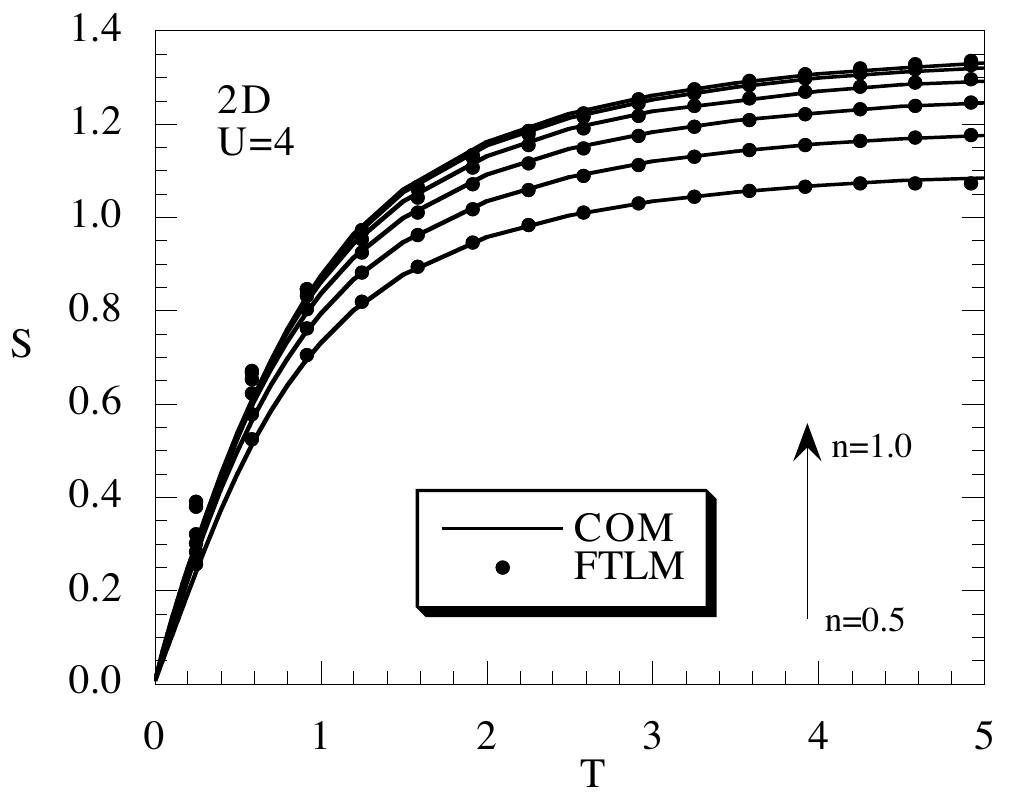}
\caption[]{Entropy $S$ versus filling $n$ for: (left) $U=4$ and various values of temperature $T$; (right) temperature $T$ for $U=4$ and $n=0.5\to 1$. Lanczos data are taken from Ref.~\cite{Bonca_02}.}
\label{fig:S}
\end{figure}

In Fig.~\ref{fig:S} (left panel), the entropy $S$ is reported as a function of the filling at $U=4$ and various values of temperature $T$. The numerical data  are taken from Ref.~\cite{Bonca_02}. The agreement between COM and numerical results is rather good except for the lower temperature at which quite anomalous oscillations appear in the Lanczos data. In Fig.~\ref{fig:S} (right panel), we report the behavior of the entropy as a function of the temperature in comparison with some Lanczos data \cite{Bonca_02}. The agreement is very good over the whole range of temperature and for any reported value of the filling, except for low temperatures and low doping. The discrepancy can be related to the capability of our data to describe the effects related to the exchange interaction which is more and more important as low are the temperature and the doping.

\subsubsection{Two-pole solution - Bosonic Sector \label{sec:Hub-2pole-Bos}}

There is one more way to tackle the problem of computing the response functions: the two-particle excitations can be considered as a new sector in the dynamics of the system and we can choose also for them a suitable basis alike it has been done for fermions. This approach will be described in this Section. We take as bosonic basis the following one \cite{Mancini_94a,Avella_03,Mancini_04}
\begin{equation}
N^{(\mu)}(i) = \left(
\begin{array}{c}
n_\mu(i) \\
\rho_\mu(i)
\end{array}
\right)
\quad \quad \quad
\begin{array}{c}
n_\mu(i) = c^\dagger(i) \sigma_\mu c(i) \hfill \\
\rho_\mu(i) = c^\dagger(i) \sigma_\mu c^\alpha(i) - c^{\alpha\dagger}(i) \sigma_\mu c(i) \hfill
\end{array} \;.
\label{eq:Hub-Bas-Bos}
\end{equation}

This basis is directly generated by the hierarchy of the equations of motion; this will assure that the first four bosonic spectral moments have the correct functional form \cite{Mancini_98b,Mancini_04}. The field $N^{(\mu )}(i)$ obeys the equation of motion
\begin{equation}
\I\frac{\partial}{\partial t} N^{(\mu)}(i) = J^{(\mu )}(i) = \left(
\begin{array}{c}
- 2dt\rho_\mu(i) \\
- 2dt l_\mu(i) + U \kappa_\mu(i)
\end{array}
\right)
\end{equation}
where the higher-order bosonic operators are defined as $\kappa _\mu (i) = c^\dagger (i)\sigma _\mu \eta ^\alpha (i) - \eta ^\dagger (i)\sigma _\mu c^\alpha (i) + \eta ^{\dagger\alpha } (i)\sigma _\mu c(i) - c^{\dagger\alpha } (i)\sigma _\mu \eta (i)$ and $l_\mu (i) = c^\dagger (i)\sigma _\mu c^{\alpha ^2 } (i) + c^{\dagger\alpha^2}(i)\sigma _\mu c(i) - 2c^{\dagger\alpha } (i)\sigma _\mu c^\alpha (i)$ \cite{Mancini_94a,Avella_03,Mancini_04}. We are using the notation $c^{\alpha ^2 } (\mathbf{i},t) = \sum_\mathbf{j} \alpha _\mathbf{ij}^2 c(\mathbf{j},t) = \sum_\mathbf{jl} \alpha_\mathbf{il} \alpha_\mathbf{lj} c(\mathbf{j},t)$. The current $J^{(\mu )}(i)$ is projected on the basis (\ref{eq:Hub-Bas-Bos}) and the residual current $\delta J^{(\mu)}$ is neglected (truncated basis). The normalization and energy matrices have the expressions
\begin{equation}
I^{(\mu)}(\mathbf{k}) = \left(
\begin{array}{cc}
0 & I_{12}^{(\mu )} (\mathbf{k}) \\
I_{12}^{(\mu )} (\mathbf{k}) & 0
\end{array}
\right) \quad \quad
\varepsilon^{(\mu )}(\mathbf{k}) = \left(
\begin{array}{cc}
0 & -2dt \\
m_{22}^{(\mu)}(\mathbf{k})/I_{12}^{(\mu)}(\mathbf{k}) & 0
\end{array}
\right) \;,
\end{equation}
where $I_{12}^{(\mu )} (\mathbf{k}) = 4[1 - \alpha (\mathbf{k})]C^\alpha$ and $m_{22}^{(\mu )} (\mathbf{k}) = -2dt I_{l_\mu \rho_\mu} (\mathbf{k}) + U I_{\kappa_\mu \rho_\mu}(\mathbf{k})$. The parameter $C^\alpha$ is the electron correlation function $C^\alpha = \langle c^\alpha (i)c^\dagger (i) \rangle$. The quantities $I_{l_\mu \rho_\mu} (\mathbf{k})$ and $I_{\kappa_\mu \rho_\mu}(\mathbf{k})$ are defined as $I_{l_\mu \rho _\mu } (\mathbf{k}) = \mathcal{F} \langle [l_\mu (\mathbf{i},t),\rho _\mu ^\dagger (\mathbf{j},t)] \rangle$ and $I_{\kappa _\mu \rho _\mu } (\mathbf{k}) = \mathcal{F} \langle [\kappa _\mu (\mathbf{i},t),\rho _\mu ^\dagger (\mathbf{j},t)] \rangle$. $\mathcal{F}$ stands for the Fourier transform operator. Because of the nonlocality of the bosonic composite field (\ref{eq:Hub-Bas-Bos}), the analytical form of these quantities depends on the dimensionality of the system and it is necessary to separately discuss the different cases \cite{Mancini_04}.

The causal $G^{C(\mu)}(i,j) = \langle \mathcal{T}[N^{(\mu)}(i)N^{(\mu)\dagger}(j)] \rangle$ and the correlation $C^{(\mu )} (i,j) = \langle N^{(\mu )} (i)N^{(\mu )\dagger} (j) \rangle$ functions are given by
\begin{align}
& G^{C(\mu)}(\mathbf{k},\omega ) = - 2 \I \pi \Gamma^{(\mu )}(\mathbf{k}) + \\
&+ \sum_{n = 1}^2 \sigma^{(n,\mu)}(\mathbf{k})\left[ \frac{1 + f_\mathrm{B}(\omega)}{\omega - \omega_n^{(\mu)} (\mathbf{k}) + \I\delta} - \frac{f_\mathrm{B}(\omega)}{\omega - \omega _n^{(\mu )} (\mathbf{k}) - \I\delta } \right] \\
& C^{(\mu )} (\mathbf{k},\omega ) = 2\pi \Gamma ^{(\mu )} (\mathbf{k})\delta (\omega ) + 2\pi \sum _{n = 1}^2 \delta [\omega - \omega_n^{(\mu )} (\mathbf{k})] [1 + f_\mathrm{B} (\omega )] \sigma ^{(n,\mu )} (\mathbf{k}) \;.
\end{align}

The energy spectra $\omega _n^{(\mu )} (\mathbf{k})$ are given by $\omega _n^{(\mu )} (\mathbf{k}) = ( - )^n \sqrt {\varepsilon _{12}^{(\mu )} (\mathbf{k})\varepsilon _{21}^{(\mu )} (\mathbf{k})}$ and the spectral functions $\sigma ^{(n,\mu )} (\mathbf{k})$ have the following expression
\begin{equation}
\sigma ^{(n,\mu )} (\mathbf{k}) = \frac{{I_{12}^{(\mu )} (\mathbf{k})}}
{{2\omega _n^{(\mu )} (\mathbf{k})}}\left( {\begin{array}{cc}
 {\varepsilon _{12}^{(\mu )} (\mathbf{k})} & 1 \\
 1 & {\varepsilon _{21}^{(\mu )} (\mathbf{k})}
 \end{array} } \right)
\end{equation}

The determination of $G^{C(\mu)}(\mathbf{k},\omega )$ and $C^{(\mu )} (\mathbf{k},\omega )$ requires the knowledge of a set of parameters: correlation functions of fermionic and bosonic operators and the unknown $\Gamma$ function. According to the self-consistent scheme given in Sect.~\ref{sec:Self-cons}, these parameters are fixed by means of the following constraints: (i) all fermionic correlators are calculated within the fermionic sector (see previous subsection); (ii) the conservation of the current; (iii) the condition that the spin susceptibility be a single value finite function; (iv) the local algebra constraint $\langle n_\mu (i)n_\mu (i) \rangle = \langle n \rangle + 2(2\delta _{\mu ,0} - 1)D$ where $D$ is the double occupancy. The unknown $\Gamma ^{(\mu )} (\mathbf{k})$ function can be calculated either by assuming the ergodicity of the system (e.g. $\Gamma _{11}^{(\mu )} (\mathbf{k}) = \delta _{\mu ,0} (2\pi /a)^d \delta ^{(d)} (\mathbf{k})n^2$) or opening a new bosonic sector (the pair sector) [see Ref.~\cite{Avella_03f}]. Once the self-consistent equations in the fermionic and in the bosonic sectors have been solved, we can calculate the Green's function and the charge and spin susceptibilities $\chi_\mu(\mathbf{k},\omega) = -\mathcal{F} \langle \mathcal{R}[n_\mu (i)n_\mu (j)] \rangle$ as well as the spin correlation functions $S(\mathbf{k}) = \mathcal{F} \langle n_3 (\mathbf{i},t) n_3 (\mathbf{j},t) \rangle$:
\begin{equation}
\chi _\mu (\mathbf{k},\omega ) = -\sum_{n = 1}^2 \frac{{\sigma ^{(n,\mu )} (\mathbf{k})}}
{{\omega - \omega _n^{(\mu )} (\mathbf{k}) + \I\delta }}
\quad \quad
S(\mathbf{k}) = -4dtC^\alpha \frac{{1 - \alpha (\mathbf{k})}}
{{\omega_2 ^{(3)} (\mathbf{k})}}\coth \frac{{\beta \omega_2 ^{(3)} (\mathbf{k})}}{2}
\end{equation}

\paragraph{Results and comparisons \label{sec:Hub-2pole-Bos-Res}}

\subparagraph{Spin correlation function \label{sec:Hub-2pole-Bos-Res-Sk}}

\begin{figure}[tbp]
\centering
\includegraphics[width=.49\textwidth]{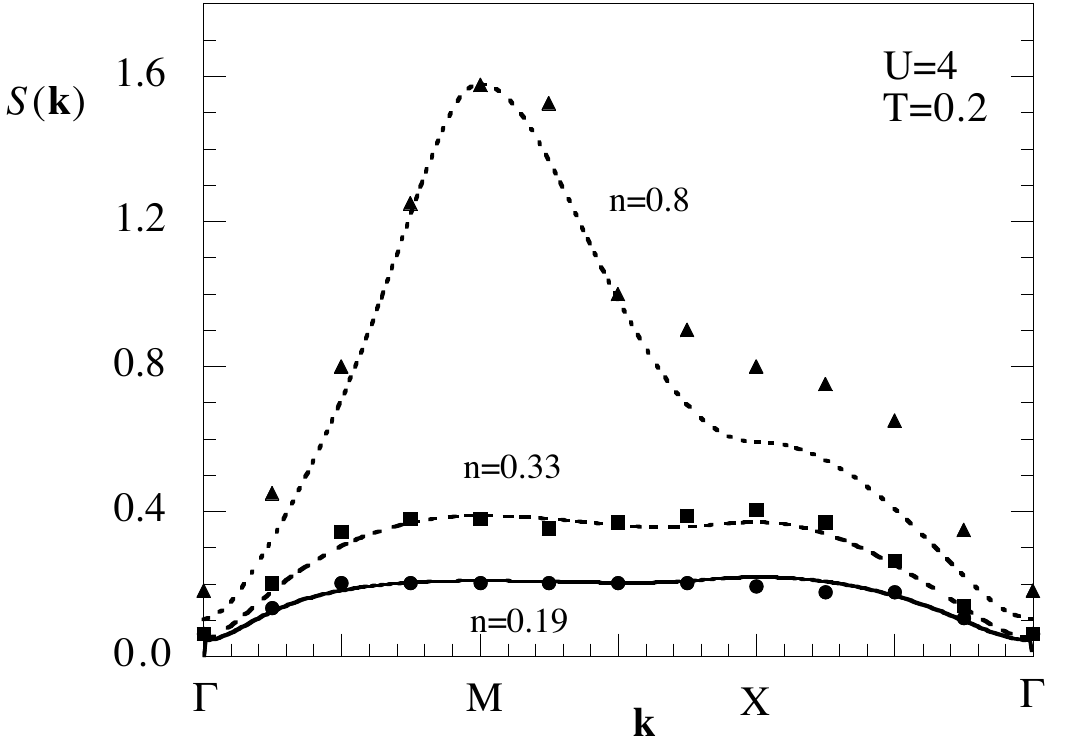}
\includegraphics[width=.49\textwidth]{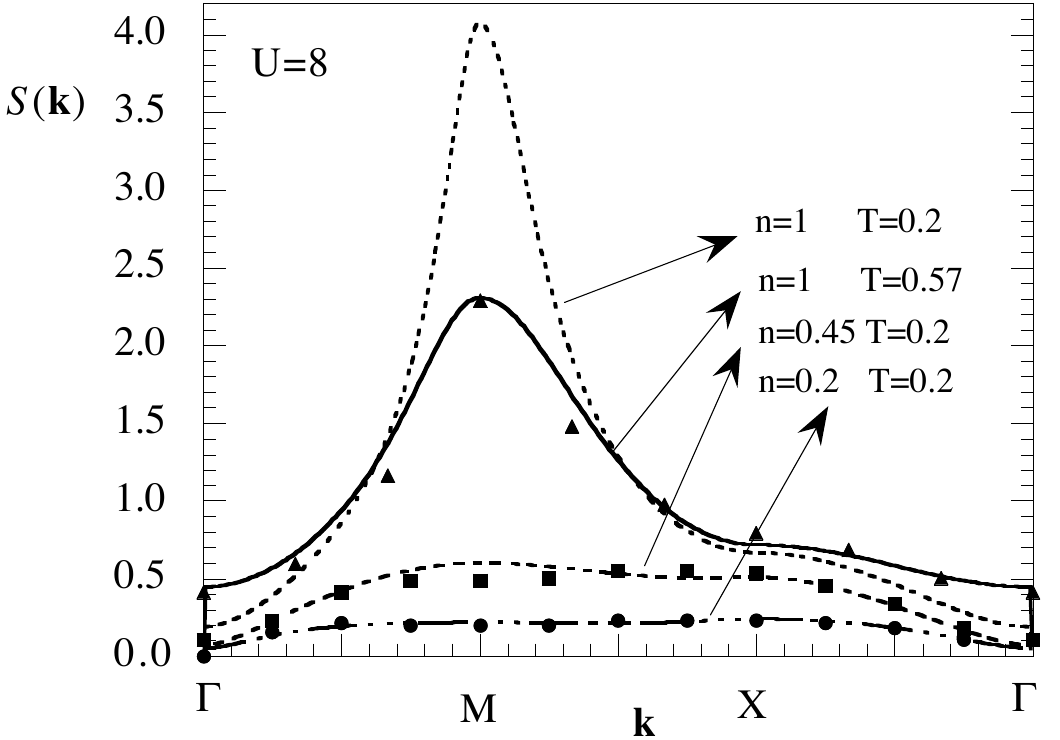}
\caption[]{Spin correlation function $S({\bf k})$ versus momentum ${\bf k}$ for: (left) $U=4$, $T=0.2$ and $n=0.8$, $0.33$ and $0.19$; (right) $U=8$, $T=0.2$ and $T=0.57$ and $n=1$, $0.45$ and $0.2$. qMC data ($8\times8$) are taken from Ref.~\cite{Vilk_94}.}
\label{fig:Sk}
\end{figure}

The behavior of $S ({\bf k} )$, as a function of the momentum, is reported in Fig.~\ref{fig:Sk} in comparison with some numerical data \cite{Vilk_94} for different values of filling, Coulomb repulsion and temperature. We have a very good agreement with the numerical results for sufficiently high values of the doping for all shown values of the Coulomb repulsion. In the proximity of half-filling the numerical data suffer from a saturation of the antiferromagnetic correlation length \cite{White_89} that becomes comparable with the size of the cluster. For $U=4$ and $n=0.8$ (see Fig.~\ref{fig:Sk} (left)), the correlation length is slightly smaller than the size of the cluster: our solution results capable to describe this situation fairly well (the height of the peak at ${\bf Q} $ is exactly reproduced and the momentum dependence is qualitatively correct, again practically exact along the diagonal) except for the exact value of the numerical data along the main axes. This is probably due again to an overestimation of the correlations by the numerical analysis owing to the finite size of the cluster. For $U=8$ and $n=1$ (see Fig.~\ref{fig:Sk} (right)), in order to reproduce the numerical data we need to increase the temperature as to decrease our value of the correlation length and match that of the numerical analysis, which is stuck at the saturation value due to the finiteness of the clusters. The results of such a procedure are astonishing, we manage to exactly reproduce the numerical data for any value of the momentum, and not only at ${\bf Q} $, revealing the correctness and power of our approach and the limitations of the numerical analysis.

\subparagraph{Charge correlation function \label{sec:Hub-2pole-Bos-Res-Nk}}

\begin{figure}[tbp]
\sidecaption[t]
\centering
\includegraphics[width=.49\textwidth]{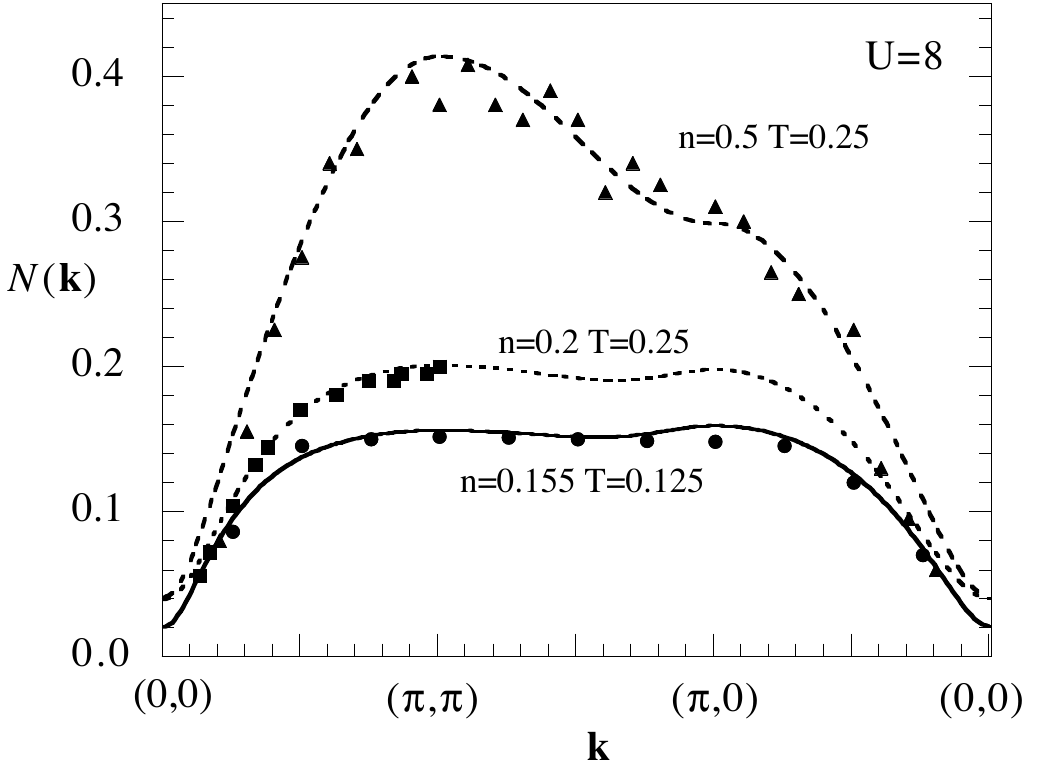}
\caption[]{Charge correlation function $N({\bf k})$ versus momentum ${\bf k}$ for $U=8$, $T=0.125$ and $0.25$ and $n=0.155$, $0.2$ and $0.5$; qMC data ($8\times8$, $12\times12$, $16\times16$) are from Ref.~\cite{Chen_94}.}
\label{fig:Nk}
\end{figure}

$N ({\bf k} )$ is reported in Fig.~\ref{fig:Nk}, as a function of the momentum, for various fillings and temperatures and $U=8$. We have again a very good agreement with quantum Monte Carlo results \cite{Chen_94} for all shown values of the external parameters and of the momentum. The enhancement at ${\bf k} ={\bf Q} =M$ for $n=0.5$ can be interpreted as the manifestation of a rather weak ordering of the charge with a checkerboard pattern. COM results manage to reproduce such double peak structure showing a capability to quantitatively describe, also in a translational invariant phase, rather strong charge correlations.

\subparagraph{Comparison with the experimental data for $La_2CuO_4$ \label{sec:Hub-2pole-Bos-Res-LCO}}

\begin{figure}[tbp]
\centering
\includegraphics[width=.49\textwidth]{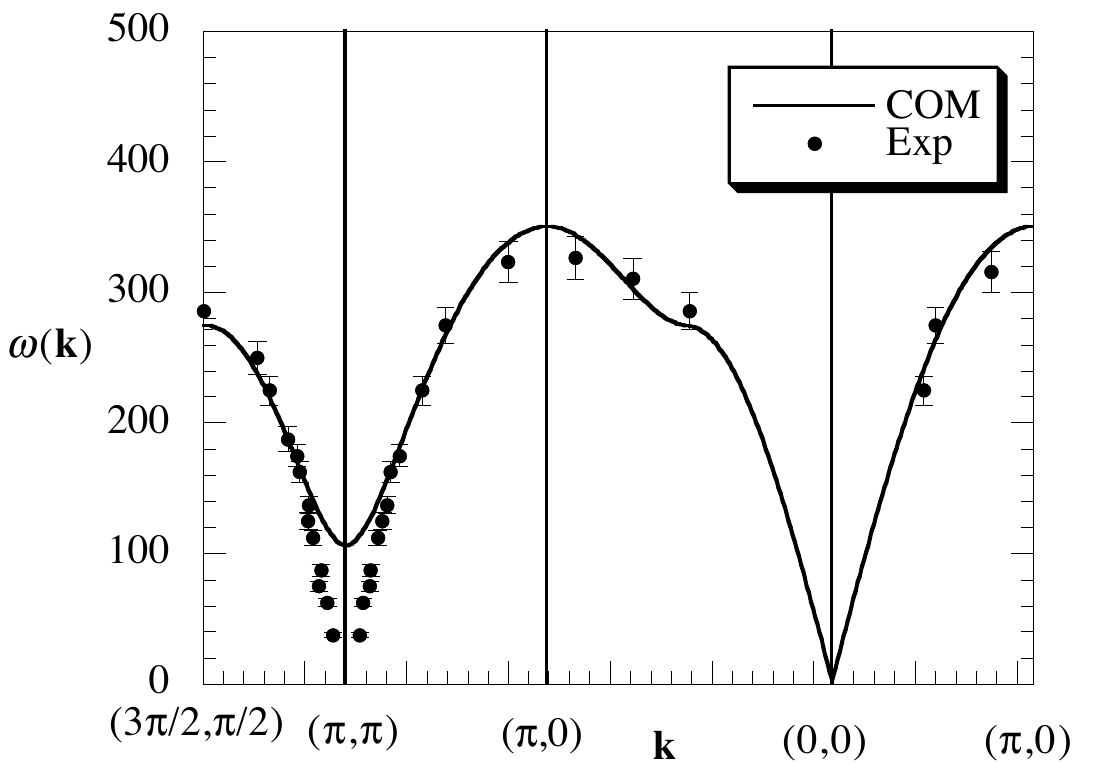}
\includegraphics[width=.49\textwidth]{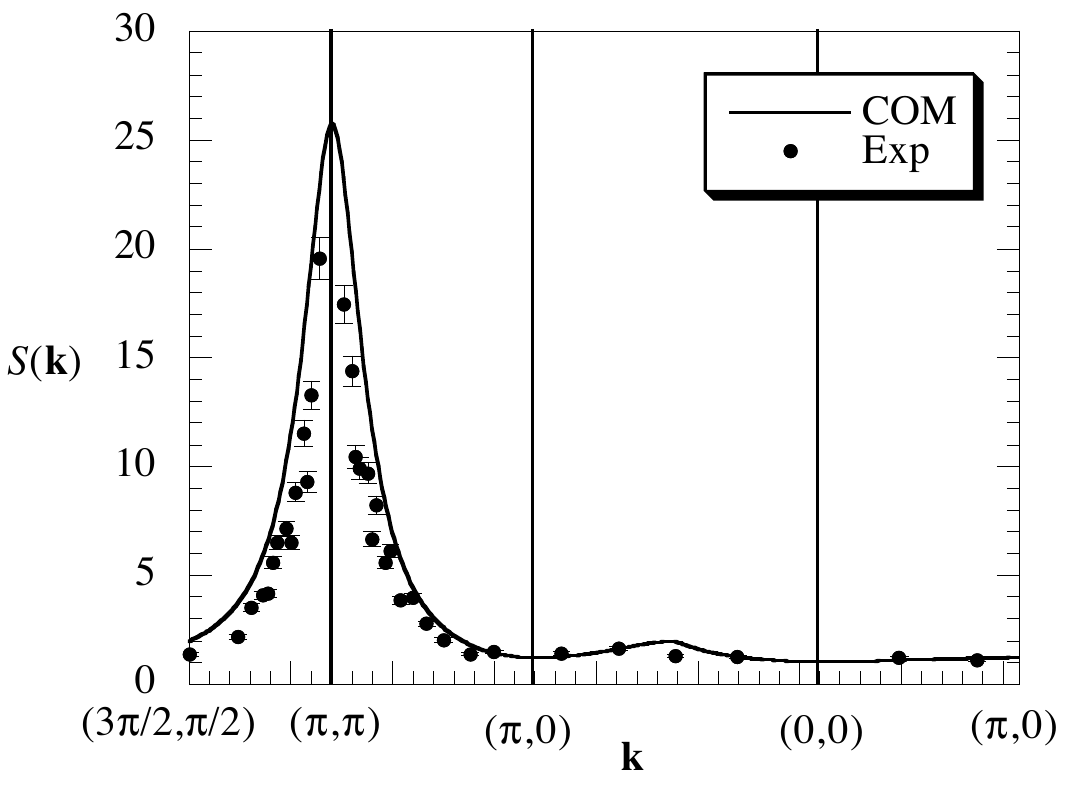}
\caption[]{(left) Spin spectrum $\omega({\bf k})$ along the PD for $n=1$, $T=10K \approx 0.003t$ and $U=8.8t$ ($t\approx 0.3eV$). (right) The spin-wave intensity along the PD for $n=1$, $T=295K \approx 0.08t$ and $U=8.8t$ ($t\approx 0.3eV$). Experimental data are taken from Ref.~\cite{Coldea_01}.}
\label{fig:LCO}
\end{figure}

In Fig.~\ref{fig:LCO} (left), we report the energy spectrum of the spin-spin propagator along the Principal Directions and compare with the experimental data of Ref.~\cite{Coldea_01} obtained for $La_2CuO_4$ by means of inelastic magnetic neutron scattering. We have fixed the temperature at $T=10K$ according to the experimental value; the value of the transfer integral ($t=0. 3eV$) and of the Coulomb repulsion ($U=8.8t$) have been chosen in order to fit the experimental points and they are within the ranges ($t=0.3 \pm 0.02eV$; $U=2.2 \pm 0.4eV$) suggested in Ref.~\cite{Coldea_01}. The agreement with the experimental data is very good all over the momentum space except around $(\pi,\pi)$. As a matter of fact, the experimental data refer to the antiferromagnetic phase of the material [the experimental spectrum gets completely soft at $ (\pi, \pi )$] and our paramagnetic solution obviously cannot fully describe such behavior. However, it is worth noting that the Hubbard model at half filling presents so strong antiferromagnetic correlations that they also show in the paramagnetic phase through a quite deep minimum. The spin-wave intensity is reported in Fig.~\ref{fig:LCO}  (right) and again compared with the experimental data. The agreement is again very good over all the momentum space and shows once more the capability of the Hubbard model within our formulation to catch the physics of such a strongly correlated system.

\subparagraph{Incommensurability \label{sec:Hub-2pole-Fer-Res-I}}

\begin{figure}[tbp]
\centering
\includegraphics[width=.49\textwidth]{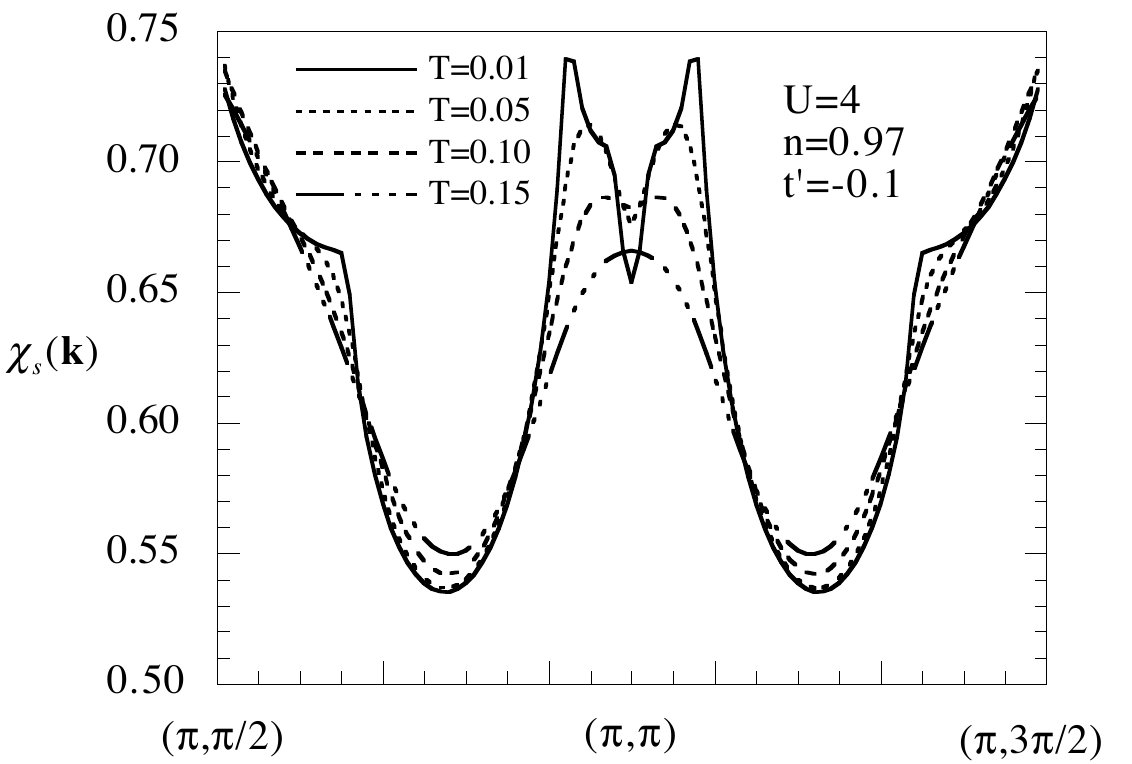}
\includegraphics[width=.49\textwidth]{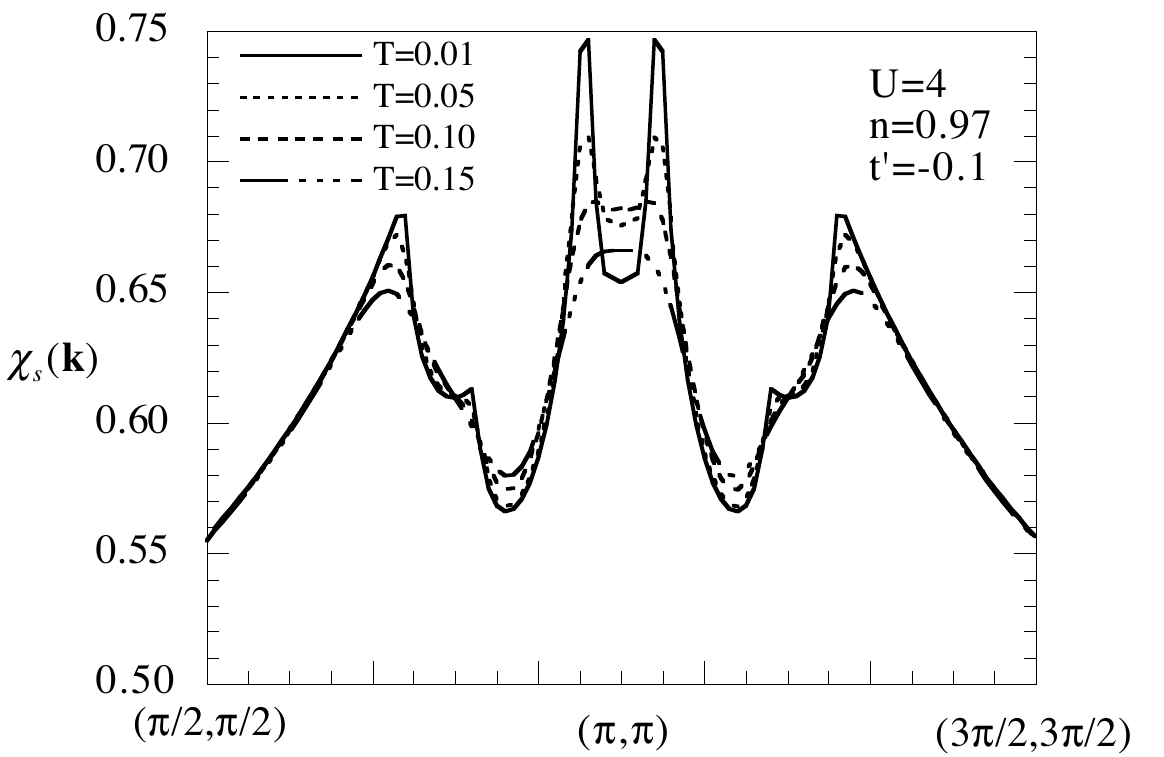}
\caption[]{Static spin susceptibility $\chi_s({\bf k})$ for various values of temperature $T$, $U=4$, $t'=-0.1$ and $n=0.97$ along (left) ${\bf k}=(\pi,k)$ and (right) ${\bf k}=(k, k)$.}
\label{fig:I1}
\end{figure}

\begin{figure}[tbp]
\centering
\includegraphics[width=.49\textwidth]{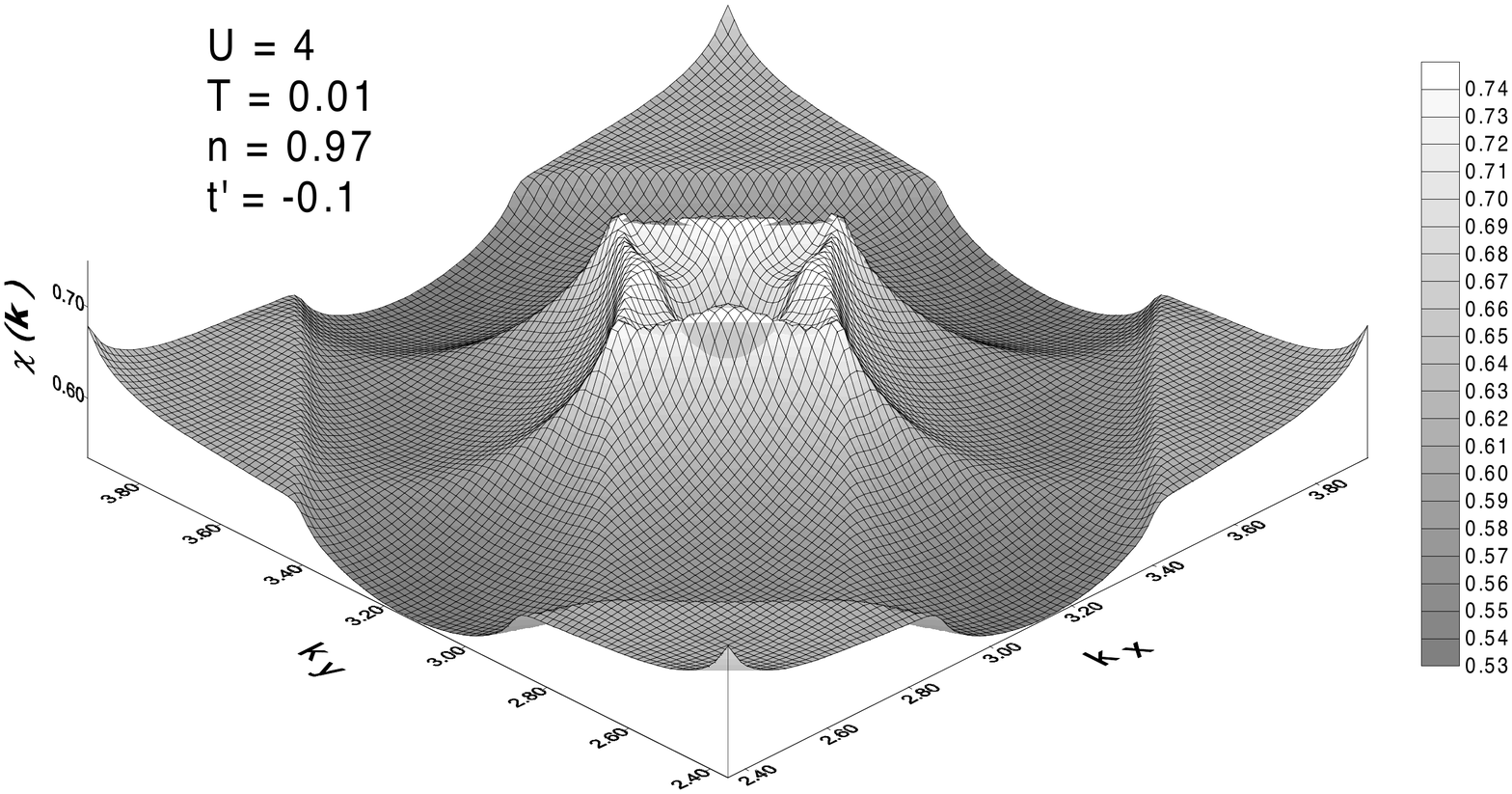}
\includegraphics[width=.49\textwidth]{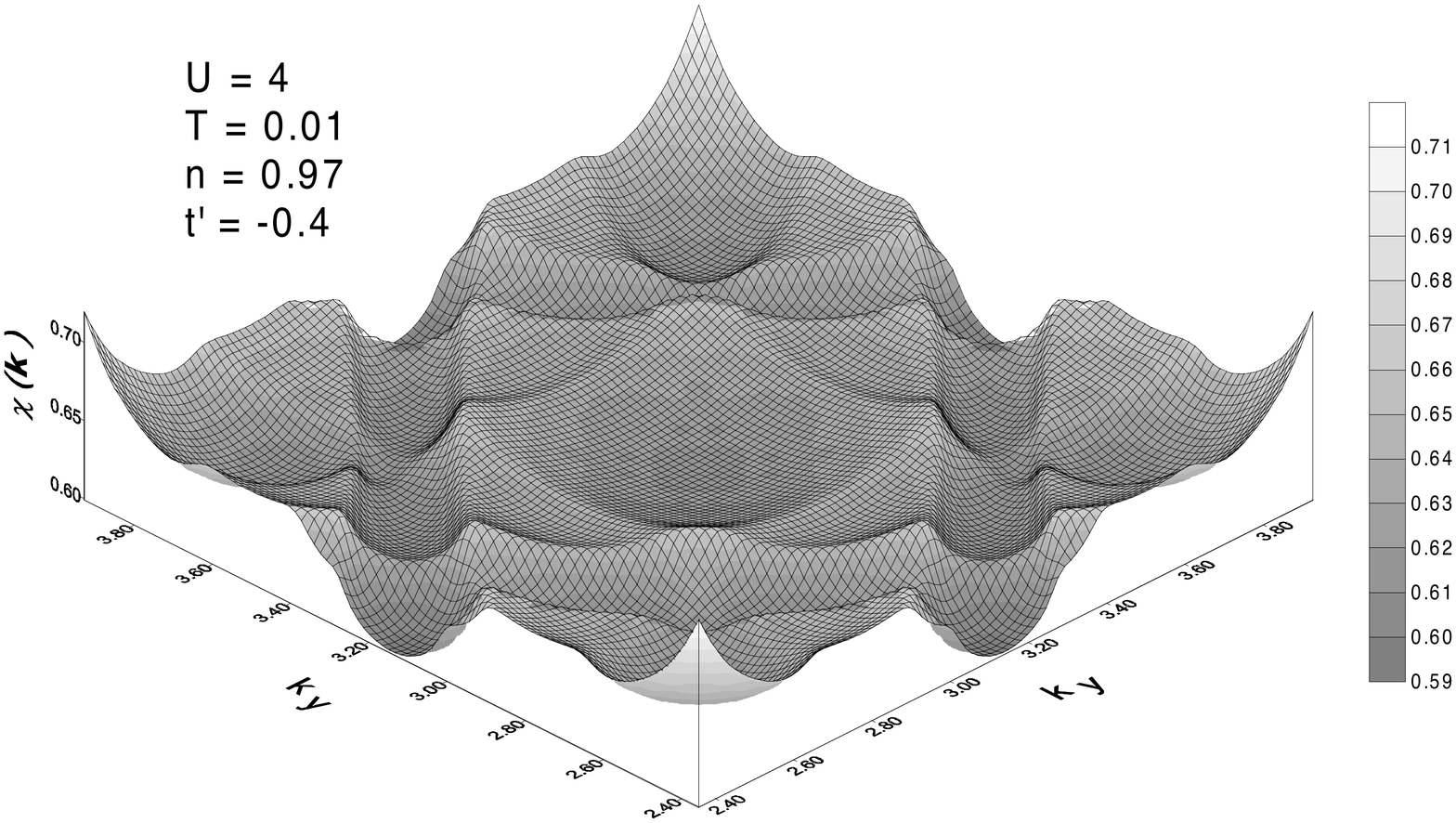}
\caption[]{Static spin susceptibility $\chi_s({\bf k})$ for $U=4$, $n=0.97$ and $T=0.01$ and (left) $t'=-0.1$ [(right) $t'=-0.4$].}
\label{fig:I2}
\end{figure}

\begin{figure}[tbp]
\centering
\includegraphics[width=.49\textwidth]{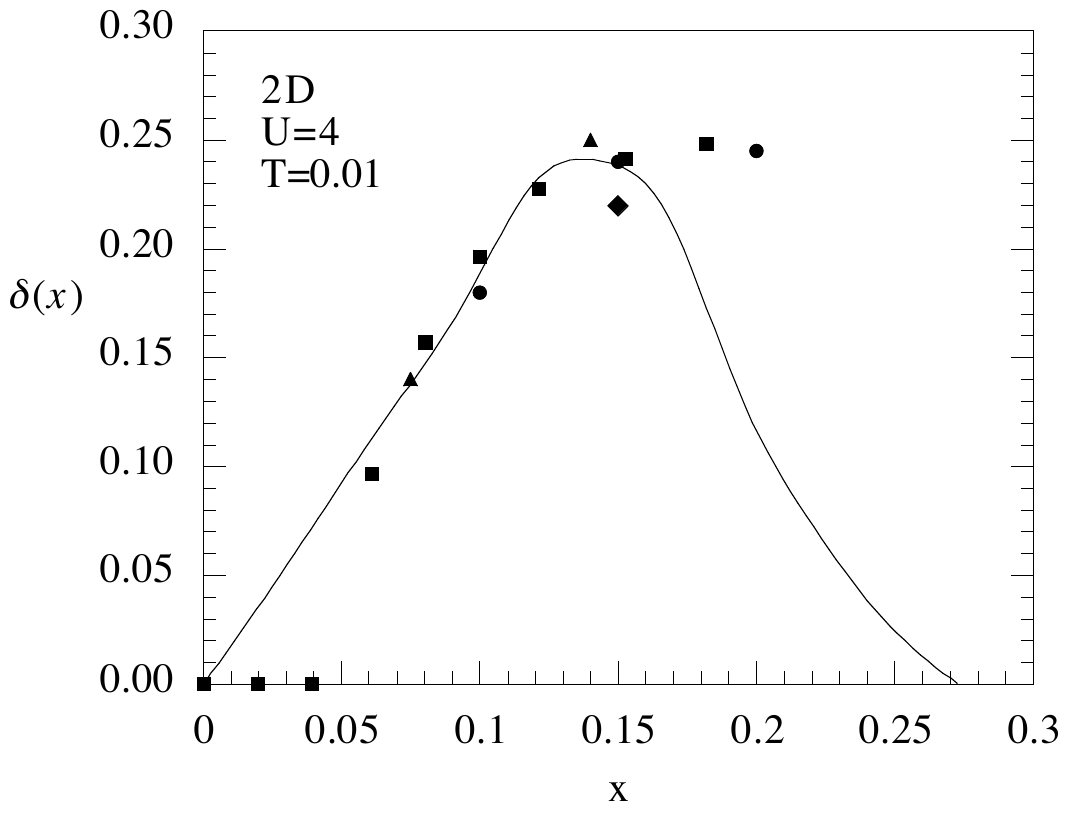}
\includegraphics[width=.49\textwidth]{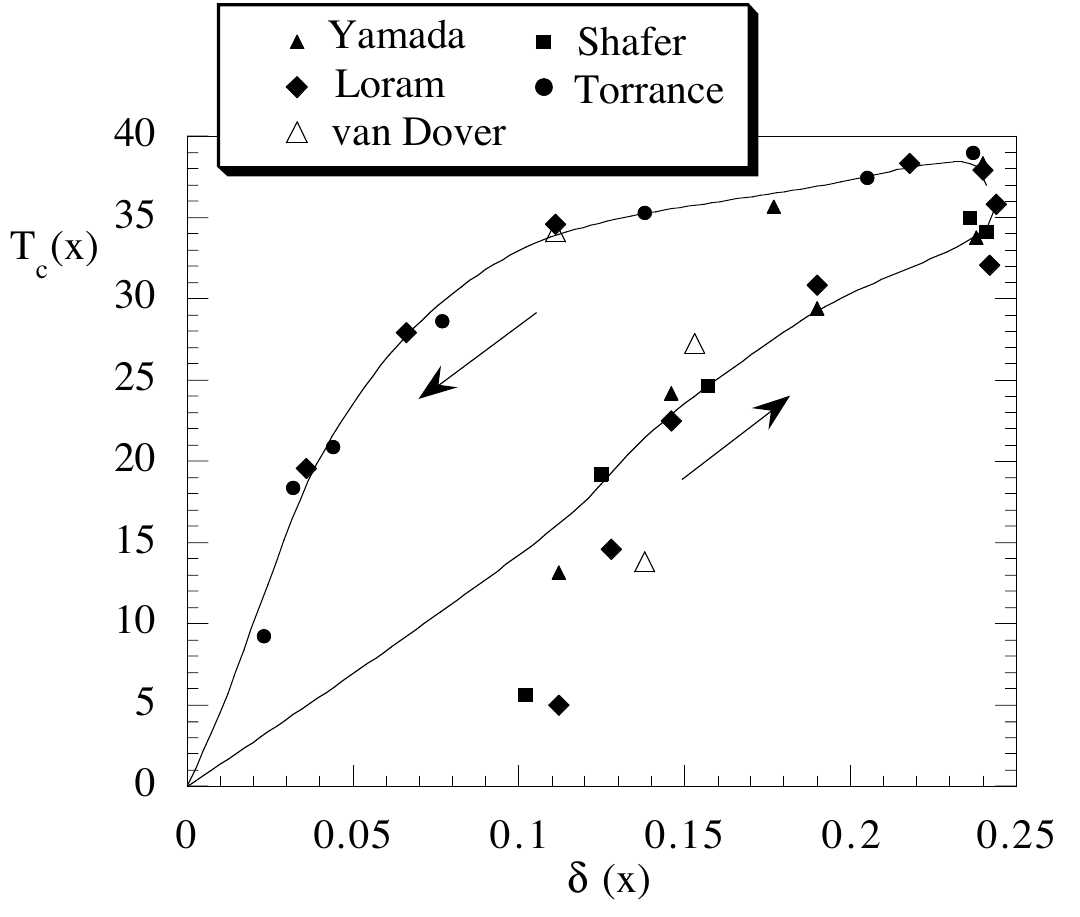}
\caption[]{(left) Incommensurability amplitude $\delta (x)$ versus doping $x$. COM results (solid line) refer to $U=4$ and $T=0.01$. (right) Experimental values of critical temperature $T_c(x)$ for $La_{2-x}Sr_xCuO_4$, taken from Yamada \cite{Yamada_96}, Loram \cite{Loram_89}, van Dover \cite{vanDover_87}, Shafer \cite{Shafer_87} and Torrance \cite{Torrance_88}, versus calculated incommensurability amplitude $\delta(x)$. The solid line is a guide to the eyes.}
\label{fig:I3}
\end{figure}

In Fig.~\ref{fig:I1}, we present the static spin susceptibility $\chi _s ({\bf k} )$ for various temperatures. The choice for the parameters is $U=4$, $t'=-0.1$ and the particle filling has been fixed as $n=0.97$. $t'$ is the next-nearest-hopping integral and the related COM formulation (i.e., for the $t$-$t'$-$U$ Hubbard model) in the $2$-pole approximation can be found in Refs.~\cite{Avella_97,Avella_97b,Avella_98d,Munzner_00,Avella_99,Avella_00b,Avella_02g,Avella_03g}. In (left) and (right), $\chi _s ({\bf k} )$ is reported along the lines ${\bf k} = (\pi, k)$ and ${\bf k} = (k, k)$, respectively. In both cases by increasing the temperature the incommensurate double-peak structure becomes a broad maximum centered at $ (\pi, \pi )$. The intensity over the whole Brillouin zone can be clearly observed in Fig.~\ref{fig:I2}. The important features of the data are: the overall square symmetry of the scattering with the sides of the square parallel to the $ (k, k)$ and to the $ (k, -k)$ lines and the accumulation of intensity near the corners of the square. These features reproduce the experimental situation for $La_{2-x} (Ba,Sr)_xCuO_4$ \cite{Thurston_89,Shirane_89,Birgenau_88,Cheong_91,Shirane_94,Yamada_95,Yamada_96,Petit_96}. In In Fig.~\ref{fig:I3} (left panel), the incommensurability amplitude $\delta (x)$ is shown as a function of doping. For comparison we report the experimental data of Refs.~\cite{Cheong_91,Shirane_94,Yamada_96,Petit_96}. The linear behavior of $\delta (x)$, observed in the low-doping region, agrees exceptionally well with the experimental data. One of the most striking features of the results presented in Fig.~\ref{fig:I3} (left panel) is the resemblance between the incommensurability amplitude $\delta(x)$ and the critical temperature $T_c$. $\delta(x)$ is maximum in the region of optimal doping where $T_c$ is maximum. It has already been experimentally observed in Ref.~\cite{Yamada_96} that there is a linear relation between $\delta(x)$ and $T_c$ up to the optimal doping level $x\approx 0.15$. Our theoretical results confirm this behavior and show that a close similarity between $\delta(x)$ and $T_c$ exists in the entire region of doping. This can be seen in Fig.~\ref{fig:I3} (right panel) where experimental values for $T_c$, taken from Refs. \cite{Yamada_96,Loram_89,vanDover_87,Shafer_87,Torrance_88}, are reported versus our calculated incommensurability amplitude $\delta(x)$.

\subsection{The residual self-energy $\Sigma(\mathbf{k},\omega)$ \label{sec:Hub-Sigma}}

We now come to the problem of taking into account the residual self-energy $\Sigma (\mathbf{k},\omega)$ appearing in the full propagator $G(\mathbf{k},\omega )$ once the truncated basis (\ref{eq:Hub-Bas}) has been adopted. In particular, we have computed \cite{Avella_03c,Krivenko_04,Avella_07,Avella_07a,Avella_07b,Avella_07c,Avella_08,Avella_09} $\Sigma (\mathbf{k},\omega )$ in the non-crossing approximation (NCA) by noting that it was possible to exactly rewrite the residual current $\delta J(i)$ as the sums of products of a bosonic operator, embodying either charge $n(i)$, spin $\vec{n}(i)$ or pair $p(i)$ operator, and of a fermionic operator, which turned to be either $\xi(i)$ or $\eta(i)$. Then, following the prescription of NCA, we expressed the residual self-energy $\Sigma(\mathbf{k},\omega)$ as sums of products of the charge-charge and spin-spin propagators (we neglected pair-pair propagator) and of the fermonic propagator. The bosonic propagators have been computed in the $2$-pole approximation as described in the previous section.

\begin{figure}[tbp]
\centering
\includegraphics[width=.7\textwidth]{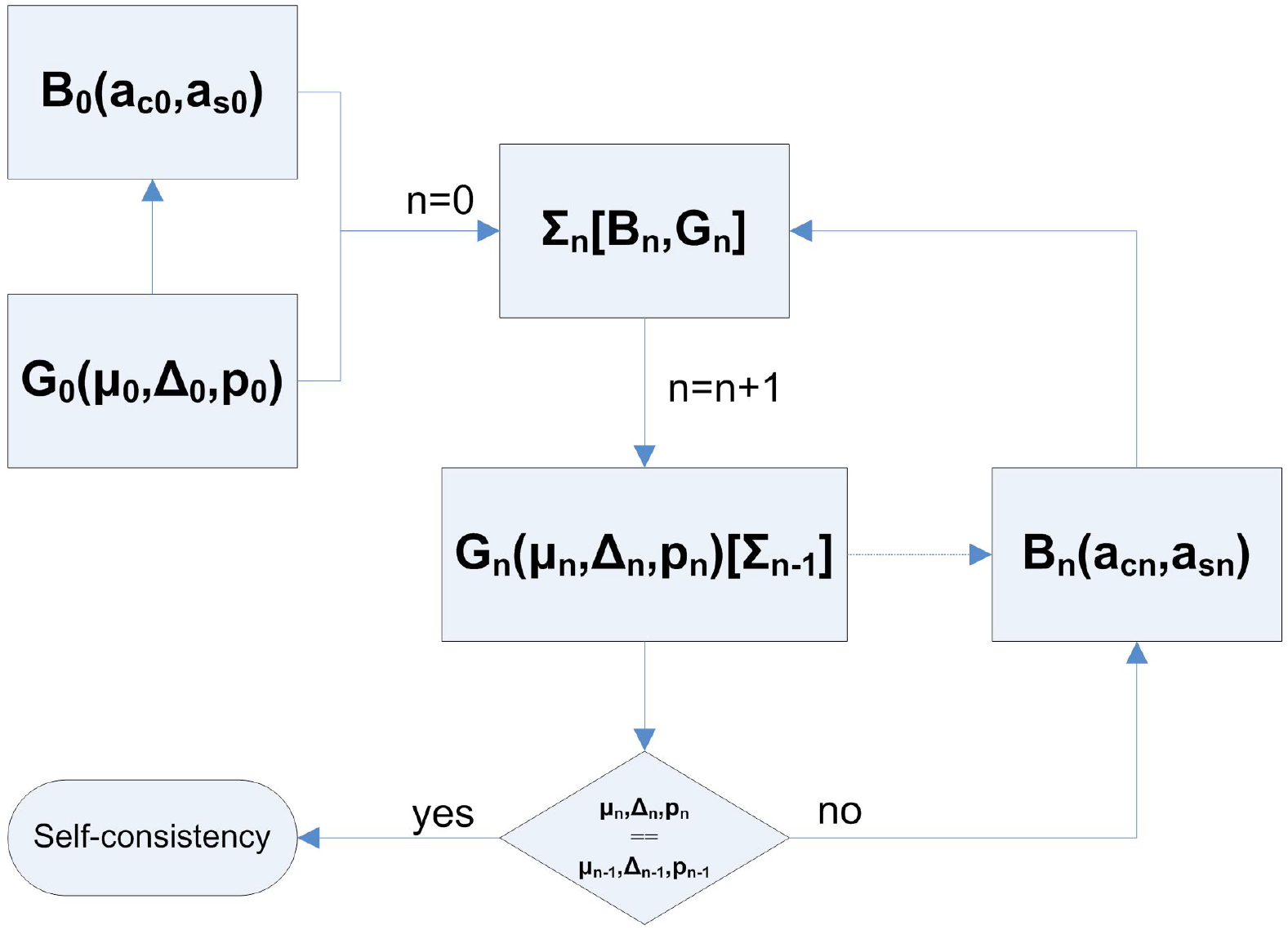}
\caption[]{Self-consistency scheme to compute the fermionic full propagator $G$ in terms of the charge-charge and spin-spin $2$-pole propagator $B$ and the residual self-energy $\Sigma$ in the COM(2p+NCA(2p)).}
\label{fig:NCAI}
\end{figure}

The whole framework closes the self-consistency cycle reported in Fig~\ref{fig:NCAI}, where $G_0$ and $B_0$ are the fermionic and bosonic (charge and spin) propagators in the $2$-pole approximation (in parentheses the unknown parameters they depend on), $\Sigma_n$ is the residual self-energy at the $n^{th}$ step (it depends on $G_n$ and $B_n$, which are equal to $G_0$ and $B_0$ at the startup), $G_n$ is the full fermionic propagator (it depends on the unknowns in round parentheses and on $\Sigma_{n-1}$) and, finally, $B_n$ is the bosonic (charge and spin) propagators in the $2$-pole approximation at the $n^{th}$ step (it depends on $G_n$, besides unknown parameters). The self-consistency cycle terminates when the fermionic parameters $\mu$, $\Delta$, and $p$ do not change; to get six-digit precision for the latter fermionic parameters, we usually need about ten cycles (it varies very much with doping, temperature, and interaction strength) on a $3D$ grid of $128 \times 128$ points in momentum space and $4096$ Matsubara frequencies.

\begin{figure}[tbp]
\centering
\includegraphics[width=.98\textwidth]{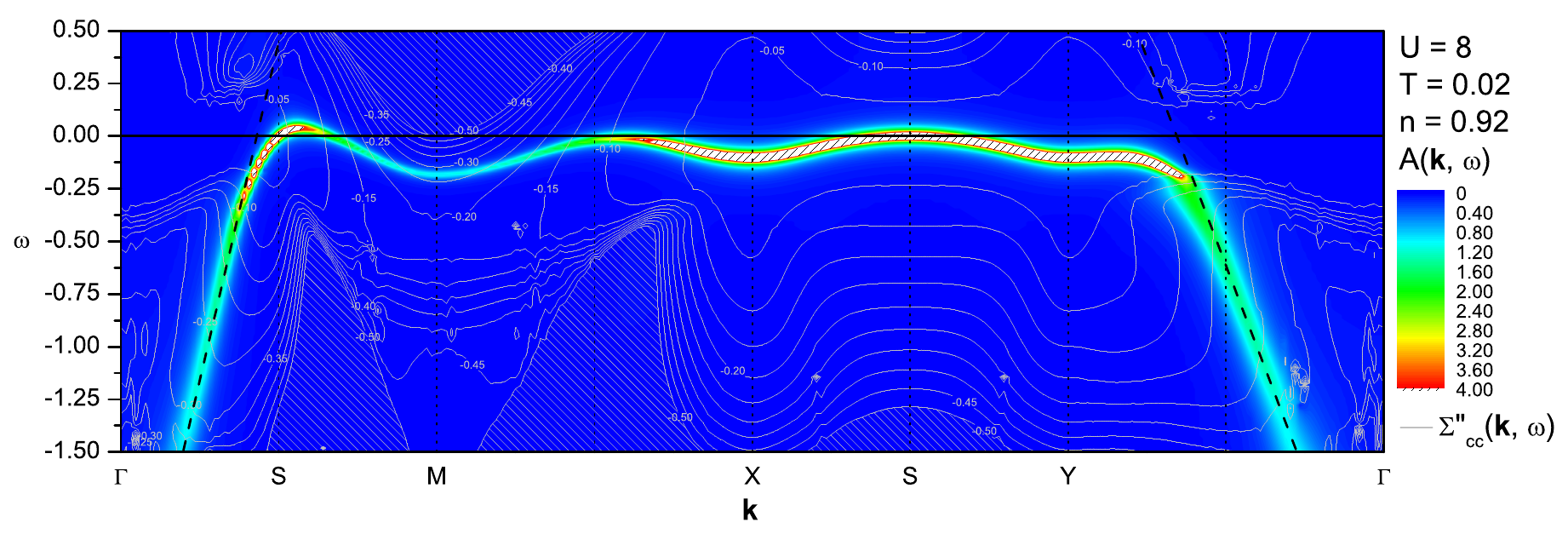}
\includegraphics[width=.49\textwidth]{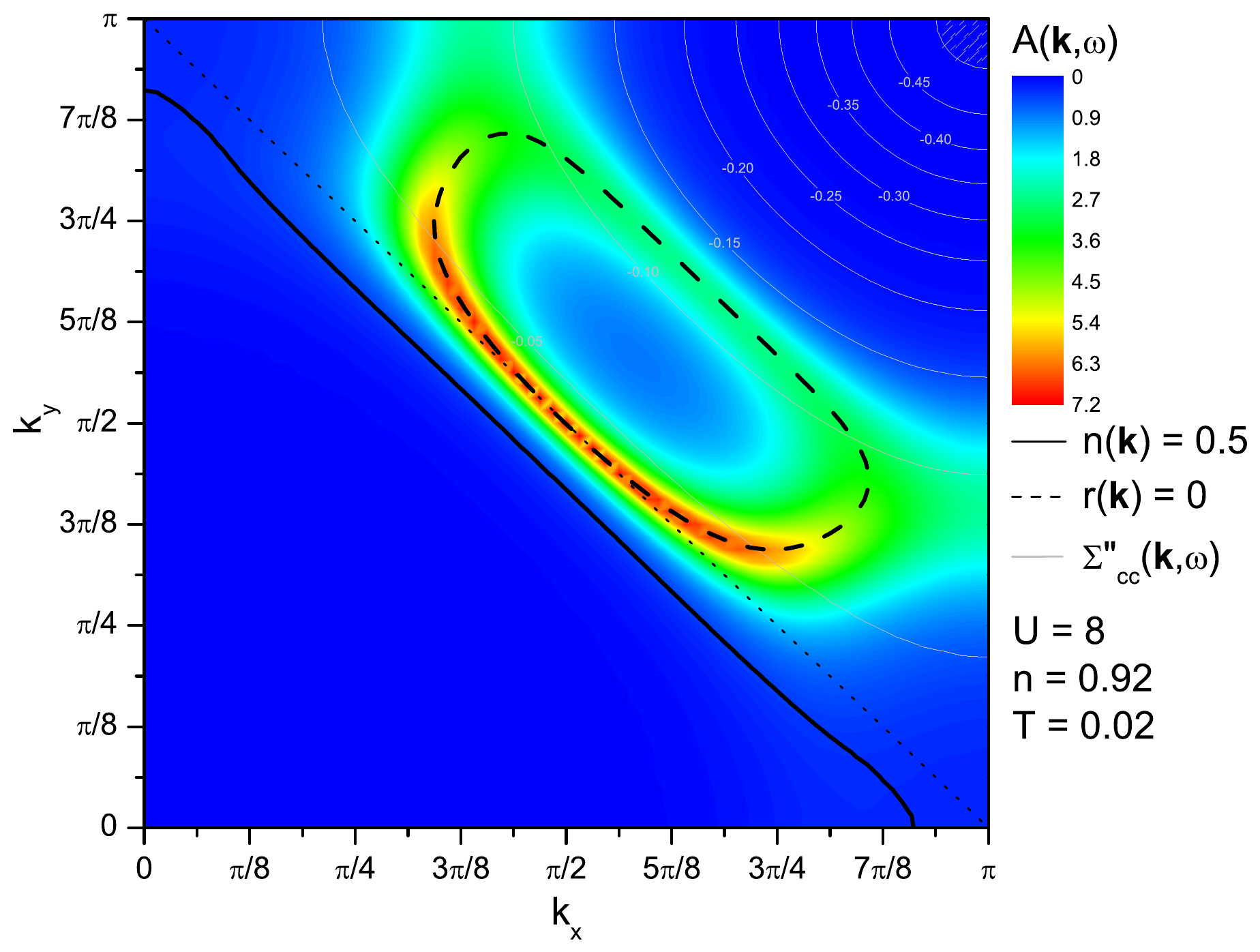}
\includegraphics[width=.49\textwidth]{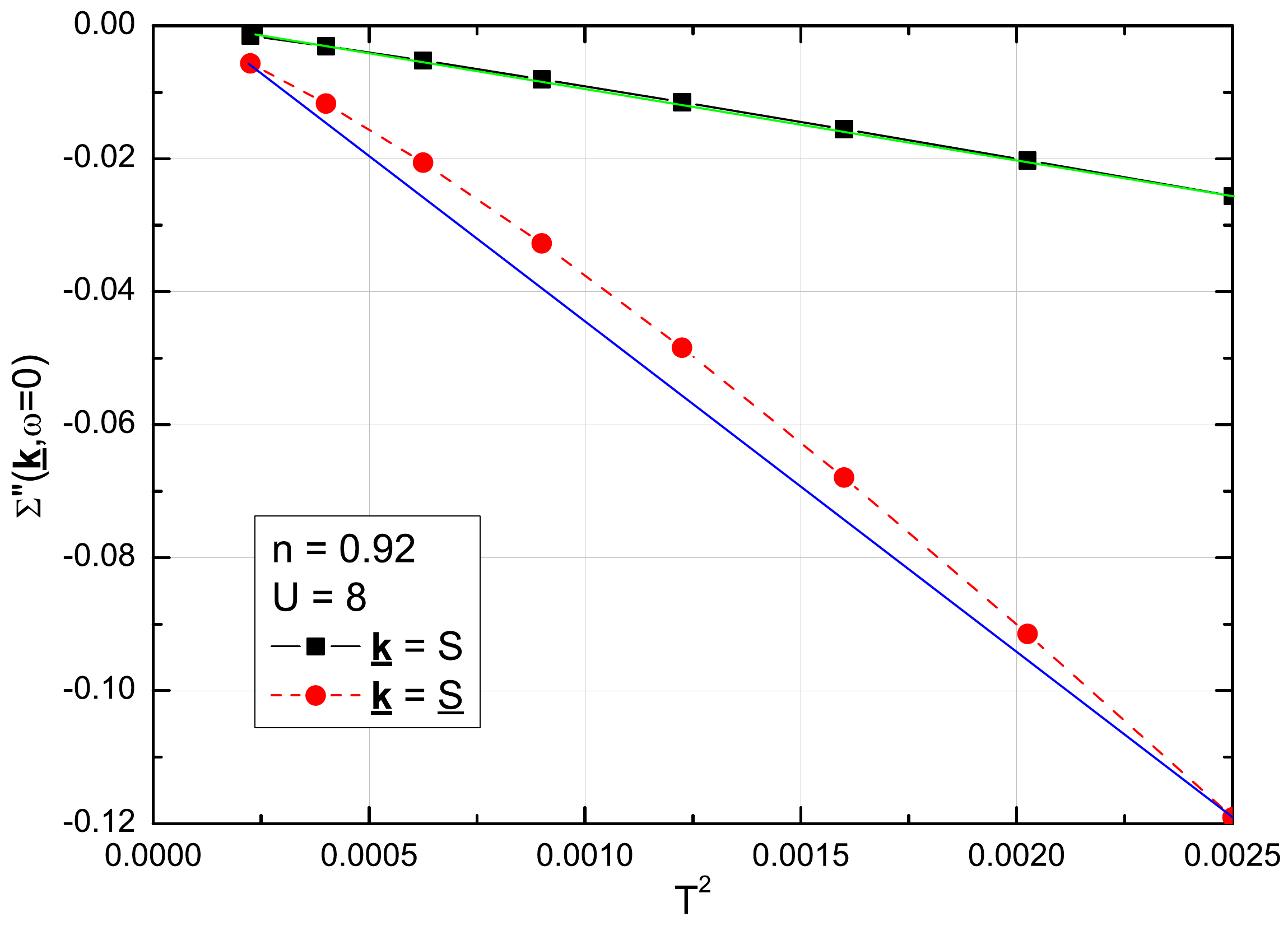}
\caption[]{$U=8$, $n=0.92$ and $T=0.02$. (top) Spectral function $A(\mathbf{k},\omega)$ along the principal directions. (bottom left) Spectral function at the chemical potential $A(\mathbf{k},\omega=0)$ as a function of momentum $\mathbf{k}$. (bottom right) Imaginary part of the self-energy at the chemical potential $\Sigma''(\mathbf{k},\omega=0)$ as a function of temperature squared $T^2$ at two momenta: $S=(\pi/2,\pi/2)$ and $\underline{S}$, where the main diagonal of Brillouin zone touches the \emph{phantom} portion of the hole pocket.}
\label{fig:NCAII}
\end{figure}

In Fig.~\ref{fig:NCAII} (top panel), the spectral function $A(\mathbf{k},\omega)$ along the principal directions ($\Gamma=(0,0) \to S=(\pi/2,\pi/2) \to M=(\pi,\pi) \to X=(\pi,0) \to S=(\pi/2,\pi/2) \to Y=(0,\pi) \to \Gamma=(0,0)$) is reported for Coulomb repulsion $U=8$, filling $n=0.92$ and temperature $T=0.02$ in the frequency range in the proximity of the chemical potential ($\omega=0$). We can clearly see may unconventional features: kinks in the dispersion along the main diagonal ($\Gamma \to M$) and the side of the Brillouin zone ($M \to X,Y$) at energies comparable with the effective exchange energy in the system ($J=4t^2/U$); extended regions in momentum where the imaginary part of the self-energy is strong enough to selectively suppress the spectral weight; almost doubling of the Brillouin zone according to the rather high intensity of the not-so-short-range anti-ferromagnetic correlations; formation of a hole pocket centred along the main diagonal ($\Gamma \to M$); formation of almost-closed electron pockets centred at $X=(\pi,0)$ and $Y=(0,\pi)$; high-intensity of the spectral weight at $X=(\pi,0)$ and $Y=(0,\pi)$ (van Hove points) although well below the Fermi surface; no-flat dispersion along the anti-diagonal ($X \to Y$) generated dynamically ($t'$ is not present in the original Hamiltonian); strong suppression of the spectral weight in proximity of $M=(\pi,\pi)$ (again, due to the rather high intensity of the not-so-short-range anti-ferromagnetic correlations) which will lead to the appearance of a pseudogap. This scenario is just the one claimed for high-$T_c$ cuprates in the underdoped region by ARPES and many other experimental techniques.

In Fig.~\ref{fig:NCAII} (bottom left panel), the spectral function at the chemical potential $A(\mathbf{k},\omega=0)$ as a function of momentum $\mathbf{k}$ is reported for Coulomb repulsion $U=8$, filling $n=0.92$ and temperature $T=0.02$. The hole pocket and the electron quasi-pocket are now clearly visible. The Fermi surface is deconstructed: there is coexistence of a small Fermi surface (the hole pocket) and of a large one partially solving the hoary dichotomy signalled by main experimental techniques; the actual Fermi surface, the locus formed by the momenta where the spectral weight reaches the highest intensity (this also the only Fermi surface detectable by ARPES), is open (it is just an arc -- half of the hole pocket) in contrast to what Fermi liquid picture would require; it is just the imaginary part of the residual self-energy to \emph{eat up} a portion of the hole pocket and make this latter just a \emph{phantom} arc. The residual self-energy results stronger and stronger close to $M=(\pi,\pi)$ as one should expect for high-intensity not-so-short-range anti-ferromagnetic correlations.

In Fig.~\ref{fig:NCAII} (bottom right panel), the imaginary part of the self-energy at the chemical potential $\Sigma''(\mathbf{k},\omega=0)$ as a function of the temperature squared $T^2$ at two momenta: $S=(\pi/2,\pi/2)$ and $\underline{S}$. The latter corresponds the point where the main diagonal ($\Gamma \to M$) touches the \emph{phantom} arc of the hole pocket; it is clearly visible in Fig.~\ref{fig:NCAII} both (top panel) and (bottom left panel). It is striking the qualitative difference in the dependence of the imaginary part of the residual self-energy on $T^2$ in the two points: $S$ \emph{lives} in an ordinary Fermi liquid, where the dependence on temperature is exactly quadratic; $\underline{S}$ belongs to a portion of the system interested to a non-Fermi liquid description according to its dependence far from quadratic. Also the dependence of frequency (not shown) turn out to be almost perfectly quadratic in $S$ and with strong linear components in $\underline{S}$. This unconventional behaviour can be at the basis of the anomalous features shown by resistivity in these materials.

\subsection{Four-pole solution \label{sec:Hub-4pole}}

In the previous section, we have shown how the 2-pole solution can be improved by taking into account residual dynamical corrections. However, as discussed in Sect.~\ref{sec:Sigma}, the 2-pole solution can be also improved by enlarging the basis through the addition of higher-order composite operators. This aspect is illustrated in this Section.

As shown in Sect.~\ref{sec:Hub-2pole-Fer}, the higher order field $\pi (i)$ appears in the equations of motion of $\xi (i)$ and $\eta(i)$. The effect of this field was approximately taken into account by projecting it on the basis (2.2). Here we promote this field to the rank of new composite operator. Actually, we divide $\pi (i)$ into two operators $\pi (i) = \xi _s (i) + \eta _s (i)$, similarly to what we have done for the electronic field [$c(i) = \xi (i) + \eta (i)$], defined as 
\begin{equation}
\xi _s (i) = \frac{1}
{2}\sigma ^\mu n_\mu (i)\xi ^\alpha (i) + \eta ^{\dagger\alpha} (i)c(i)c(i)\quad \quad \eta _s (i) = \frac{1}
{2}\sigma ^\mu n_\mu (i)\eta ^\alpha (i) + \xi ^{\dagger\alpha} (i)c(i)c(i)
\end{equation}
and take as new basis
\begin{equation}
\psi (i) = \left(
\begin{array}{c}
\xi (i) \\
\eta (i) \\
\xi _s (i) \\
\eta _s (i)
\end{array}
\right)
\end{equation}

The reason for such a choice is that this basis is the fermionic closed basis for the Hubbard model on two sites \cite{Avella_01}. In such a case, $\xi _s (i)$ and $\eta _s (i)$ are eigenoperators of the Hamiltonian with eigenenergies given by $E_3 (\mathbf{k}) = - \mu + t\alpha (\mathbf{k}) - J_U$ and $E_4 (\mathbf{k}) = - \mu + t\alpha (\mathbf{k}) + U + J_U$. It is worth noting that the energy term $J_U = [\sqrt {U^2 + 16t^2 } - U]/2$ returns, in the limit $U/t \gg 1$, the virtual exchange energy $J = 4t^2 /U$.

COM, as formulated in Sect.~\ref{sec:COM}, requires as first step the calculation of the normalization $I$ and energy $\varepsilon$ matrices ($4 \times 4$ matrices). The detailed expression of the matrix $I$ is reported in Ref.~\cite{Odashima_05}. In order to effectively perform calculations, the elements of $I$ which contain three-site correlation functions of the form
\begin{multline}
\langle A^\alpha(\mathbf{i}) B^\alpha(\mathbf{i}) C(\mathbf{i}) D(\mathbf{i}) \rangle = \frac{1}{4} \langle (A(\mathbf{i})B(\mathbf{i}))^\alpha C(\mathbf{i}) D(\mathbf{i}) \rangle +\\
+ \sum_{\mathbf{j} \ne \mathbf{k}} {} \alpha_\mathbf{ij} \alpha _\mathbf{ik} \langle A(\mathbf{j}) B(\mathbf{k}) C(\mathbf{i}) D(\mathbf{i}) \rangle
\end{multline} 
have been computed through the following decoupling procedure: (i) the first term is reducible into two-site correlation functions, which can be directly evaluated in terms of the propagators under analysis; (ii) the second one has been decoupled in terms of two-site correlation functions. This procedure preserves the particle-hole symmetry enjoined by the Hamiltonian.

The energy matrix can be straightforwardly calculated by means of the anticommuting algebra given in Tab.~\ref{tab:comm_eta_xi}. However, due to the complexity of the equations of motion of the fields $\xi_s(i)$ and $\eta _s(i)$, a direct calculation will lead to the appearance of an enormous number of unknown correlation functions. In order to determine all correlation functions, we would then be forced to use some uncontrolled decoupling procedure and we would completely lose every control over the approximation procedure. Hence, we have opted for a controlled approximation at the level of equations of motion by neglecting irreducible three-site operators and paying attention to evaluating exactly all one- and two- site components. By means of this approximation procedure, the matrix $\varepsilon$ can be easily computed. Then, we have all ingredients, following the procedure reported in Sect.~\ref{sec:Self-cons} and \ref{sec:Hub-2pole-Fer}, to calculate all single-particle and thermodynamic properties.

\paragraph{Results and comparisons \label{sec:Hub-4pole-Res}}

\begin{figure}[tbp]
\centering
\includegraphics[width=.49\textwidth]{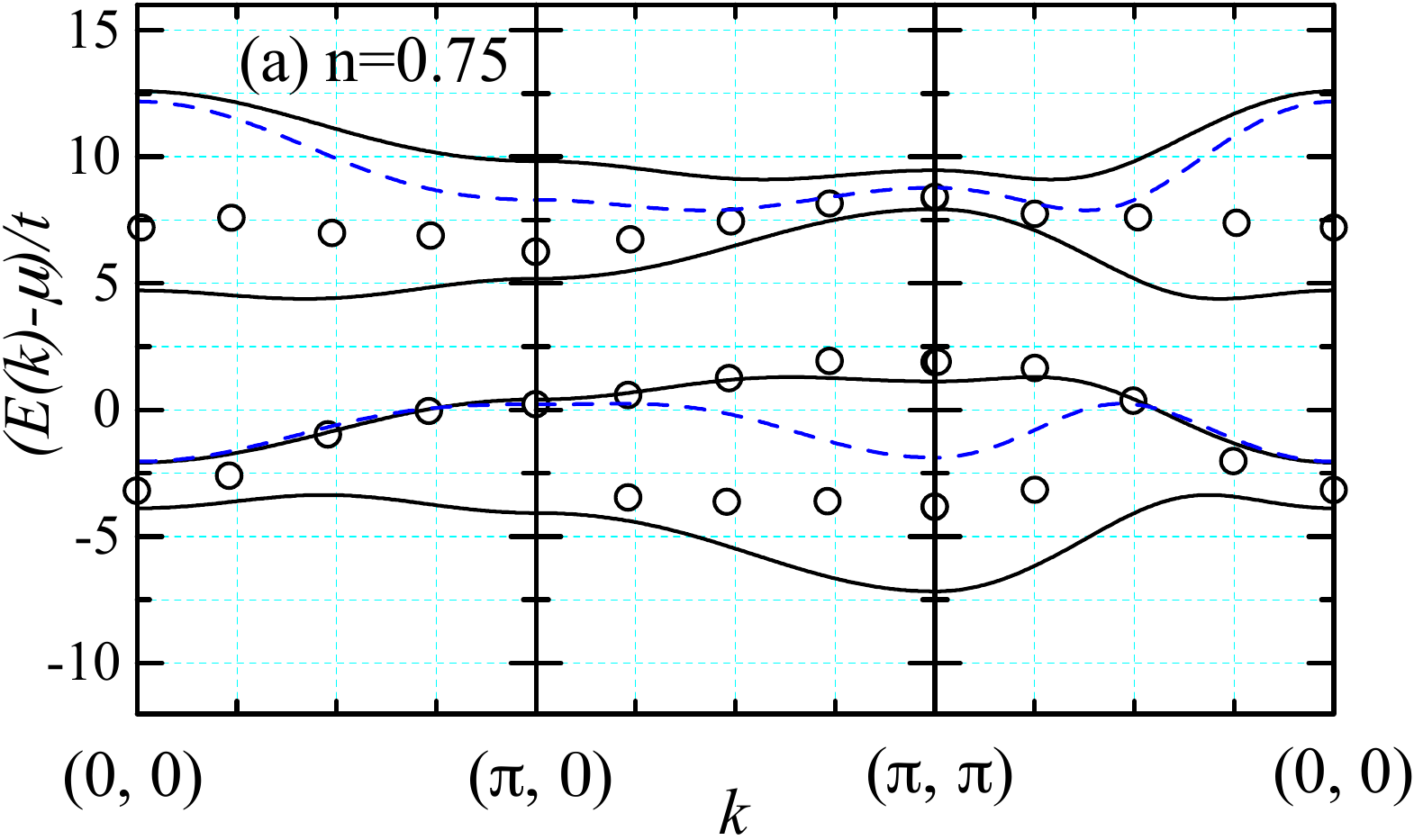}
\includegraphics[width=.49\textwidth]{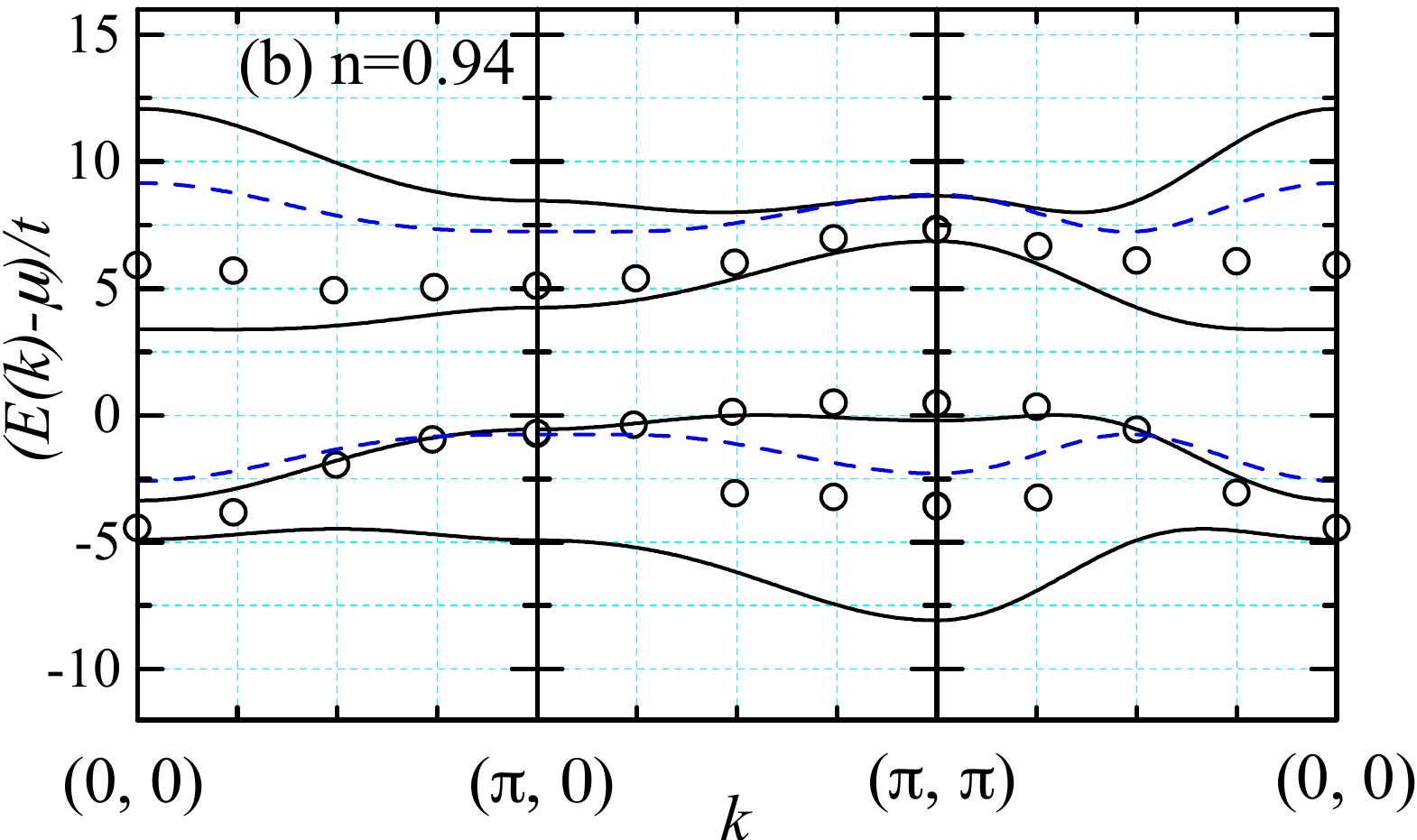}
\caption[]{The dispersion relation at $U/t=8$, $T/t=0.5$ and (left) $n=0.75$ and (right) $0.94$. The 2-pole solution (dashed line) and QMC data (circle) of Ref.~\cite{Bulut_94} are also reported.}
\label{fig:4-1}
\end{figure}

\begin{figure}[tbp]
\centering
\includegraphics[width=.49\textwidth]{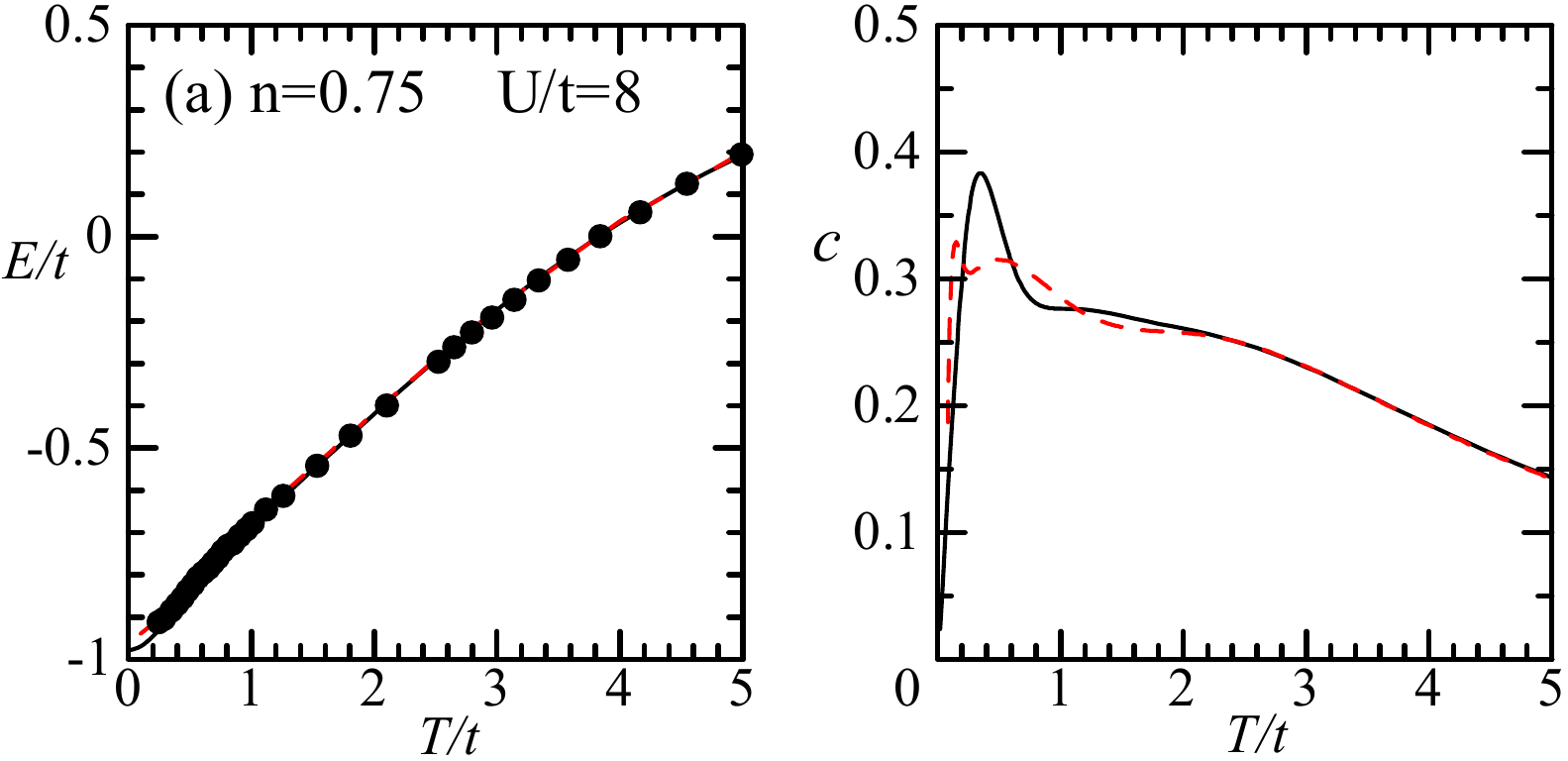}
\includegraphics[width=.49\textwidth]{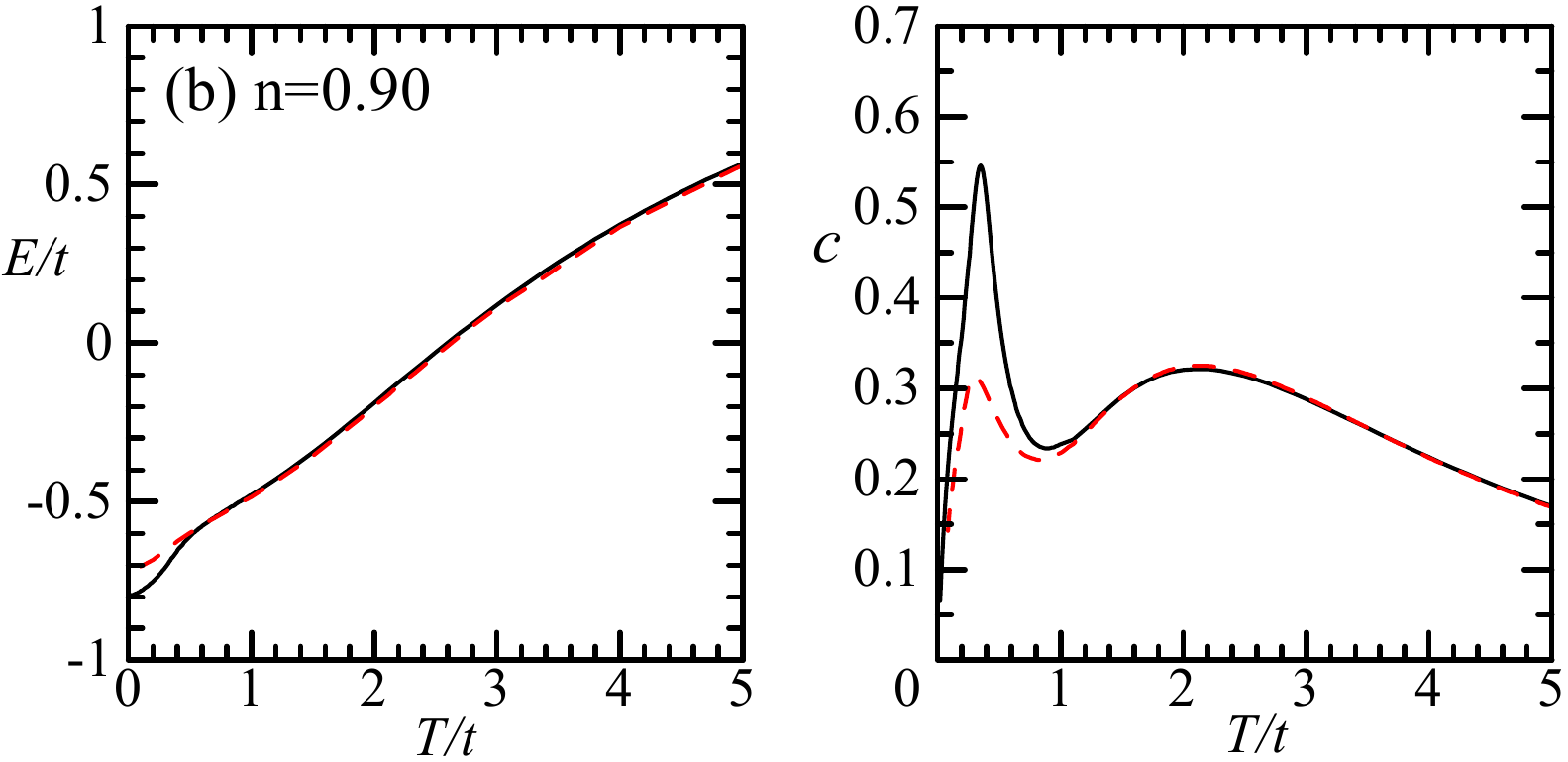}
\caption[]{Internal energy $E$ and specific heat $C$ per site at $U/t=4$ and $8$ and (left) $n=0.75$ and (right) $0.90$. Data from finite temperature Lanczos \cite{Bonca_02} and QMC \cite{Duffy_97,Duffy_97a} are provided for comparison.}
\label{fig:4-2}
\end{figure}

In Fig.~\ref{fig:4-1}, we provide a detailed comparison of the band structure obtained by the present formulation with QMC results \cite{Bulut_94}. As can be easily seen, we have a good agreement with the QMC data, especially for the low-energy band around the Fermi level. Internal energy $E$ and specific heat $C$ per site at $U/t=4$ and $8$ and $n=0.75$ and $0.90$ are reported in Fig.~\ref{fig:4-2}. Data from finite temperature Lanczos \cite{Bonca_02} and QMC \cite{Duffy_97,Duffy_97a} are provided for comparison. As regards the internal energy, the agreement with the Lanczos data is excellent except for the low temperature region at $U/t=8$. As regards the specific heat, we observe a sharp peak around $T/t=0.3$ and a fairly broad peak in the higher temperature region $T/t=1$. The two-peak structure is more pronounced for $U/t=8$, but not so	much for $U/t=4$. This tendency	is also	observed in	several numerical simulations \cite{Bonca_02,Duffy_97,Duffy_97a,Paiva_01}. Usually, the sharp peak at lower temperatures and the broad peak at higher temperatures are interpreted as consequences of spin and charge fluctuations related to the energy scales of $J$ and $U$, respectively. The main difference between our results and numerical ones regards the height of the peak in the specific heat	around	$T/t=0.3$ that comes from the decrease in the internal energy. This is an indication of well established spin ordering which cannot be correctly evaluated on a small cluster. Numerical simulation cannot describe spin and charge ordering in the case that the correlation lengths exceed the cluster size. On the other hand, in our formulation, there is a tendency to have too pronounced spin and charge correlations.

\subsection{Superconducting solution \label{sec:Hub-dwave}}

The various solutions of the Hubbard model presented up to now refer to a paramagnetic homogeneous state. However, the same formulation can be easily applied to states with completely different symmetry properties. Ferromagnetic, antiferromagnetic and superconducting solutions of the model have been reported in Ref.~\cite{Mancini_04}. As an example, we here summarize the superconducting solution of the model in the d-wave channel.

We consider the Nambu representation and introduce the 4-vector composite field \cite{DiMatteo_97,Avella_97a,Mancini_04}
\begin{equation}
\psi (i) = \left(
\begin{array}{c}
\xi _ \uparrow (i) \\
\eta _ \uparrow (i) \\
\xi _ \downarrow ^\dagger (i) \\
\eta _ \downarrow ^\dagger (i)
\end{array}
\right)
\end{equation}

At a first level of approximation, the current of this operator is projected on the basis and the residual current is neglected. The effects of the residual current have also been analysed, but they will not be presented here. Although the model admits various superconducting phases with different symmetries of the order parameter (see Ref.~\cite{Mancini_04}), on the basis of experimental evidence for high T$_c$ superconductors, we restrict the analysis to singlet pairing and d-wave symmetry for a two-dimensional lattice. Then, we impose as boundary conditions $\langle \psi _ \uparrow (i)\psi _ \downarrow (i) \rangle = 0$ and $\langle \psi _ \uparrow (i)\psi _ \downarrow ^\alpha (i) \rangle = 0$. Furthermore, we assume translational and spin rotational invariance. Accordingly, the normalization and m matrices have the expressions
\begin{equation}
I(\mathbf{k}) = \left(
\begin{array}{cccc}
I _{11} & 0 & 0 & 0 \\
0 & I _{22} & 0 & 0 \\
0 & 0 & I _{11} & 0 \\
0 & 0 & 0 & I _{22}
\end{array}
\right) \quad \quad
\begin{array}{l}
I_{11} = 1 - \frac{n}{2} \\
I_{22} = \frac{n}{2}
\end{array} 
\end{equation}
\begin{equation}
m = \left(
\begin{array}{cc}
{m _1 } & {m _2 } \\
{m _2 } & { - m _1 }
\end{array}
\right) \quad \quad
m _1 = \left(
\begin{array}{cc}
{m _{11} } & {m _{12} } \\
{m _{12} } & {m _{22} }
\end{array}
\right) \quad \quad
m _2 = m _{13} \left(
\begin{array}{cc}
1 & { - 1} \\
{ - 1} & 1
\end{array}
\right) \;,
\end{equation}
with
\begin{align}
& m _{11} (\mathbf{k}) = - \mu I_{11} - 4t\Delta - 4t\alpha (\mathbf{k})(1 - n + p) \\
& m _{12} (\mathbf{k}) = 4t\Delta + 4t\alpha (\mathbf{k})(p - I_{22} ) \\
& m _{13} (\mathbf{k}) = 4t\gamma (\mathbf{k})f \\
& m _{22} (\mathbf{k}) = (U - \mu )I_{22} - 4t\Delta - 4t\alpha (\mathbf{k})p
\end{align} 
and $\gamma (\mathbf{k}) = [\cos (k_x a) - \cos (k_y a)]/2$. The parameters $\Delta$ and $p$ have been defined in Sect.~\ref{sec:Hub-2pole-Fer}, the parameter $f$ plays the role of order parameter and is defined as $f = \langle c_ \downarrow ^\dagger (i)c_ \downarrow (i)[c_ \downarrow (i)c_ \uparrow (i)]^\gamma \rangle + \langle [c_ \uparrow ^\dagger (i)c_ \uparrow (i)]^\gamma c_ \downarrow (i)c_ \uparrow (i) \rangle$. The retarded and correlation functions, which are now $4 \times 4$ matrices containing the anomalous components, are then known functions of the internal parameters $\mu$, $\Delta$, $p$ and $f$. The self-consistent determination of these parameters is performed by means of the equations given in Sect.~\ref{sec:Hub-2pole-Fer} and by using a simple decoupling procedure for the latter.

\paragraph{Results and comparisons \label{sec:Hub-2pole-Bos-Res}}

\begin{figure}[tbp]
\sidecaption[t]
\centering
\includegraphics[width=.49\textwidth]{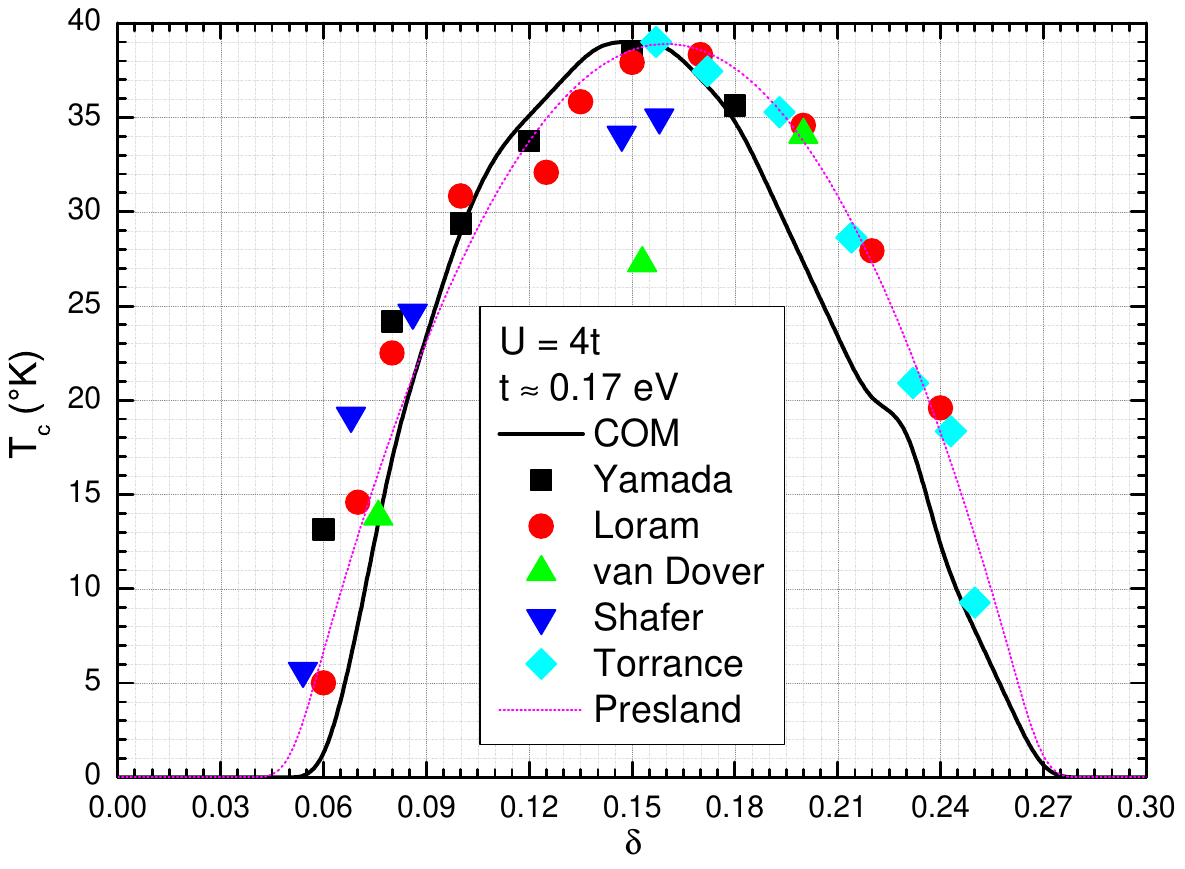}
\caption[]{Critical temperature $T_c$ as a function of the doping $\delta$ for $U=4t$ and $t\approx 0.17 eV$. Experimental data for $La_{2-\delta}Sr_{\delta}CuO_4$ from Yamada \cite{Yamada_96}, Loram \cite{Loram_89}, van Dover \cite{vanDover_87}, Shafer \cite{Shafer_87} and Torrance \cite{Torrance_88}. Phenomenological curve from Presland \cite{Presland_91}.}
\label{fig:Tc}
\end{figure}

In Fig.~\ref{fig:Tc}, we report the critical temperature $T_c$ as a function of the doping $\delta$ for $U=4t$ and $t\approx 0.17 eV$. The experimental data for $La_{2-\delta}Sr_{\delta}CuO_4$ are taken from Yamada \cite{Yamada_96}, Loram \cite{Loram_89}, van Dover \cite{vanDover_87}, Shafer \cite{Shafer_87} and Torrance \cite{Torrance_88}. The phenomenological curve suuggested by Presland \cite{Presland_91} is also reported. The agreement is really very good: the position and presence of \emph{endpoints} (critical fillings at zero temperature), the position of the \emph{peak} (critical filling at highest temperature) and the overall shape. This clearly shows that, together with the very remarkable capability to describe the very unconventional physics of the underdoped region (see Sect.~\ref{sec:Hub-Sigma}), COM manages to catch many relevant aspects of cuprate physics.

\section{Conclusions and Outlook}

Developing new methods and techniques to properly take into account electron-electron interaction in many-body systems is certainly one of the most challenging problems in theoretical physics in the last decades. Among the various methods proposed and reported in this volume, the Composite Operator Method is a general method to study strongly correlated systems. The formalism, presented in a systematic way, is based on two main concepts: the use of the propagators of composite operators as building blocks of a perturbation calculation; use of algebra constraints to fix the representation of the Green's functions in order to keep the algebraic and symmetry properties of the Hamiltonian. Although the method has been used for the study of several models, we preferred just to present its application to the Hubbard model. The presentation has been given by going through a scheme of successive approximations, reporting at each stage a comparison with the results of exact and numerical data. Some comparison with the experimental data for high T$_c$ cuprates has also been reported. It emerges that the Composite Operator Method can be considered as a powerful method to obtain information regarding the features of complex strongly correlated Hamiltonian systems and materials mimed by them.

\printindex

\end{document}